\documentclass[10pt]{iopart}
\usepackage{graphicx,setstack}
\usepackage[colorlinks,linkcolor=blue,citecolor=blue,urlcolor=blue ]{hyperref}
\begin{document}

\newcommand{\be}{\begin{equation}}
\newcommand{\ee}{\end{equation}}
\newcommand{\bq}{\begin{eqnarray}}
\newcommand{\eq}{\end{eqnarray}}
\newcommand{\bsq}{\begin{subequations}}
\newcommand{\esq}{\end{subequations}}
\newcommand{\bc}{\begin{center}}
\newcommand{\ec}{\end{center}}

\review[The status of varying constants]{The status of varying constants: a review of the physics, searches and implications}
\author{C J A P Martins}
\address{Centro de Astrof\'{\i}sica da Universidade do Porto, Rua das Estrelas, 4150-762 Porto, Portugal\\
and Instituto de Astrof\'{\i}sica e Ci\^encias do Espa\c co, CAUP, Rua das Estrelas, 4150-762 Porto, Portugal}
\ead{Carlos.Martins@astro.up.pt}
\vspace{10pt}
\begin{indented}
\item[]24 January 2017
\end{indented}

\begin{abstract}
The observational evidence for the recent acceleration of the universe demonstrates that canonical theories of cosmology and particle physics are incomplete---if not incorrect---and that new physics is out there, waiting to be discovered. A key task for the next generation of laboratory and astrophysical facilities is to search for, identify and ultimately characterize this new physics. Here we highlight recent developments in tests of the stability of nature's fundamental couplings, which provide a direct handle on new physics: a detection of variations will be revolutionary, but even improved null results provide competitive constraints on a range of cosmological and particle physics paradigms. A joint analysis of all currently available data shows a preference for variations of $\alpha$ and $\mu$ at about the two-sigma level, but inconsistencies between different sub-sets (likely due to hidden systematics) suggest that these statistical preferences need to be taken with caution. On the other hand, these measurements strongly constrain Weak Equivalence Principle violations. Plans and forecasts for forthcoming studies with facilities such as ALMA, ESPRESSO and the ELT, which should clarify these issues, are also discussed, and synergies with other probes are briefly highlighted. The goal is to show how a new generation of precision consistency tests of the standard paradigm will soon become possible.
\end{abstract}

\pacs{98.80.-k, 98.80.Es, 95.36.+x, 98.62.Ra, 95.55.-n}
\vspace{2pc}
\noindent{\it Keywords}: Observational cosmology, Fundamental physics, Fundamental couplings, Dark energy, Astronomical facilities
\submitto{\RPP}
\maketitle
\ioptwocol

\section{Introduction\label{intro}}

These are fortunate times for those working in observational cosmology or particle physics. On the one hand, both fields have standard models, which agree with a wide and fast-growing range of observational and experimental data. On the other hand, it is precisely our confidence in these models that leads us to expect that they are incomplete, and that currently unknown physics is waiting to be discovered. Importantly, we also have an unprecedented range of new tools with which to search for this new physics.

To put this into a wider perspective it is worth looking for a minute into the history of astronomy. In the middle of the XIX century Urbain Le Verrier and others mathematically discovered two new Solar System planets by insisting that the observed orbits of Uranus and Mercury should agree with the predictions of Newtonian physics. The first of these---which we now call Neptune (and was also predicted by John Couch Adams)---was soon observed in Berlin by Johann Galle and Heinrich d'Arrest. However, the second (dubbed Vulcan) was never found. We now know that the discrepancies in Mercury's orbit were a consequence of the fact that Newtonian physics can't fully describe Mercury's orbit. Explaining these discrepancies was the first major success of Einstein's General Relativity. Accounts of these two fascinating stories can be found in \cite{Neptune,VUlcan}. They carry an important lesson that is particularly relevant today: apparently similar questions can have totally different answers, so no matter how successful one particular approach has proven in tackling past problems there is no guarantee that it will work when facing a new problem.

Over the past several decades, cosmologists have mathematically discovered two new components of the universe, which we now call dark matter \cite{Rubin1,Rubin2} and dark energy \cite{SN1,SN2}, but so far these have not been directly detected in the laboratory---we don't yet have an unambiguous smoking gun. Whether each of them will turn out to be a new Neptune or a new Vulcan remains to be seen, but their mathematical discovery alone highlights the fact that the standard $\Lambda$CDM paradigm, despite its phenomenological success, is at least incomplete. Incidentally, a similar point could be made about inflation (in the sense that it also implies new and unknown physics).

Something similar applies to particle physics, whose standard model is older and (arguably) on even firmer ground than that of cosmology: neutrino masses, dark matter and the size of the baryon asymmetry of the universe all require new physics beyond the current model. It is particularly striking that all of these have obvious astrophysical and cosmological implications. One of the key lessons learned in the field in recent years is that further progress in fundamental particle physics will increasingly depend on progress in cosmology.

Before proceeding, it is useful to introduce a working definition of the term 'fundamental physics'. In this review we will follow \cite{Shaver} and define it to include two distinct but nevertheless inter-related aspects:
\begin{itemize}
\item {\bf Tests of fundamental laws and symmetries}: this includes tests of the Equivalence Principle in its various forms (see the detailed discussion in \cite{Will}), probing the behavior of gravity on all scales, understanding the structure and dimensionality of spacetime, and testing the foundations of quantum mechanics. Note that many of these principles are necessarily violated in extensions of the standard model: the spacetime structure is modified (violating Lorentz Invariance), fundamental couplings become dynamical, violating the Einstein Equivalence Principle (which we will discuss in detail in what follows), and gravity laws are modified at large and/or small scales.
\item {\bf Searches for Nature's fundamental constituents}: this includes scalar fields as an explanation for dark energy, new particles for dark matter, magnetic monopoles or fundamental strings. It also includes characterizing the constituents we already know, such as the Higgs or the masses of neutrinos. Most of these issues are reviewed in great detail in \cite{PDG}.
\end{itemize}

After a quest of several decades, the recent LHC discovery of the Higgs particle \cite{LHC1,LHC2} finally shows that fundamental scalar fields are part of Nature's building blocks. A pressing follow-up question is then whether the Higgs field has some cosmological role, or indeed if there are additional scalar fields that do have such a role. The fact that we don't yet know the answer to the latter question has not prevented cosmologists and particle physicists from speculating, and it is now a challenging task to find one active researcher in the field that has never used a scalar field at any point in his or her career.

Apart from their comparative simplicity, scalar fields are popular in cosmology because they can take a vacuum expectation value while preserving Lorentz Invariance. (By contrast, other degrees of freedom such as vector fields or fermions would break Lorentz Invariance, very quickly leading to conflicts with Special Relativity.) They often imply the presence of additional particles, which may lead to new physical effects (an example will be discussed in the following section) but can also be used to experimentally or observationally constrain the models. For this reason scalar fields now play a key role in most paradigms of modern cosmology, including
\begin{itemize}
\item The period of exponential expansion of the early universe (inflation) which is believed to have seeded the primordial density fluctuations that led to the large-scale cosmic structures we now observe \cite{Lyth}.
\item The dynamics of cosmological phase transitions and of their unavoidable relics known as cosmic defects, such as cosmic strings, monopoles, domain walls and various hybrid defects, as well as cosmic superstrings \cite{Vilenkin}.
\item Dynamical dark energy, an alternative to Einstein's cosmological constant for powering the current acceleration of the universe (and, arguably, a more likely one than a cosmological constant itself) \cite{Weinberg,Amendola}.
\item The spacetime variation of nature's fundamental dimensionless couplings, which is unavoidable in many extensions of the current standard model, and for which there is currently some tentative evidence. This will be the subject of this report.
\end{itemize}

If scalar fields are indeed behind any of these paradigms, this is of course reason enough to study them. But even more important than each of these is the fact that they don't occur alone: whenever a scalar field plays one of the above roles, it will unavoidably also leave imprints in other contexts that one can look for. Although this complementary point is often overlooked, it will be crucial for the future of precision cosmology, since it can be exploited in the form of consistency tests. Three simple examples will suffice to illustrate the point:
\begin{itemize}
\item In realistic models of inflation, the inflationary phase ends with a phase transition at which cosmic defects---most often strings---will form \cite{Sarangi,Jeannerot}, and the energy scales of both will therefore be unavoidably related.
\item Conformally stretched defect networks would in principle be natural dark energy or dark matter candidates \cite{Bucher}, although improved  observational constraints together with a better understanding of the evolution of these networks have now ruled out such scenarios.
\item In realistic models of dark energy, where the acceleration of the universe is due to a dynamical scalar field, this field will naturally couple to the rest of the model (unless some unknown symmetry is postulated to suppress the couplings) and thus lead to potentially observable variations of nature's fundamental couplings \cite{Carroll,Dvali,Chiba}.
\end{itemize}

A detailed exploration of the possibilities afforded by the latter connection (which will come to fruition with forthcoming astronomical facilities) is among the key recent developments in the field, and will be discussed in detail in the second half of this review. We all know that fundamental couplings {\it run} with energy, and in many (or arguably most?) extensions of current canonical models they will equally naturally {\it roll} in time and {\it ramble} in space (meaning that they will depend on the local environment). Therefore astrophysical as well as local (laboratory) tests of their stability provide us with key probes of fundamental cosmology, and in particular they can---by themselves or in combination with other cosmological probes---shed light on the enigma of dark energy. Whether current and future tests detect variations or simply find more strong upper limits, the improved sensitivity of the measurements will yield improved constraints on a range of cosmology and fundamental physics paradigms.

The aim of this review is to present a concise overview of the current status of these tests and their implications, as well as to highlight currently open issues or bottlenecks and describe forthcoming developments in the field that are likely to have a major impact (to the extent that such forecasts are possible). A canonical and far more extensive review of this topic has been written by Uzan in 2011 \cite{Uzan}, and for this reason I will make no attempt to provide an exhaustive 'historical' overview of the field. Instead I will concentrate on the key developments in the field in these last few years, both at the theoretical and the observational level, and in doing so I will occasionally rely on material of a shorter and more focused review which I wrote in 2014 \cite{GRG}.

As we will see in Section \ref{uveslp}, stability tests of several dimensionless couplings (as well as various combinations thereof) can be made. This review will mainly focus on the fine-structure constant
\be
\alpha\equiv\frac{e^2}{\hbar c}\,,
\ee
a measure of the strength of the electromagnetic interaction, both on the grounds that it is the more actively explored one and that a short but thorough review of tests of the stability of the proton-to-electron mass ratio,
\be
\mu\equiv\frac{m_p}{m_e}\,,
\ee
by Ubachs {\it et al.} appeared very recently \cite{Ubachs}. (Beware that some authors use the opposite definition for $\mu$, namely $m_e/m_p$.) Reading all of these is recommended for those wanting to gain a wider perspective of the field, not least because it will make it clear how active the field is and how fast progress has been in recent years.

Last but not least, the review will often highlight the work of my own Dark Side Team, so the reader may want to keep in mind that this is a possible bias. Similarly, in discussing forecasts for the performance of future ground and space-based astrophysical facilities, I will mostly focus on the ESO and ESA ones in which we are directly involved, such as ESPRESSO \cite{ESPRESSO}, the ELT \cite{EELT} (and in particular its high-resolution spectrograph, provisionally dubbed ELT-HIRES \cite{HIRES}), and more occasionally Euclid \cite{Euclid} or CORE \cite{COREgen}. While it is uncontroversial that these are world-class facilities that will play a leading role in future developments in the field, the reader should note that other non-European facilities can certainly also provide significant contributions---and indeed, it is important that they do.

Briefly, the structure of the rest of the review is as follows. Section 2 has a generic and conceptual introduction to possible variations of nature's fundamental couplings, and in particular discusses how variations of different couplings may be related. Section 3 describes in some detail the recent advances in spectroscopic tests of the stability of these couplings, including a meta-analysis of all currently available data allowing both for possible time (redshift) and space variations of these couplings. Section 4 provides a shorter discussion of other probes of the stability of these couplings, including local tests with atomic clocks, compact astrophysical objects, the cosmic microwave background and big bang nucleosynthesis. Sections 5 and 6 review possible models for varying couplings, as well as their current observational constraints. Section 5 describes three important and representative classes of models: Bekenstein and Runaway Dilaton models (both of which lead to time variations) and Symmetron models (which lead to environmental variations). Section 6 focuses on models where the same dynamical degree of freedom is responsible for both the varying couplings and the dark energy, and discussing both canonical models and an example of a non-canonical one. Section 7 briefly considers three complementary probes of the aforementioned tests, namely the evolution of the cosmic microwave background temperature, the distance duality relation, and the redshift drift. Finally Section 8 discusses some Fisher Matrix based forecasts of the improvements in sensitivity expected for forthcoming astrophysical facilities, and Section 9 offers some conclusions.

\section{Fundamental couplings\label{consts}}

Nature is characterized by a set of physical laws and fundamental dimensionless couplings, which historically we have assumed to be spacetime-invariant. For the former this is a cornerstone of the scientific method (indeed it's hard to imagine how one could do science at all if it were not the case), but for the latter it is only a simplifying assumption without further justification. These couplings ultimately determine the properties of atoms, cells, humans, planets and the universe as a whole, so it's remarkable how little we know about them. We have no 'theory of constants' that describes their role in physical theories or even which of them are really fundamental---see for example the 'trialogue' on the subject by Duff, Okun and Veneziano \cite{Duff}. In any case one thing is clear: if they do vary, all the physics we know is incomplete.

\subsection{Introducing varying couplings}

Fundamental couplings are indeed expected to vary in most extensions of the current standard model. In particular, this will be the case in theories with additional spacetime dimensions, such as string theory \cite{Polchinski1,Polchinski2}. In such paradigms the true fundamental constants of nature are defined in higher dimensions, while the $(3+1)$-dimensional 'constants' are only effective quantities, typically related to the true constants via characteristic sizes of the extra dimensions. Many explicit illustrations of these concepts have been discussed, including in 
\begin{itemize}
\item Kaluza-Klein models \cite{Chodos},
\item superstring theories \cite{WuWang}, and
\item brane world models \cite{Kiritsis}.
\end{itemize}

As an historical remark, it is interesting that the first generation of string theorists had the hope that the theory would ultimately predict a unique set of laws and couplings for low-energy physics. However, following the discovery of the evidence for the acceleration of the universe this claim was swiftly and pragmatically replaced by (to put it somewhat crudely) an 'anything goes' approach, often referred to as the multiverse and sometimes combined with anthropic arguments. Regardless of the merit of such approaches (which in the author's mind remains to be demonstrated), is clear that experimental and observational tests of the stability of fundamental couplings are probably their best route---and possibly even the only one---towards a testable prediction.

Clearly a detection of varying fundamental couplings will be revolutionary: it will immediately prove that the Einstein Equivalence Principle is violated (and therefore that gravity can't be a purely geometric phenomenon) and that there is a fifth force of nature \cite{Damour}. But even improved null results are important and indeed extremely useful. This can be simply understood by realizing that the natural scale for the cosmological evolution of one of these couplings, if driven by a canonical fundamental scalar field, would be the Hubble time. We would therefore expect a relative drift rate of the order of $10^{-10}$ per year. However, current local bounds, coming from laboratory atomic clock comparison experiments (which we will discuss in Section \ref{otherobs}), are already about six orders of magnitude stronger \cite{Rosenband}. Thus any such dynamical scalar field must in some sense be 'slow-rolling'---something that has obvious analogies with dark energy and inflation.

This leads us to a key point, which is again related to the relevance of improved null results. If no variations are seen at a certain level of sensitivity, should one make an effort to tighten these bounds? An analogy with dynamical dark energy provides the clearest way to understand the answer. Let's consider the present-day value of the dark energy equation of state, $w_0$, or more specifically $(1+w_0)$ which is the dynamically relevant quantity. Recall that for a canonical scalar field this is just the ratio of the square of the field speed to the field's total energy,
\be
1+w\equiv1+\frac{p}{\rho}=\frac{{\dot\phi}^2}{\rho}\,,
\ee
thus vanishing in the limit of a cosmological constant. Naively we would expect that a dynamical scalar field would have $(1+w_0)$ of order unity, but observationally we know that this must be $(1+w_0)<0.1$ \cite{PlanckParams} (depending on what data sets and priors are used). Now, the point is that if this number is not of order unity there is no natural scale for it: either there is some fine-tuning to make it small, or there is a new (currently unknown) symmetry which forces it to be zero.

An analogous argument can now be made for the fine-structure constant: a dynamical scalar field coupled to the electromagnetic sector of the Lagrangian will lead to $\alpha$ variations, and we would expect the dynamically relevant parameter, which is its relative variation
\be\label{defalpha}
\frac{\Delta\alpha}{\alpha}(z)\equiv\frac{\alpha(z)-\alpha_0}{\alpha_0}\,,
\ee
with $\alpha_0$ being its present-day value, to be of order unity, but observationally we know (as will be discussed in detail in Section \ref{uveslp}) that it must be less than $10^{-5}$. So if no variations are confirmed at the $10^{-6}$ level (parts per million, henceforth ppm) which corresponds to the current state-of-the-art sensitivity, is it worth pushing to even better (that is, numerically smaller) sensitivities? Certainly the answer is yes, and the Strong CP Problem in QCD clearly illustrates why: a parameter which we would have expected (given our present knowledge of particle physics) to be of order unity is known to be smaller than $10^{-10}$, leading to the postulate of the Peccei-Quinn symmetry \cite{Peccei}, which in turn leads to a range of further interesting consequences---including axions, an interesting though currently not favored dark matter candidate. Hence a sufficiently tight bound will either imply that there are no dynamical scalar fields fields in cosmology or that the couplings of the scalar field to the rest of the model are suppressed by some currently unknown symmetry of Nature---whose existence would be as significant as that of the original field.

\subsection{Relating different couplings}

It is also important to bear in mind that in theories where a dynamical scalar field yields a varying $\alpha$, the other gauge and Yukawa couplings are also expected to vary. In particular, in Grand Unified Theories the variation of $\alpha$ will be related to that of the energy scale of Quantum Chromodynamics, whence the nucleon masses will necessarily vary when measured in an energy scale that is independent of QCD, such as the electron mass. It follows that we should expect a varying proton-to-electron mass ratio, $\mu$, which can be probed with $H_2$ \cite{Thompson} as well as other molecules.

The specific relation between $\alpha(z)$ and $\mu(z)$ will be model-dependent---indeed, highly so---but this very fact makes this a unique discriminating tool between competing models. A very useful and quite generic parametrization of joint variations has been developed in \cite{Coc,Luo}, considering a class of grand unification models with the following assumptions:
\begin{itemize}
\item The weak scale is determined by dimensional transmutation,
\item The relative variations of all the Yukawa couplings are the same, and 
\item The variation of the couplings is driven by a dilaton-type scalar field, as in \cite{Campbell}.
\end{itemize}
With these simplifying but otherwise reasonable assumptions one can obtain the following relations \cite{Coc}
\begin{equation}
\frac{\Delta m_e}{m_e}=\frac{1}{2}(1+S)\frac{\Delta\alpha}{\alpha}
\end{equation}
\begin{equation}
\frac{\Delta m_p}{m_p}=[0.8R+0.2(1+S)]\frac{\Delta\alpha}{\alpha}\,.
\end{equation}
and therefore the variations of $\mu$ and $\alpha$ are related through
\begin{equation}
\frac{\Delta\mu}{\mu}=[0.8R-0.3(1+S)]\frac{\Delta\alpha}{\alpha}\,,
\end{equation}
where $R$ and $S$ are universal dimensionless parameters, respectively related to the strong and electroweak sectors of the model in question. To give just two examples, Coc {\it et al.} \cite{Coc} suggest typical values of $R\sim36$ and $S\sim160$, while in the dilaton-type model studied by Nakashima {\it et al.} \cite{Nakashima} we have $R\sim109$ and $S\sim0$. Additional discussion can be found in the review by Uzan \cite{Uzan}. At a phenomenological level, the choice $S=-1$, $R=0$ can also describe the limiting case where $\alpha$ varies but the masses do not. Further useful relations can be obtained \cite{flambaum2} for the gyromagnetic factors for the proton and neutron
\begin{equation}
\frac{\Delta g_p}{g_p}=[0.10R-0.04(1+S)]\frac{\Delta\alpha}{\alpha}\,
\end{equation}
\begin{equation}
\frac{\Delta g_n}{g_n}=[0.12R-0.05(1+S)]\frac{\Delta\alpha}{\alpha}\,.
\end{equation}
Relative variations of other quantities, including the neutron mass and lifetime and the deuteron binding energy can also be cast in this form, as discussed in detail in \cite{Coc}.

Any model in this class may therefore be phenomenologically characterized by its values of $R$ and $S$, and thus tested using astrophysical or laboratory measurements. Alternatively, $R$ and $S$ can simply be taken as free parameters to be constrained by the available data. It follows from this discussion that from a theoretical perspective it is highly desirable to identify astrophysical systems where various constants can be simultaneously measured, or systems where a constant can be measured in several independent ways---we will see examples of these in the following section. Systems where several combinations of constants can be measured are also interesting, and can provide useful consistency tests: an example is PKS1413$+$135, an edge-on radio source at redshift $z=0.247$ \cite{Ferreira13,FerreiraJiC}. Other consistency tests will be described in Section \ref{redundancy}.

\section{Recent spectroscopic measurements\label{uveslp}}

The idea behind spectroscopic measurements of dimensionless couplings---typically the fine-structure constant $\alpha$, the proton-to-electron mass ration $\mu$, the proton gyromagnetic ratio $g_p$, or combinations thereof---is in principle quite simple. If one accurately knows the laboratory wavelength of a particular atomic or molecular transition and observes said transition in an astrophysical system, its wavelength will be changed due to redshift effects (be they cosmological, gravitational or both). If in addition the physics in the region where the absorption or emission originated was different, there will be additional shifts, which may be to the red or to the blue depending on the detailed structure of the atom or molecule in question. How much each transition shifts, for a given variation of the coupling, is known as the transition's sensitivity coefficient.

Obviously, for a single transition in a single system the two effects are completely degenerate and cannot be separated. However, if one observes two or more transitions with different sensitivities to the coupling in question, and knows that these transitions were formed at the same physical location, then the degeneracy can be broken, and one is able to determine the redshift and the value of the coupling (or its relative variation as compared to the present-day value, cf. Eq. \ref{defalpha}) at the place where the transitions were produced. To give some examples, $\alpha$ can be measured by looking at fine-structure doublets, $\mu$ by comparing molecular Hydrogen vibration and rotation modes, and various products of $g_p$ with $\alpha$ and/or $\mu$ by comparing Hydrogen hyperfine transitions with rotational, fine-structure or optical ones.

That said, there are of course several practicalities to bear in mind, including
\begin{itemize}
\item One needs to have accurate measurements of the rest wavelengths of relevant transitions (indeed, up to a few years ago uncertainties in laboratory wavelengths provided the dominant part of the error budget of many measurements), and knowing the relative isotopic abundances of some species is also important; a recent compilation of this data for $\alpha$ measurements can be found in \cite{MurBer}.
\item Atomic physics calculations are needed to determine how much a transition will shift for a given variation of the relevant coupling(s); these quantities are known as sensitivity coefficients, and are typically known with uncertainties which are much better than one percent for $\mu$ (at least for $H_2$ and other common molecules) and in the range of less than one percent to about ten percent for $\alpha$, depending on the transition.
\item Not all transitions are sufficiently sensitive to variations, and only relatively few astrophysical systems are clean enough to provide accurate measurements: more than one hundred lines of sight enable $\alpha$ measurements, though only a fraction of them are ideal (more on this in Section \ref{forecasts}) and in any case atomic sensitivity coefficients are typically small; conversely molecular Hydrogen or other molecules enabling $\mu$ measurements are far less common, but several molecular transitions are highly sensitive to variations \cite{Kozlov}. 
\item Emission line measurements are in general more straightforward than absorption ones (and possibly less vulnerable to some systematics), but much less sensitive---the best available emission constraints on $\alpha $ are in \cite{Albareti}; for this reason, in most of what follows we will focus on absorption line measurements.
\end{itemize}

Much of the recent interest and activity in this field emerged as the result of the work of Webb {\it et al.} over the last to decades. In particular, their most recent work suggests, at more than four-sigma level of statistical significance, a ppm spatial variation of the fine-structure constant $\alpha$ at low redshifts (roughly $1<z<4$) \cite{Webb}. Their data set contains a total of 293 archival measurements from the HIRES and UVES spectrographs, respectively at the Keck and VLT telescopes. The data is unable to distinguish between a purely spatial dipole and one with an additional dependence on look-back time (both provide equally good statistical fits to the data, at just above the four-sigma level). Although theoretical models that may explain such a result seem to require some amount of fine-tuning (a point to which we will return in Section \ref{models}), there is also no identified systematic effect that is able to fully explain it. Nevertheless some concerns do exist \cite{Whitmore}, since it is clear that there are some systematics in the data that at present have not been fully modeled or corrected for---though work on this is ongoing \cite{Dumont}.

\subsection{The UVES Large Program and beyond}

A specific cause for caution regarding the results of \cite{Webb} is that it is based on archival data, meaning that the data was originally taken for other purposes---by a large number of different observers, under a broad range of observing conditions, and over a time span of almost a decade---and subsequently re-analyzed for this purpose. Thus although the data set is quite large with 293 absorption systems in total, roughly half coming from each telescope (a few of them were observed by both of them), the data acquisition procedures were far form ideal, particularly regarding the key issue of wavelength calibration.

Trying to confirm these results was the main motivation for an ESO UVES Large Program (Program L 185.A-0745, PI: Paolo Molaro). This is so far the only large program dedicated to tests of the stability of fundamental couplings, with an optimized sample and methodology. The program consisted of about 40 VLT nights, with observations in the period 2010-13, partly in service and partly in visitor mode. Key improvements in the data acquisition include obtaining calibration lamp exposures attached to science exposures (without resetting the cross-disperser encoding the position for each exposure) and observing bright (magnitude 9-11) asteroids at twilight, to monitor the radial velocity accuracy of UVES and the optical alignments \cite{Asteroids}. The collaboration includes members from all active observational groups---another of our key goals is to compare, check and optimize the different analysis pipelines currently being used by different groups, including the introduction of blind analysis techniques.

With 40 VLT nights one can only observe a relatively small sample. Criteria for the sample selection included the presence of multiple absorption systems, brightness, relatively high redshift, simplicity of the spectrum, narrow components at sensitive wavelengths, and no line broadening/saturation. The preference for high redshift stems both from observational reasons (so that the FeII1608 transition can be observed: this is desirable because it has a large negative sensitive coefficient, while other Iron lines have large positive sensitivity coefficients) and from theoretical reasons (other things being equal, it leads to stronger constraints on dark energy, as will be seen in Sections \ref{darkside} and \ref{forecasts}). Typically the spectral resolution of the data is around $R\sim60000$ and the signal-to-noise per pixel $S/N\sim100$. This led us to an expectation of a potential accuracy of $1-2$ ppm per system, where photon noise and calibration errors are comparable, and thus an overall goal of 2 ppm per system and 0.5 ppm for the full sample.

The list of astrophysical targets to be observed in the Large Program was selected was before the dipole indications of Webb {\it et al.} were known. Therefore, the selected sample is not optimized to test this dipole (at least in the strict sense that none of the observed targets is near the north pole of the best-fit dipole direction). The sample consists of 13 lines of sight for $\alpha$ measurements, and 2 lines of sight for $\mu$ measurements. Note that in the former case the lines of sight often include several absorption systems at different redshifts, each of which may lead to a separate measurement. These are particularly useful for testing for hypothetical dependencies on look-back time. A more detailed description of the Large Program sample may be found in \cite{Bonifacio}.

The first complete quasar spectrum analyzed was that of HE 2217-2818 \cite{LP1}, which includes 5 absorption systems at redshifts $z_{\rm abs} = 0.787$, 0.942, 1.556, 1.628 and 1.692. It was found that the most precise result is obtained for the absorber at $z_{\rm abs} = 1.692$, where 3 Fe II transitions and Al II ${\lambda}$1670 have high S/N and provide a wide range of sensitivities to $\alpha$. The final result for the relative variation in this system is
\begin{equation}
\frac{\Delta\alpha}{\alpha}=+1.3\pm2.4_{\rm stat}\pm{1.0}_{\rm sys}\, {\rm ppm}\,, \label{paperlp1}
\end{equation}
one of the tightest current bounds from an individual absorber. There is no evidence for variation in $\alpha$ at the 3 ppm precision level (at the 1${\sigma}$ confidence level). If the dipolar variation of Webb {\it et al.} \cite{Webb} is correct, the expectation at this sky position is $(3.2-5.4)\pm1.7$ ppm depending on whether one assumes a pure spatial dipole or one with a further dependence on look-back time. The above constraint is not inconsistent with this expectation. 

The second Large Program result was an accurate analysis of the $H_2$ absorption lines from the $z_{\rm abs}=2.402$ damped Ly${\alpha}$ system towards HE 0027-1836 to constrain the variation of $\mu$ \cite{LP2}. A detailed cross-correlation analysis between 19 individual exposures, taken over three years, as well as the combined spectrum, was carried out to check the wavelength calibration stability. The presence of possible wavelength dependent velocity drifts was noticed, and available asteroid spectra taken with UVES close to these observations were used to confirm, quantify and correct for this effect. Using both linear regression analysis and Voigt profile fitting where ${\Delta}{\mu}/{\mu}$ is explicitly considered as an additional fitting parameter, the final corrected result was
\begin{equation}
\frac{\Delta\mu}{\mu}=-7.6\pm8.1_{\rm stat}\pm{6.3}_{\rm sys}\, {\rm ppm}\,, \label{paperlp2}
\end{equation}
consistent with the null result. It should be noted that intra-order and long-range distortions are not exclusive to the UVES spectrograph at the VLT, but have also been identified in HIRES at Keck and (to a lesser extent) in HARPS---a more detailed discussion of the impact of these distortions can be found in \cite{Whitmore}. 

In order to gain a better understanding of these distortions, the equatorial quasar HS 1549$+$1919 was observed with world's three largest optical telescopes: the VLT, Keck and, for the first time in such analyses, Subaru \cite{LP3}. By directly comparing these spectra to each other, and by `supercalibrating' them using asteroid and iodine-cell tests, long-range distortions of the quasar spectra's wavelength scales which would have caused significant systematic errors in the $\alpha$ measurements were detected and removed. For each telescope $\Delta\alpha/\alpha$ was measured in 3 absorption systems at redshifts $z_{\rm abs}=1.143$, 1.342, and 1.802. The nine measurements of $\Delta\alpha/\alpha$ were all found to be consistent with zero at the 2-$\sigma$ level, with 1-$\sigma$ statistical (systematic) uncertainties in the range 5.6--24 (1.8--7.0) ppm. They were also found to be consistent with each other at the 1-$\sigma$ level, allowing the calculation of a combined value for each telescope and, finally, a single value for this line of sight:
\begin{equation}
\frac{\Delta\alpha}{\alpha}=-5.4\pm3.3_{\rm stat}\pm{1.5}_{\rm sys}\, {\rm ppm}\,, \label{paperlp3}
\end{equation}
which again is consistent with both zero and with the best-fit dipole predictions for this line of sight. If one averages all the Large Program $\alpha$ results published so far, we obtain
\begin{equation}
\left(\frac{\Delta\alpha}{\alpha}\right)_{LP}=-0.6\pm1.9_{\rm stat}\pm{0.9}_{\rm sys}\, {\rm ppm}\,. \label{paperlpall}
\end{equation}
Thus while a full analysis of this sample is still in progress, the results so far already demonstrate the robustness and reliability at the 3 ppm level afforded by supercalibration techniques and the direct comparison of spectra from different telescopes. Analysis of the rest of the Large Program data set is currently ongoing.

Before moving on, let us pause for a moment to ask why these spectroscopic measurements of $\alpha$ and $\mu$ are so difficult, and why the issue of systematics features quite frequently in the discussion. The fact is that, while to some extent the spectroscopic velocity measurements in question are akin to finding exoplanets, they are  much harder in the fundamental physics context, both because one is dealing with much fainter sources (QSOs with magnitude 16 or fainter, rather than very bright nearby stars) and because only a few absorption lines are clean enough to be useful. In a nutshell, spectroscopic measurements of fundamental couplings require observing procedures---and indeed instruments---beyond current facilities. Despite their obvious success in other fields, spectrographs such as UVES, HARPS or Keck-HIRES were not built with this science case in mind and are far from optimal for it. One also needs customized data reduction pipelines, as well as careful wavelength calibration procedures. In particular, one must calibrate with laser frequency combs \cite{Li,Steinmetz}, rather than Th-Ar lamps or $I_2$ cells which are the currently standard methods.

These issues highlight the need for future more precise measurements, to be provided by a new generation of high-resolution, ultra-stable spectrographs like ESPRESSO for the VLT \cite{ESPRESSO} and ELT-HIRES for the ELT \cite{HIRES,Community}, which have these tests as a key science and design driver: they will significantly improve the precision of these measurements and, crucially, have a much better control over possible systematics. At lower redshifts, there will also be complemented by ALMA measurements---two recent white papers discussing the ALMA role are \cite{ALMA1,ALMA2}.

Despite these difficulties, significant progress is being made. In addition to the work in the Large Program (whose main long-term legacy will probably be a clearer understanding of what features an absorber should have in order to yield precise measurements and of how to optimally analyze the data), other improvements in sensitivity are being achieved. Chromium an Zinc transitions, which are not as common as Iron or Magnesium ones but are highly sensitive to variations and less vulnerable to long-range distortions in the wavelength calibration, can now provide competitive constraints on their own \cite{MalecNew}. A sensitivity better than 1 ppm has recently been achieved for an individual absorber in the line of sight of the bright quasar HE0515-4414 \cite{Kotus}, although this is likely to be the only target for which this is feasible until ESPRESSO becomes available. Finally, genetic algorithms are being used to develop automated analysis pipelines \cite{Bainbridge}, which should lead to significantly faster (and possibly also more objective) processing of the data.

\begin{figure*}
\begin{center}
\includegraphics[width=2.6in]{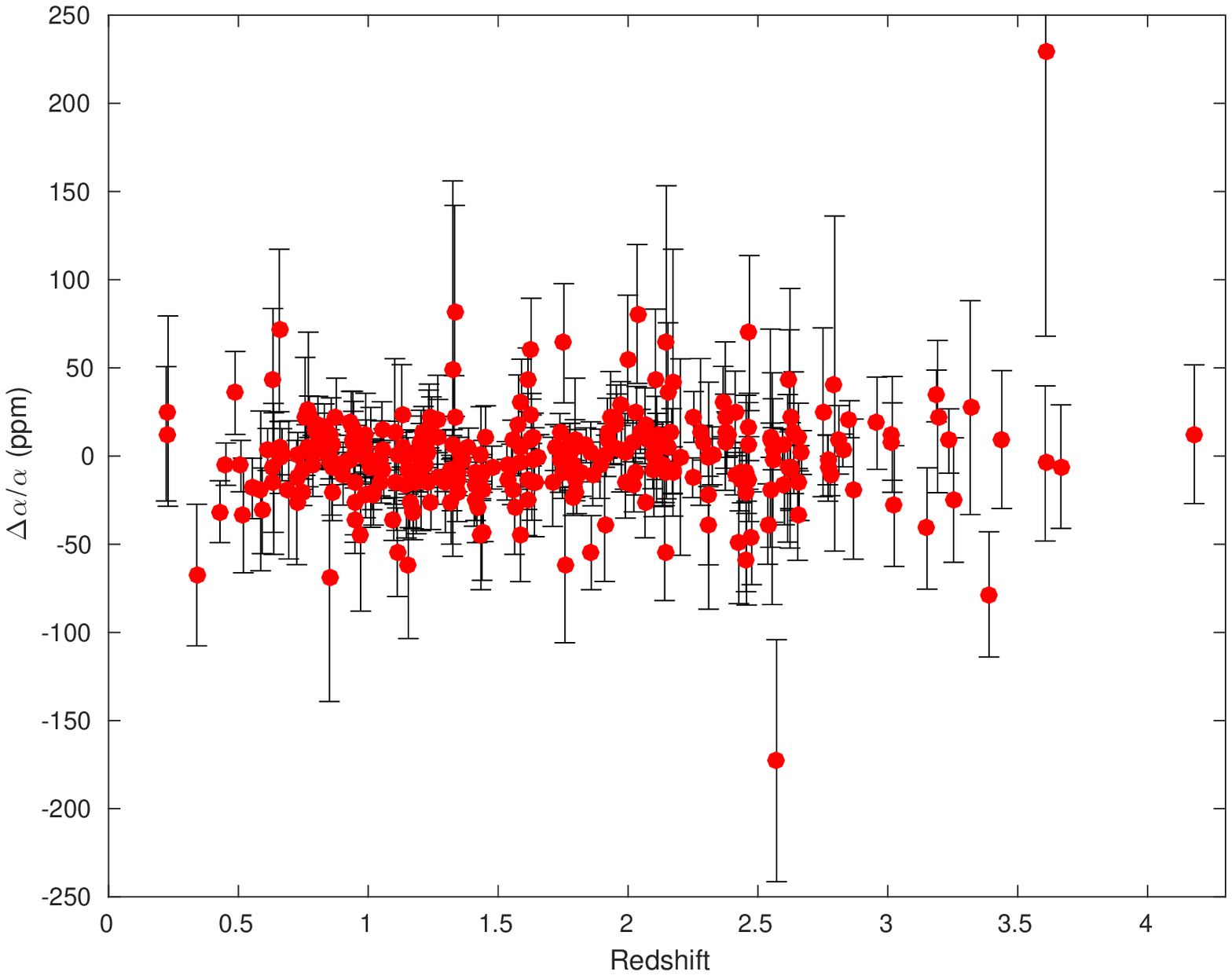}
\hskip0.2in
\includegraphics[width=2.6in]{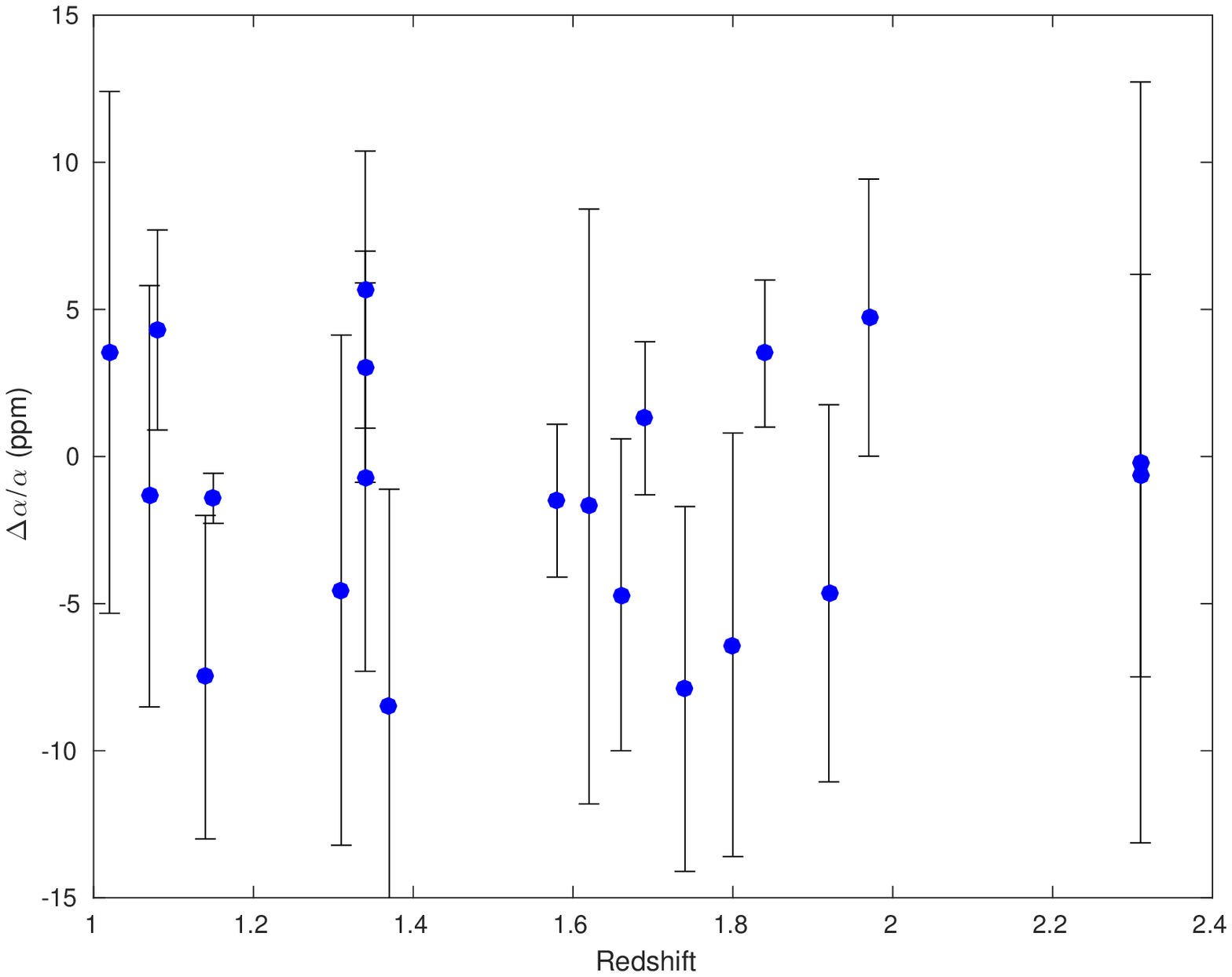}
\vskip0.25in
\hskip-0.1in
\includegraphics[width=2.6in]{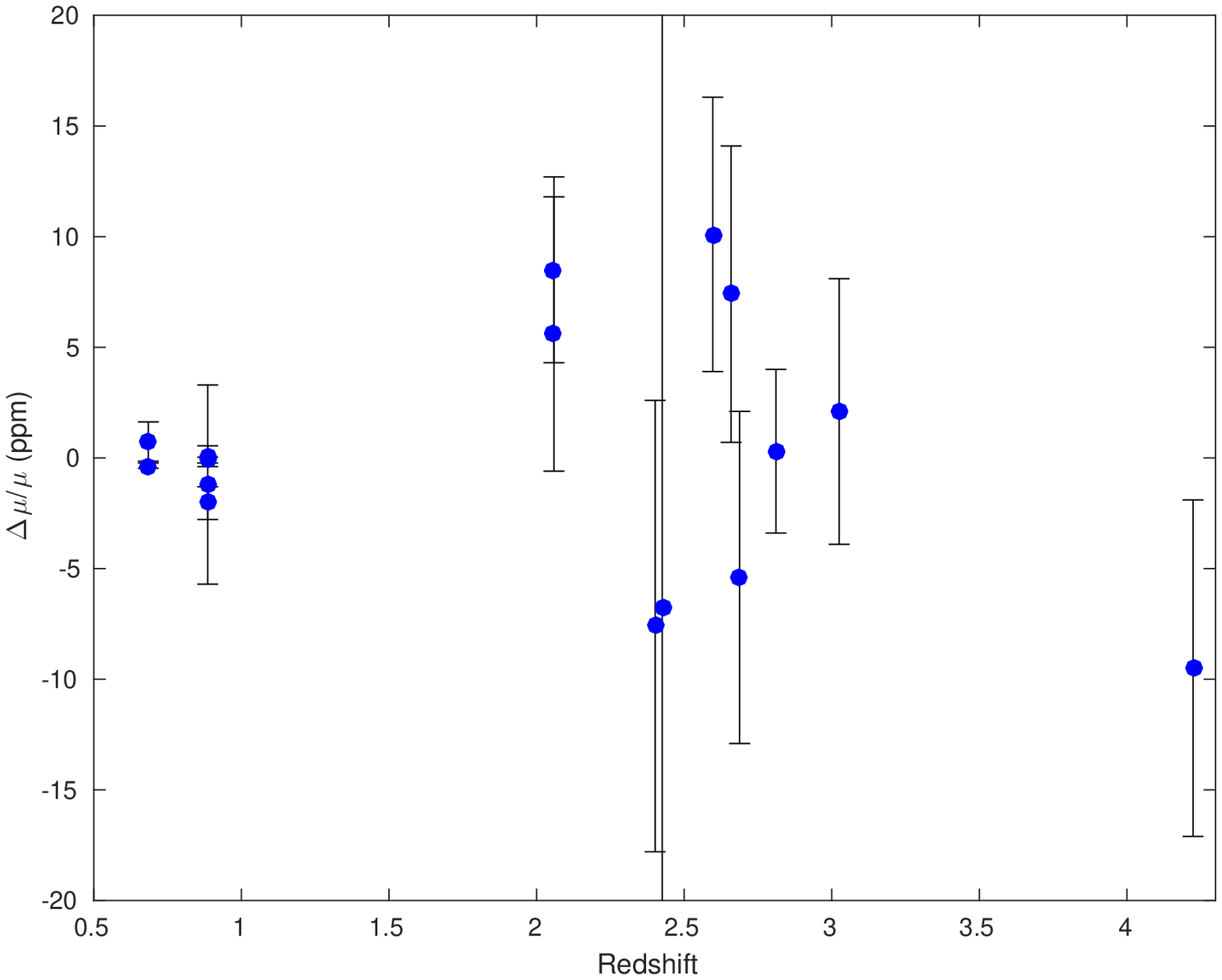}
\hskip0.2in
\includegraphics[width=2.6in]{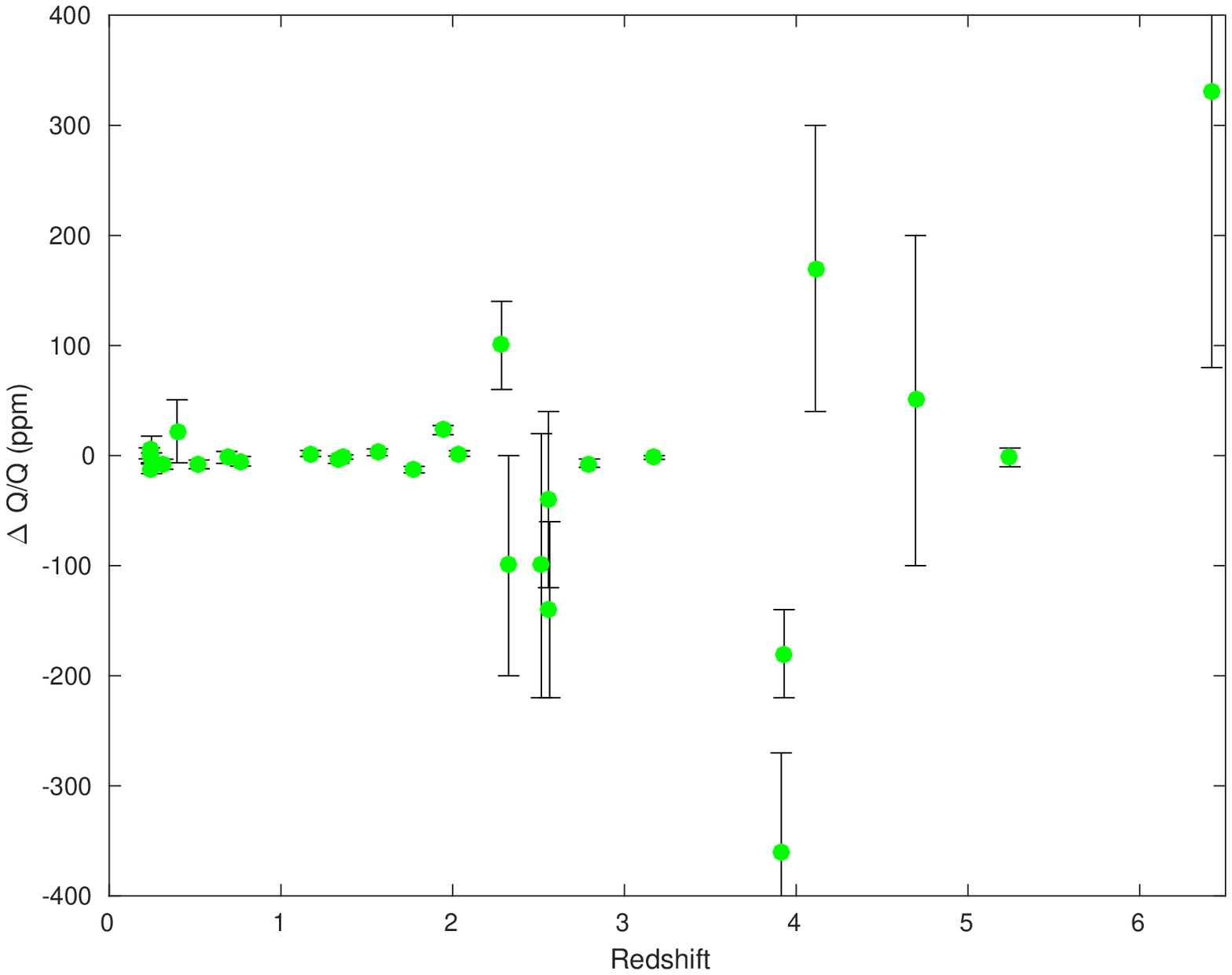}
\end{center}
\caption{\label{fig01}Currently available astrophysical measurements of fundamental couplings: the archival data set of Webb {\it et al.} \protect\cite{Webb} (top left), and dedicated measurements of $\alpha$ (top right, see Table \protect\ref{tab01}), $\mu$ (bottom left, see Table \protect\ref{tab02}) and combinations of parameters, generically denoted $Q$ (bottom right, see Table \protect\ref{tab03}). Note that both the horizontal and the vertical scales are different in each panel.}
\end{figure*}

\begin{table*}
\begin{center}
\begin{tabular}{|c|c|c|c|c|}
\hline
 Object & z & ${ \Delta\alpha}/{\alpha}$ (ppm) & Spectrographs & Reference \\
\hline\hline
J0026$-$2857 & 1.02 & $3.5\pm8.9$ & UVES & Murphy {\it et al.} (2016) \protect\cite{MalecNew} \\
\hline
J0058$+$0041 & 1.07 & $-1.4\pm7.2$ & HIRES & Murphy {\it et al.} (2016) \protect\cite{MalecNew} \\
\hline
3 sources & 1.08 & $4.3\pm3.4$ & HIRES & Songaila \& Cowie (2014) \protect\cite{Songaila} \\
\hline
HS1549$+$1919 & 1.14 & $-7.5\pm5.5$ & UVES/HIRES/HDS & Evans {\it et al.} (2014) \protect\cite{LP3} \\
\hline
HE0515$-$4414 & 1.15 & $-1.4\pm0.9$ & UVES & Kotus {\it et al.} (2017) \protect\cite{Kotus} \\
\hline
J1237$+$0106 & 1.31 & $-4.5\pm8.7$ & HIRES & Murphy {\it et al.} (2016) \protect\cite{MalecNew} \\
\hline
HS1549$+$1919 & 1.34 & $-0.7\pm6.6$ & UVES/HIRES/HDS & Evans {\it et al.} (2014) \protect\cite{LP3} \\
\hline
J0841$+$0312 & 1.34 & $3.0\pm4.0$ & HIRES & Murphy {\it et al.} (2016) \protect\cite{MalecNew} \\
J0841$+$0312 & 1.34 & $5.7\pm4.7$ & UVES & Murphy {\it et al.} (2016) \protect\cite{MalecNew} \\
\hline
J0108$-$0037 & 1.37 & $-8.4\pm7.3$ & UVES & Murphy {\it et al.} (2016) \protect\cite{MalecNew} \\
\hline
HE0001$-$2340 & 1.58 & $-1.5\pm2.6$ &  UVES & Agafonova {\it et al.} (2011) \protect\cite{alphaAgafonova}\\
\hline
J1029$+$1039 & 1.62 & $-1.7\pm10.1$ & HIRES & Murphy {\it et al.} (2016) \protect\cite{MalecNew} \\
\hline
HE1104$-$1805 & 1.66 & $-4.7\pm5.3$ & HIRES & Songaila \& Cowie (2014) \protect\cite{Songaila} \\
\hline
HE2217$-$2818 & 1.69 & $1.3\pm2.6$ &  UVES & Molaro {\it et al.} (2013) \protect\cite{LP1}\\
\hline
HS1946$+$7658 & 1.74 & $-7.9\pm6.2$ & HIRES & Songaila \& Cowie (2014) \protect\cite{Songaila} \\
\hline
HS1549$+$1919 & 1.80 & $-6.4\pm7.2$ & UVES/HIRES/HDS & Evans {\it et al.} (2014) \protect\cite{LP3} \\
\hline
Q1103$-$2645 & 1.84 & $3.5\pm2.5$ &  UVES & Bainbridge \& Webb (2016) \protect\cite{Bainbridge}\\
\hline
Q2206$-$1958 & 1.92 & $-4.6\pm6.4$ &  UVES & Murphy {\it et al.} (2016) \protect\cite{MalecNew}\\
\hline
Q1755$+$57 & 1.97 & $4.7\pm4.7$ & HIRES & Murphy {\it et al.} (2016) \protect\cite{MalecNew} \\
\hline
PHL957 & 2.31 & $-0.7\pm6.8$ & HIRES & Murphy {\it et al.} (2016) \protect\cite{MalecNew} \\
PHL957 & 2.31 & $-0.2\pm12.9$ & UVES & Murphy {\it et al.} (2016) \protect\cite{MalecNew} \\
\hline
\end{tabular}
\caption{\label{tab01}Available dedicated measurements of $\alpha$. Listed are, respectively, the object along each line of sight, the redshift of the measurement, the measurement itself (in parts per million), the spectrograph, and the original reference. The third measurement is the weighted average from 8 absorbers along the lines of sight of HE1104-1805A, HS1700+6416 and HS1946+7658, reported in \cite{Songaila} without the values for individual systems.}
\end{center}
\end{table*}

\subsection{A meta-analysis of all current data}

We now provide an overall summary of the current status of measurements of the various couplings. Whenever several measurements of the same source exist we list only the most recent one (which is almost always the most sensitive one), except in cases where they are done using different telescopes or different molecular species.

Table \ref{tab01} lists the recent dedicated measurements of $\alpha$, which are also plotted in the top right panel of Fig. \ref{fig01}; for comparison, the archival data of Webb {\it et al.} is shown in the top left panel of the same figure. We note that the weighted mean of the 21 measurements on the table is
\begin{equation}\label{prioralpha}
\left(\frac{\Delta\alpha}{\alpha}\right)_{New}=-0.64\pm0.65\, ppm\,,
\end{equation}
and thus compatible with the null result, unlike the archival data set, for which the weighted mean of its 293 measurements is nominally \cite{Webb}
\begin{equation}\label{priorwebb}
\left(\frac{\Delta\alpha}{\alpha}\right)_{Webb}=-2.16\pm0.86\, ppm\,.
\end{equation}

\begin{table*}
\begin{center}
\begin{tabular}{|c|c|c|c|c|}
\hline
 Object & z & ${\Delta\mu}/{\mu}$ & Method & Reference \\ 
\hline\hline
B0218$+$357 & 0.685 & $0.74\pm0.89$ & $NH_3$/$HCO^+$/$HCN$ & Murphy {\it et al.} (2008) \protect\cite{Murphy2} \\
B0218$+$357 & 0.685 & $-0.35\pm0.12$ & $NH_3$/$CS$/$H_2CO$ & Kanekar (2011) \protect\cite{Kanekar3} \\
\hline
PKS1830$-$211 & 0.886 & $0.08\pm0.47$ &  $NH_3$/$HC_3N$ & Henkel {\it et al.} (2009) \protect\cite{Henkel}\\
PKS1830$-$211 & 0.886 & $-1.2\pm4.5$ &  $CH_3NH_2$ & Ilyushin {\it et al.} (2012) \protect\cite{Ilyushin}\\
PKS1830$-$211 & 0.886 & $-2.04\pm0.74$ & $NH_3$ & Muller {\it et al.} (2011) \protect\cite{Muller}\\
PKS1830$-$211 & 0.886 & $-0.10\pm0.13$ &  $CH_3OH$ & Bagdonaite {\it et al.} (2013) \protect\cite{Bagdonaite2}\\
\hline
J2123$-$005 & 2.059 & $8.5\pm4.2$ & $H_2$/$HD$ (VLT)& van Weerdenburg {\it et al.} (2013) \protect\cite{vanWeerd} \\
J2123$-$005 & 2.059 & $5.6\pm6.2$ & $H_2$/$HD$ (Keck)& Malec {\it et al.} (2010) \protect\cite{Malec} \\
\hline
HE0027$-$1836 & 2.402 & $-7.6\pm10.2$ & $H_2$ & Rahmani {\it et al.} (2013) \protect\cite{LP2} \\
\hline
Q2348$-$011 & 2.426 & $-6.8\pm27.8$ & $H_2$ & Bagdonaite {\it et al.} (2012) \protect\cite{Bagdonaite} \\
\hline
Q0405$-$443 & 2.597 & $10.1\pm6.2$ & $H_2$ & King {\it et al.} (2008) \protect\cite{King} \\
\hline
J0643$-$504 & 2.659 & $7.4\pm6.7$ & $H_2$ & Albornoz-V\'asquez {\it et al.} (2014) \protect\cite{Albornoz} \\
\hline
J1237$+$0648 & 2.688 & $-5.4\pm7.5$ & $H_2$/$HD$ & Dapr\`a {\it et al.} (2015) \protect\cite{Dapra} \\
\hline
Q0528$-$250 & 2.811 & $0.3\pm3.7$ & $H_2$/$HD$ & King {\it et al.} (2011) \protect\cite{King2} \\
\hline
Q0347$-$383 & 3.025 & $2.1\pm6.0$ & $H_2$ & Wendt \& Reimers (2008) \protect\cite{Wendt} \\
\hline
J1443$+$2724 & 4.224 & $-9.5\pm7.6$ & $H_2$ & Bagdonaite {\it et al.} (2015) \protect\cite{Bagdonaite4} \\
\hline
\end{tabular}
\caption{\label{tab02}Available measurements of $\mu$. Listed are, respectively, the object along each line of sight, the redshift of the measurement, the measurement itself, the molecule(s) used, and the original reference. Low-redshift measurements were obtained with various facilities in the radio/mm band, while high-redshift ones were obtained in the UV/optical with the UVES spectrograph.}
\end{center}
\end{table*}

Table \ref{tab02} contains individual $\mu$ measurements, which are shown in the bottom left panel of Fig. \ref{fig01}. For a more detailed discussion of these measurements see also the review by Ubachs {\it et al.} \cite{Ubachs}. Note that several different molecules can be used \cite{Kozlov}, and in the case of the gravitational lens PKS1830$-$211 there are actually four independent measurements, with different levels of sensitivity. Currently ammonia is the most common molecule at low redshift, though others such as methanol \cite{Methanol1,Methanol2}, peroxide \cite{Peroxide}, hydronium \cite{Hydronium} and methanetiol (also known as methyl mercaptan) \cite{Methanetiol} have a greater potential in the context of facilities like ALMA, due to their large sensitivity coefficients \cite{Kozlov}. At higher redshifts, optical/near UV measurements are done using molecular hydrogen as first suggested in \cite{Thompson}. Carbon monoxide is less common but has sensitivity coefficients similar to those of molecular hydrogen \cite{Salumbides}, and can certainly provide important independent tests. The ultimate goal here is to find other molecules at higher redshifts, enabling optical and radio measurements in the same targets. Efforts in this direction are currently ongoing.

The tightest available constraint on $\mu$ comes precisely from PKS1830$-$211, from observations of methanol transitions \cite{Bagdonaite2}. We can similarly calculate the weighted mean of the low and high-redshift samples ($z<1$ and $z>2$ respectively), finding
\begin{equation}\label{priormulo}
\left(\frac{\Delta\mu}{\mu}\right)_{Radio}=-0.24\pm0.09\, ppm\,
\end{equation}
\begin{equation}\label{priormuhi}
\left(\frac{\Delta\mu}{\mu}\right)_{Optical}=2.9\pm1.9\, ppm\,;
\end{equation}
in both cases this is weak evidence for a variation, although note the preferred sign of this variation is different at high and low redshifts. Importantly, in this case the division between low and high redshift measurements is also a division between radio/mm and UV/optical measurements.

It is also worthy of note that while for molecular hydrogen one is indeed measuring $\mu$, for more complex molecules (which are often far more sensitive to $\mu$ variations than $H_2$ itself) one is actually measuring a ratio of an effective nucleon mass to the electron mass, and the relative variation of this quantity will only equal that of $\mu$ if there are no composition-dependent forces---in other words, if protons and neutrons have identical couplings to putative scalar fields. A test of this hypothesis could thus by carried out by finding a system where $\mu$ can be separately measured from different molecules with different numbers of protons and neutrons: for example $H_2$, $HD$, and perhaps also ammonia, methanol or carbon monoxide, which are all (comparatively) common molecules. This would be a revolutionary direct astrophysical test of the Weak Equivalence Principle, and it could in principle be done by ELT-HIRES \cite{HIRES}, provided it has a suitable wavelength coverage in the blue part of the spectrum.

\begin{table*}
\begin{center}
\begin{tabular}{|c|c|c|c|c|}
\hline
Object & z & $Q_{AB}$  & ${ \Delta Q_{AB}}/{Q_{AB}}$ & Reference \\ 
\hline\hline
J0952$+$179 & 0.234 & ${\alpha^{2}g_{p}/\mu}$ & $2.0\pm5.0$ & Darling (2012) \protect\cite{DarlingNew} \\
\hline
PKS1413$+$135 & 0.247 & ${\alpha^{2\times1.85}g_{p}\mu^{1.85}}$  & $-11.8\pm4.6$ & Kanekar {\it et al.} (2010) \protect\cite{Kanekar2} \\
PKS1413$+$135 & 0.247 & ${\alpha^{2\times1.57}g_{p}\mu^{1.57}}$  & $5.1\pm12.6$ & Darling (2004) \protect\cite{Darling} \\
PKS1413$+$135 & 0.247 & ${\alpha^{2}g_{p}}$  & $-2.0\pm4.4$ & Murphy {\it et al.} (2001) \protect\cite{Murphy} \\
\hline
J1127$-$145 & 0.313 & ${\alpha^{2}g_{p}/\mu}$ & $-7.9\pm4.6$ & Darling (2012) \protect\cite{DarlingNew} \\
\hline
J1229$-$021 & 0.395 & ${\alpha^{2}g_{p}/\mu}$ & $20.1\pm28.7$ & Darling (2012) \protect\cite{DarlingNew} \\
\hline
J0235$+$164 & 0.524 & ${\alpha^{2}g_{p}/\mu}$ & $-8.0\pm3.9$ & Darling (2012) \protect\cite{DarlingNew} \\
\hline
B0218$+$357 & 0.685 & ${\alpha^{2}g_{p}}$ & $-1.6\pm5.4$ & Murphy {\it et al.} (2001) \protect\cite{Murphy} \\
\hline
J0134$-$0931 & 0.765 & ${\alpha^{2\times1.57}g_{p}\mu^{1.57}}$  &  $-5.2\pm4.3$ & Kanekar {\it et al.} (2012) \protect\cite{Kanekar} \\
\hline
J2358$-$1020 & 1.173 & ${\alpha^{2}g_{p}/\mu}$ & $1.8\pm2.7$ & Rahmani {\it et al.} (2012) \protect\cite{Rahmani} \\
\hline
J1623$+$0718 & 1.336 & ${\alpha^{2}g_{p}/\mu}$ & $-3.7\pm3.4$ & Rahmani {\it et al.} (2012) \protect\cite{Rahmani} \\
\hline
J2340$-$0053 & 1.361 & ${\alpha^{2}g_{p}/\mu}$ & $-1.3\pm2.0$ & Rahmani {\it et al.} (2012) \protect\cite{Rahmani} \\
\hline
J0501$-$0159 & 1.561 & ${\alpha^{2}g_{p}/\mu}$ & $3.0\pm3.1$ & Rahmani {\it et al.} (2012) \protect\cite{Rahmani} \\
\hline
J1381$+$170 & 1.776 & ${\alpha^{2}g_{p}/\mu}$ & $-12.7\pm3.0$ & Darling (2012) \protect\cite{DarlingNew} \\
\hline
J1157$+$014 & 1.944 & ${\alpha^{2}g_{p}/\mu}$ & $23.1\pm4.2$ & Darling (2012) \protect\cite{DarlingNew} \\
\hline
J0458$-$020 & 2.040 & ${\alpha^{2}g_{p}/\mu}$ & $1.9\pm2.5$ & Darling (2012) \protect\cite{DarlingNew} \\
\hline
J1024$+$4709 & 2.285 & ${\alpha^{2}\mu}$ & $100\pm40$ & Curran {\it et al.} (2011) \protect\cite{Curran} \\
\hline
J2135$-$0102 & 2.326 & ${\alpha^{2}\mu}$ & $-100\pm100$ & Curran {\it et al.} (2011) \protect\cite{Curran} \\
\hline
J1636$+$6612 & 2.517 & ${\alpha^{2}\mu}$ & $-100\pm120$ & Curran {\it et al.} (2011) \protect\cite{Curran} \\
\hline
H1413$+$117 & 2.558 & ${\alpha^{2}\mu}$ & $-40\pm80$ & Curran {\it et al.} (2011) \protect\cite{Curran} \\
\hline
J1401$+$0252 & 2.565 & ${\alpha^{2}\mu}$ & $-140\pm80$ & Curran {\it et al.} (2011) \protect\cite{Curran} \\
\hline
J0911$+$0551 & 2.796 & ${\alpha^{2}\mu}$ & $ -6.9\pm3.7$ & Weiss {\it et al.} (2012) \protect\cite{Weiss} \\
\hline
J1337$+$3152  & 3.174 & ${\alpha^{2}g_{p}/\mu}$ & $-1.7\pm1.7$ & Srianand {\it et al.} (2010) \protect\cite{Petitjean1} \\
\hline
APM0828$+$5255 & 3.913 & ${\alpha^{2}\mu}$ & $-360\pm90$ & Curran {\it et al.} (2011) \protect\cite{Curran} \\
\hline
MM1842$+$5938 & 3.930 & ${\alpha^{2}\mu}$ & $-180\pm40$ & Curran {\it et al.} (2011) \protect\cite{Curran} \\
\hline
PSS2322$+$1944 & 4.112 & ${\alpha^{2}\mu}$ & $170\pm130$ & Curran {\it et al.} (2011) \protect\cite{Curran} \\
\hline
BR1202$-$0725 & 4.695 & ${\alpha^{2}\mu}$ & $50\pm150$ & Lentati {\it et al.} (2013) \protect\cite{Lentati} \\
\hline
J0918$+$5142 & 5.245 & ${\alpha^{2}\mu}$ & $-1.7\pm8.5$ & Levshakov {\it et al.} (2012) \protect\cite{Levshakov} \\
\hline
J1148$+$5251 & 6.420 & ${\alpha^{2}\mu}$ & $330\pm250$ & Lentati {\it et al.} (2013) \protect\cite{Lentati} \\
\hline
\end{tabular}
\caption{\label{tab03}Available measurements of several combinations of the dimensionless couplings $\alpha$, $\mu$ and $g_p$. Listed are, respectively, the object along each line of sight, the redshift of the measurement, the dimensionless parameter being constrained, the measurement itself (in parts per million), and its original reference.}
\end{center}
\end{table*}

While direct measurements of $\alpha$ and $\mu$ are most commonly obtained in the ultraviolet/optical, in the radio band one can more often measure combinations of them. Typically combinations of HI 21cm absorption lines, conjugate 18cm OH lines and molecular rotation lines are sensitive to various combinations of $\alpha$, $\mu$ and $g_p$. Here a ppm sensitivity is nominally easier to reach ({\it inter alia} because sensitivity coefficients tend to be larger), though usually only at significantly lower redshifts. Recent measurements are listed in Table \ref{tab03}, and also plotted in the bottom right panel of Fig \ref{fig01}. Note that measurements have been made beyond redshift $z=6.4$, and that PKS1413$+$135, an edge-on radio source at redshift $z=0.247$, allows measurements of all three couplings \cite{Ferreira13}, though currently only at a modest level of sensitivity. 

\begin{table*}
\begin{center}
\begin{tabular}{|c|c|c|c|c|}
\hline
 Sample & ${ \Delta\alpha}/{\alpha}$ (ppm) & ${ \Delta\mu}/{\mu}$ (ppm) & ${\Delta g_p}/{g_p}$ (ppm) & $\chi^2_\nu$  \\
\hline
Table 3 only &  $-3.5\pm2.2$ & $-0.6\pm1.7$ & $5.4\pm5.7$ & 3.83 \\
\hline
Table 3 + Webb &  $-2.3\pm0.8$ &  $-1.4\pm1.2$ & $2.4\pm2.4$ & 1.28 \\
Table 3 + 1 &  $-0.9\pm0.6$ & $-2.3\pm1.1$ & $-1.4\pm2.0$ & 2.58 \\
Table 3 + 1 + Webb &  $-1.4\pm0.5$ &  $-2.1\pm1.1$ & $-0.2\pm1.7$ & 1.26 \\
\hline
Table 3 + 2 & $-3.9\pm1.3$ & $-0.2\pm0.1$ & $6.6\pm2.9$ & 2.95 \\
Table 3 + 2 + 1 & $-1.3\pm0.6$ & $-0.2\pm0.1$ & $1.0\pm1.5$ & 2.33 \\
\hline
All data & $-1.6\pm0.5$ & $-0.2\pm0.1$ & $1.7\pm1.3$ & 1.27 \\
\hline
\end{tabular}
\caption{\label{tab04}One-dimensional marginalized one-sigma constraints for $\alpha$, $\mu$ an $g_p$, for various combinations of data sets. All constraints are in parts per million. The last column has the reduced chi-square for the maximum of the 3D likelihoods.}
\end{center}
\end{table*}

It is interesting to carry out a global statistical analysis of these data sets. These results are displayed in Table \ref{tab04}. At face value there is a mild preference, at the level of two to three standard deviations, for negative variations of $\alpha$ and $\mu$. However, the most noteworthy result of this analysis are the very large values of the reduced chi-square at the maximum of the three-dimensional likelihoods. This is mostly due to the combined measurements data set, but the issue remains when they are combined with direct measurements of $\alpha$ or $\mu$. One possible explanation is that the uncertainties of some of the measurements have been underestimated. The problem persists if the data is divided into different redshift bins, which could more accurately account for redshift dependencies in the variations, an issue first noticed in \cite{Further}. A more detailed analysis and discussion can be found in \cite{Combined}.

\subsection{Spatial variations?}

As previously mentioned, the Webb {\it et al.} analysis of their large archival data set provided evidence for spatial variations of $\alpha$ at the level of a few ppm, at a statistical level of significance of more than four standard deviations. A recent analysis \cite{Pinho0}, combining this with the then-existing set of 11 dedicated measurement found that the dipole was still a reasonable fit, although the preferred amplitude was reduced by twenty percent. It is worth revisiting and updating this analysis, given that there are now 21 dedicated $\alpha$ measurements in Table \ref {tab01}, significantly increasing the sky coverage, and that some of the previously existing measurements have also been improved. We note that the third measurement listed in Table \ref{tab01} is the weighted average from measurements along three lines of sight which contain absorbers at roughly similar redshifts but are in fact widely separated on the sky (specifically, HE1104-1805A, HS1700+6416 and HS1946+7658), reported in \cite{Songaila}. (The authors only report this average and not the individual measurements.) For this reason this measurement has been removed from spatial variations analysis, leaving the dedicated data set with 20 measurements.

The simplest approach in the modeling of spatial variations is to fit the $\alpha$ measurements to two different phenomenological parametrizations. The first is a pure spatial dipole for the relative variation of $\alpha$, which on a sphere has the form
\begin{equation}\label{puredipole}
\frac{\Delta\alpha}{\alpha}(A,\Psi)=A\cos{\Psi}\,,
\end{equation}
which depends on the orthodromic distance $\Psi$ to the North Pole of the dipole (the locus of maximal positive variation) given by
\begin{equation}\label{ortho}
\cos{\Psi}=\sin{\theta_i}\sin{\theta_0}+\cos{\theta_i}\cos{\theta_0}\cos{(\phi_i-\phi_0)}\,,
\end{equation}
where $(\theta_i,\phi_i)$ are the Declination and Right Ascension of each measurement and $(\theta_0,\phi_0)$ those of the North Pole. These latter two coordinates, together with the overall amplitude of the variation, $A$, are our free parameters. An additional monopole term is not included, both because there is no significant statistical preference for it in previous analyses \cite{Webb} and because it is physically clear that any such term would be understood as being due to the assumption of terrestrial isotopic abundances, in particular for Magnesium---we refer the interested reader to \cite{Monopole} for a detailed technical discussion of this point.

\begin{figure}
\begin{center}
\includegraphics[width=2.6in]{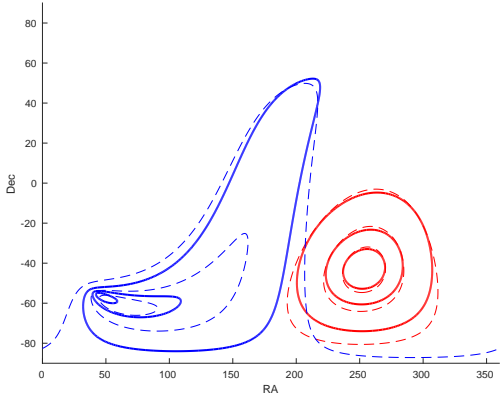}
\end{center}
\caption{\label{fig02}Two-dimensional Right Ascension--Declination likelihood, with the amplitude marginalized, for a pure spatial dipole (solid lines) and for one with a dilaton-like redshift dependence (dashed lines). Red lines correspond to all the available $\alpha$ data, while blue lines corresponds to the $\mu$ data. One, two and three sigma contours are shown in all cases. Note that the two dipoles point to different directions on the sky.}
\end{figure}

As a means to check the impact of the above choices (and hence the ability of the data to discriminate between different models for spatial variations), one can also consider a parametrization where there is an implicit time dependence in addition to the overall spatial variation. Previous analyses considered the case of a dependence on look-back time, but this requires the assumption of a cosmological model and moreover it's not clear how such a dependence would emerge from realistic varying $\alpha$ models. We can instead assume a logarithmic dependence on redshift
\begin{equation}\label{redshiftdipole}
\frac{\Delta\alpha}{\alpha}(A,z,\Psi)=A\, \ln{(1+z)}\, \cos{\Psi}\,.
\end{equation}
This has the practical advantage of not requiring any additional free parameters, but as we will see in Section \ref{models} such logarithmic redshift dependencies are also typical of well motivated dilaton-type models \cite{Dilaton}.

\begin{table*}
\begin{center}
\begin{tabular}{|c|c|c|c|}
\hline
Data set \& C.L. & Amplitude ($ppm$) & Right Ascension ($h$) & Declination (${}^\circ$) \\
\hline
Webb {\it et al.} \protect\cite{Webb} ($68.3\%$) & $9.4\pm2.2$ & $17.2\pm1.0$ & $-61\pm10$ \\
Webb {\it et al.} \protect\cite{Webb} ($99.7\%$) & $9.4\pm6.4$ & $17.2^{+4.4}_{-5.3}$ & $<-28$ \\
\hline
Table 1 ($68.3\%$) & $<2.3$ & $14.1^{+3.4}_{-5.8}$ & $>17$ \\
Table 1 ($99.7\%$) & $<6.4$ & N/A & N/A \\
\hline
All data ($68.3\%$) & $5.6\pm1.8$ & $16.9\pm0.8$ & $-43\pm7$ \\
All data ($99.7\%$) & $<10.9$ & $16.9^{+3.4}_{-3.2}$ & $-43^{+34}_{-31}$ \\
\hline
\end{tabular}
\caption{\label{tab05}One- and three-sigma constraints on the Amplitude and Equatorial sky coordinates of maximal variation (Right Ascension and Declination) for a pure spatial dipole variation of $\alpha$. The 'All Data' case corresponds to using the data of Webb {\it et al.} \protect\cite{Webb} together with the 20 individual measurements presented in Table \protect\ref{tab01}; this is plotted with solid red contours in Fig. \protect\ref{fig05}.}
\end{center}
\end{table*}

The results of this analysis are summarized in the red contours in Fig. \ref{fig02}, and in Table \ref{tab05}. For the Webb {\it et al.} data we recover the statistical preference for a dipole at more than four standard deviations, while there is no preference for a dipole in the more recent data. Combining the two data sets, the statistical preference for a dipole is reduced to only 2.3 standard deviations, and the best-fit amplitude is less than 6 ppm. As for the direction of maximal variation, we note that the preferred Declination is significantly changed by the addition of the most recent data, moving by about 18 degrees, while the Right Ascension is comparatively less affected.

Comparing the results for the pure spatial dipole and the redshift-dependent one, we see that they are very similar (with the constraints on the latter being very slightly weaker); this is visually clear in Fig. \ref{fig02}, where the results for both models are represented, and for this reason Table \ref{tab05} only reports the results for the pure spatial case. The current sensitivity and redshift distribution of the measurements is not sufficient to distinguish between these models. In any case, it may be argued that while assuming such simple dipole parametrizations is phenomenologically legitimate, it is not quite realistic, since in models where there are environmental dependencies the observational behavior would be more complex that that. This is a legitimate point, and in Section \ref{models} we will discuss a way to address it.

An additional independent test of possible spatial variations can be done with the sample of 13175 emission line measurements of $\alpha$ from the SDSS-III/BOSS DR12 quasar sample of Albareti {\it et al.} \cite{Albareti}. While the sensitivity of each of their individual measurements of the relative variation of $\alpha$ is much worse than the ones reported in Table \ref{tab01} (ranging from $2.4\times10^{-4}$ to $1.5\times10^{-2}$, to be compared to parts-per-million), the much large number of measurements covering a significant fraction of the sky still allows for a worthwhile test of spatial variations. For comparison, the weighted mean of the 13175 measurements, which span the redshift range $0.041<z<0.997$ is
\begin{equation}\label{albaretimean}
\left(\frac{\Delta\alpha}{\alpha}\right)_{BOSS}=9\pm18\, ppm\,.
\end{equation}
One can also use this data set to test for possible spatial variations. A detailed analysis can be found in \cite{Combined}, with the result that there is no preference for a particular direction on the sky, and specifically with the following three-sigma (99.7\% C.L.) upper bound for the amplitude of a putative dipole
\begin{equation}\label{sdss}
A_{SDSS}<7\times 10^{-4} \,.
\end{equation}
This bound on the amplitude is about 64 times weaker than the one discussed above from the absorption line measurements (cf. bounds on the amplitude in Table \ref{tab05}), but it is independent from it. Moreover, it is stronger than recent bounds on spatial variations coming from the combination of Sunyaev-Zel'dovich cluster measurements and Planck satellite data (and even stronger than analogous bounds from the Planck cosmic microwave background data alone) \cite{SZclusters}.

Since the number of available measurements of $\mu$ in Table \ref{tab02} is not much smaller than those of $\alpha$ in Table \ref{tab01}, one may also ask whether there is any evidence for a dipole in the $\mu$ measurements. This issue was briefly addressed in \cite{Ubachs}, but that work only considered the high-redshift molecular hydrogen measurements. Doing the analysis with all the data in Table \ref{tab02} we find that there is no strong preference for it: the statistical significance of a possible dipole is less than two sigma. Specifically, the two-sigma ($95.4\%$ confidence level) upper limit for the dipole amplitude in the case of a pure spatial dipole is
\be
A_\mu<1.9\, ppm\,,
\ee
while for the redshift-dependent one it is
\be
A_\mu<4.3\, ppm\,.
\ee
Moreover, the directions on the sky corresponding to the North pole of such putative $\alpha$ and $\mu$ dipoles are incompatible at more than three sigma for the case of pure spatial dipoles, or at more than two sigma for the case of a dilaton-type redshift dependence---see Fig. \ref{fig02} for a visual comparison. The difference between the pure and redshift-dependent dipole analyses in somewhat larger in the $\mu$ case than in the $\alpha$ one, the reason for this being that in the former the tightest measurements are at low redshifts---compare Eqs. \ref{priormulo}--\ref{priormuhi}. In any case, the conclusion is that at present there is also no strong evidence for spatial variations, though the issue will certainly be revisited in the near future as ESPRESSO becomes available.

Finally, we should point out that the radio band sensitivity is even better for measurements within the Galaxy (thus effectively at $z=0$), where one can search for environmental dependencies since measurements can be made in regions with densities that are many orders of magnitude smaller than the local one. Here again there is no current evidence for variations, up to a sensitivity at the 0.05 ppm level for $\mu$ \cite{Galaxy} and at the 0.1 ppm level for $\alpha$ \cite{Joao}. Searches for environmental dependencies can also be done using compact astrophysical objects, to be briefly discussed in the next section.

\section{Other probes\label{otherobs}}

Although the QSO high-resolution spectroscopic measurements described in the previous section constitute the most actively pursued topic in this field, they are complemented by a range of other local and astrophysical measurements, which probe the stability of fundamental couplings in a vast range of cosmological epochs and physical environments. While it is not the main goal of this review to provide a thorough discussion of all of these, in this section we provide some comments and updates on a few of these, on which there has been relevant recent activity. We remind the reader that a more systematic review may be found in \cite{Uzan}.

\begin{table*}
\begin{center}
\begin{tabular}{|c|c|c c c|c|}
\hline
Clocks & ${\dot \nu_{AB}}/{\nu_{AB}}$ (yr${}^{-1}$) & $\lambda_\alpha$ & $\lambda_\mu$ & $\lambda_g$ & Reference \\ 
\hline
Hg-Al & $(5.3\pm7.9)\times10^{-17}$ & -2.95 & 0.0 &  0.000 & Rosenband {\it et al.} (2008) \protect\cite{Rosenband} \\
Dy162-Dy164 & $(-5.8\pm6.9)\times10^{-17}$ &  1.00 & 0.0 &  0.000 & Leefer {\it et al.} (2013) \protect\cite{Leefer} \\
Cs-SF${}_6$ & $(-1.9\pm2.7)\times10^{-14}$ &  2.83 & 0.5 & -1.266 & Shelkovnikov {\it et al.} (2008) \protect\cite{Shelkovnikov} \\
Cs-H & $(3.2\pm6.3)\times10^{-15}$ &  2.83 & 1.0 & -1.266 & Fischer {\it et al.} (2004) \protect\cite{Fischer} \\
Cs-Sr & $(1.80\pm0.55)\times10^{-16}$ &  2.77 & 1.0 & -1.266 & Abgrall {\it et al.} (2015) \protect\cite{Abgrall} \\
Cs-Hg & $(-3.7\pm3.9)\times10^{-16}$ &  5.77 & 1.0 & -1.266 & Fortier {\it et al.} (2007) \protect\cite{Fortier} \\
Cs-Yb(E2) & $(-0.5\pm1.9)\times10^{-16}$ &  1.83 & 1.0 & -1.266 & Tamm {\it et al.} (2014) \protect\cite{Tamm} \\
Cs-Yb(E3) & $(-0.2\pm4.1)\times10^{-16}$ &  8.83 & 1.0 & -1.266 & Huntemann {\it et al.} (2014) \protect\cite{Huntemann} \\
Cs-Rb & $(1.07\pm0.49)\times10^{-16}$ &  0.49 & 0.0 & -2.000 & Gu\'ena {\it et al.} (2012) \protect\cite{Bize2} \\
\hline
\end{tabular}
\caption{\label{tab06} Atomic clock constraints of varying fundamental couplings. The third, fourth and fifth columns show the sensitivity coefficients of each frequency ratio to the various dimensionless couplings.}
\end{center}
\end{table*}
\begin{figure}
\begin{center}
\includegraphics[width=2.6in]{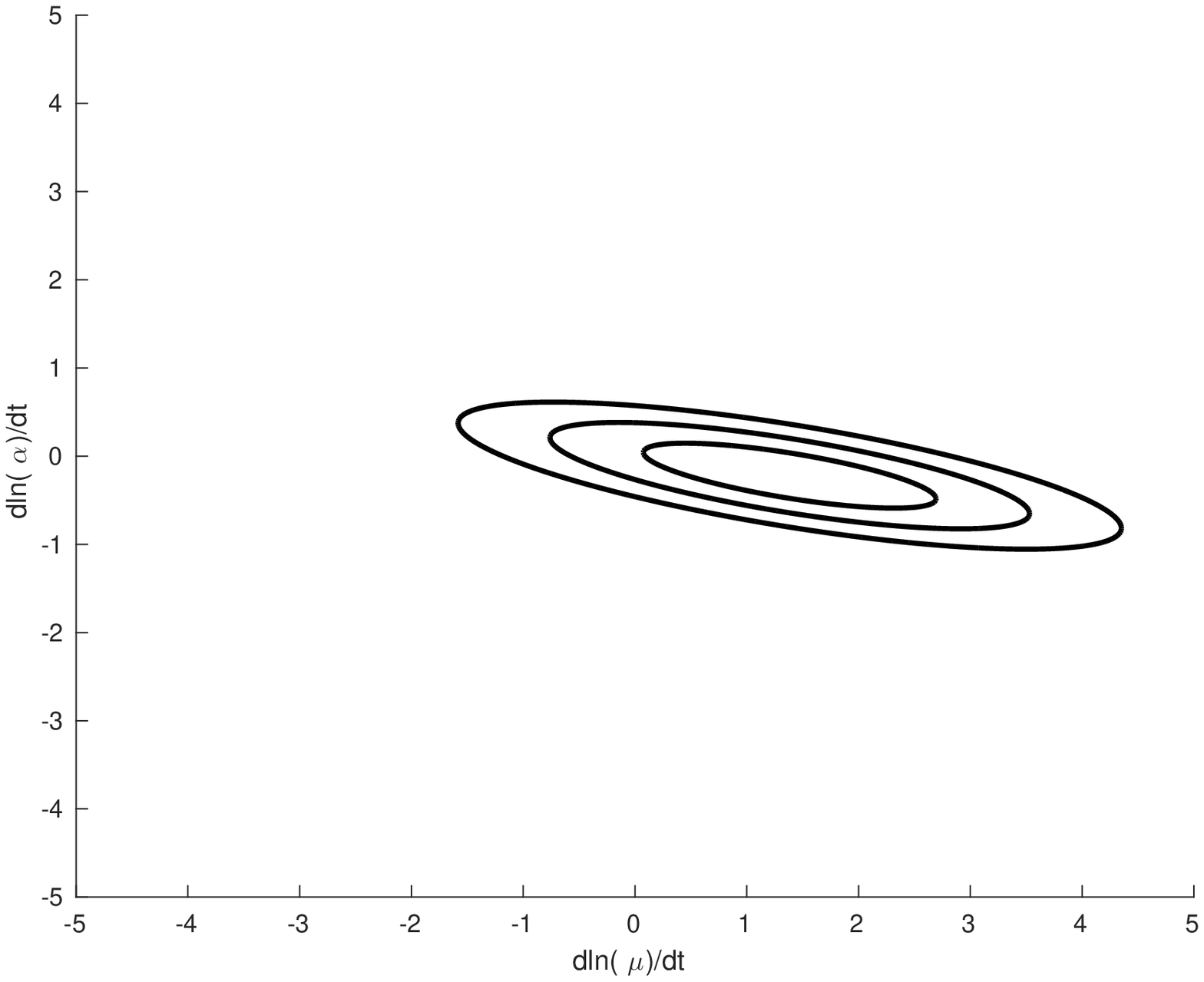}
\vskip0.2in
\includegraphics[width=2.6in]{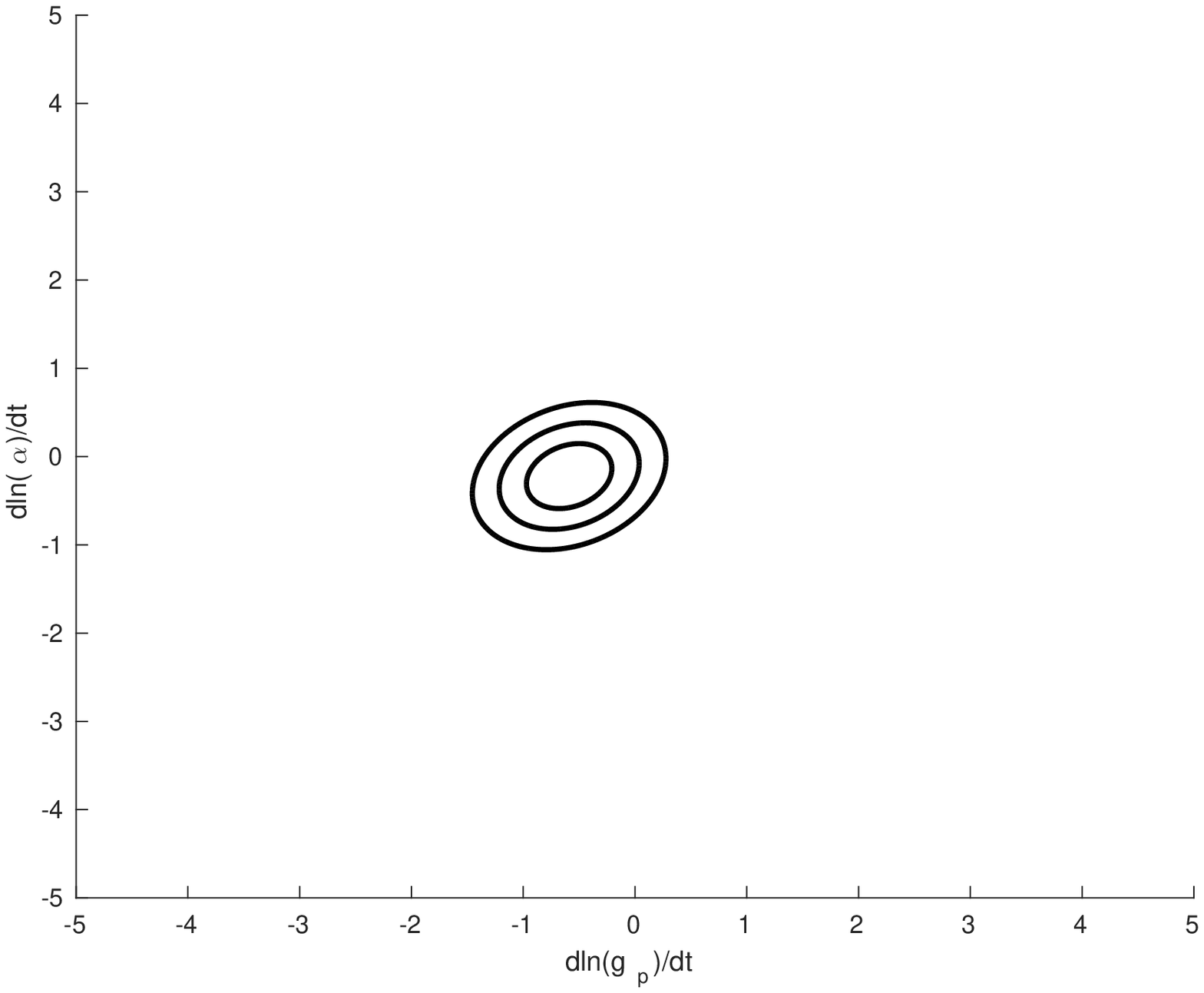}
\vskip0.2in
\includegraphics[width=2.6in]{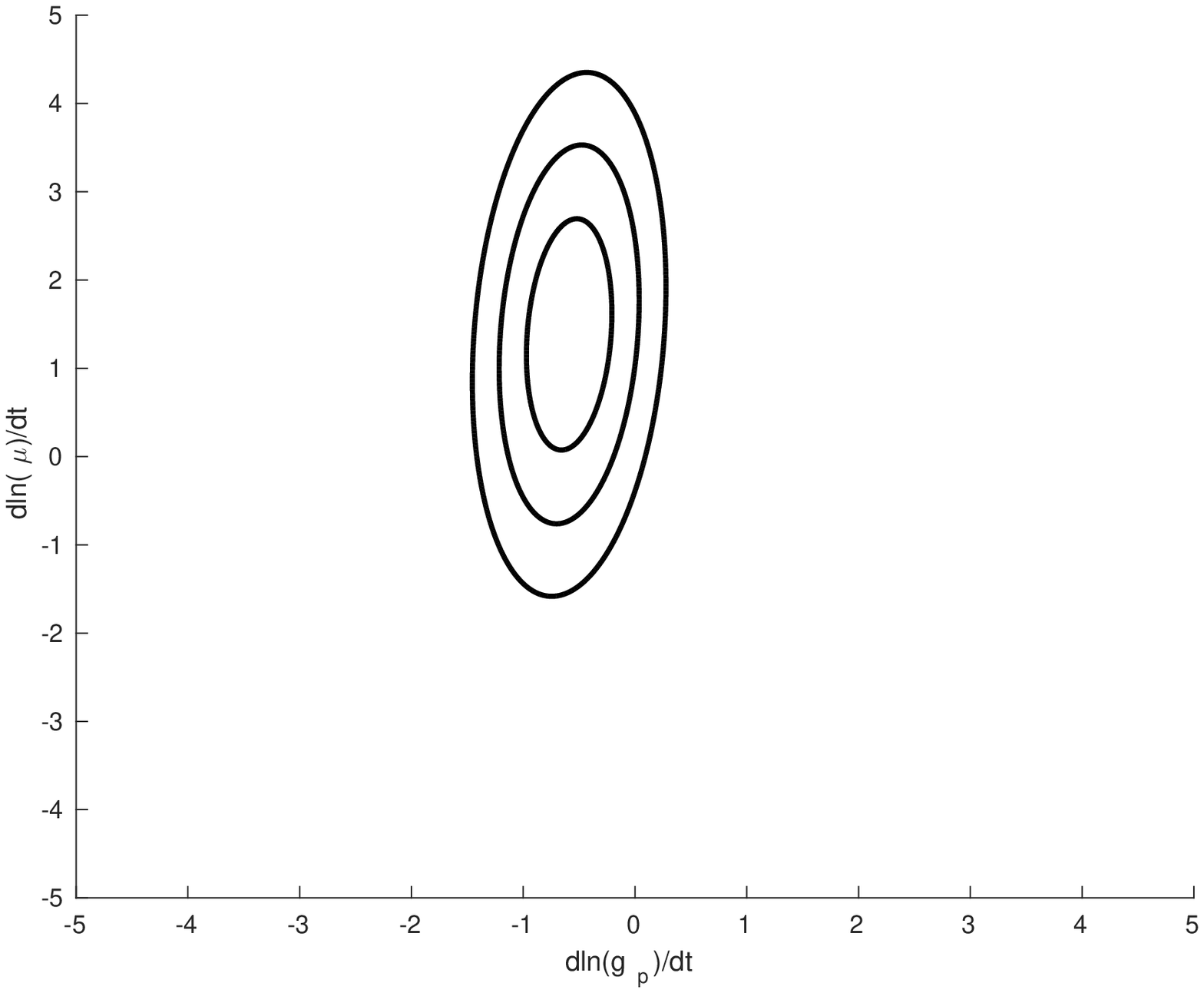}
\end{center}
\caption{\label{fig03}Two-dimensional likelihoods for each pair of couplings, marginalizing over the other, from a global analysis of the data in Table \protect\ref{tab06}. All the axes are in units of $10^{-16}$ yr${}^{-1}$, and one, two and three sigma contours are shown in all cases. Note that all axis ranges are the same, to highlight the different sensitivities with which the various possible drifts can currently be constrained.}
\end{figure}

In strict terms of sensitivity, the other probe that is competitive with QSO spectroscopy is provided by laboratory tests using atomic clocks \cite{ClockReview}. The idea is analogous to that of QSO spectroscopy. One compares (in this case over a time span of at least a few months, but possibly several years) two atomic clocks whose characteristic frequencies have different sensitivities to certain combinations of $\alpha$, $\mu$ and $g_p$, thereby obtaining a measurement of the drift rate of this combination in the relevant time period. Typically the ratio of the two frequencies will be proportional to
\be
\nu_{AB}=\frac{\nu_A}{\nu_B}\propto \alpha^{\lambda_\alpha} \mu^{\lambda_\mu} g_p^{\lambda_g}
\ee
where the $\lambda_i$ are again the sensitivity coefficients
\be
\lambda_\alpha=\frac{d\ln{\nu_{AB}}}{d\ln\alpha}\,,
\ee
and analogously for the other couplings. Table \ref{tab06} presents the latest available laboratory measurements. Notice that it is possible to individually constrain $\alpha$ (indeed, at very high sensitivity) while the best measurements of $\mu$ and $g_p$ are obtained for pairs of clocks sensitive to their combinations with $\alpha$. The constraint of Rosenband {\it et al.} \cite{Rosenband} is currently the single most stringent constraint on $\alpha$, and has been frequently used in combination with other data to constrain several models for $\alpha$ variations---some of these constraints will be discussed in Sections \ref{models} and \ref{darkside}. A joint statistical analysis of the data in this table leads to the following best-fit values, at the one-sigma (68.3\%) confidence level (see also fig. \ref{fig03})
\be
\frac{d\ln{\alpha}}{dt}=(-2.2\pm2.4)\times 10^{-17}\, {\rm yr}^{-1}
\ee
\be\label{atomclockmu}
\frac{d\ln{\mu}}{dt}=(13.8\pm8.6)\times 10^{-17}\, {\rm yr}^{-1}
\ee
\be
\frac{d\ln{g_p}}{dt}=(-5.8\pm2.5)\times 10^{-17}\, {\rm yr}^{-1}\,;
\ee
thus there is no evidence for $\alpha$ or $\mu$ variations, while intriguingly a drift of $g_p$ is preferred at more than two standard deviations. A joint analysis also leads to useful constraints on the unification parameters $R$ and $S$, introduced in Section \ref{consts}. For example the work of \cite{Ferreira12} led to
\begin{equation}
\label{constrainatomclocks}
(S+1)-2.7R=-5 \pm 15 \,;
\end{equation}
with the newer data this constraint is improved by roughly a factor of two. Significant progress is expected in laboratory measurements in the coming years: with forthcoming molecular and nuclear clocks, particularly those based on Thorium229 \cite{Thorium}, a sensitivity as high as $10^{-21}$yr${^{-1}}$ may be achieved.

The Oklo natural nuclear reactor is another complementary probe of the stability of fundamental couplings. In particular, it nominally provides a strong constraint on $\alpha$, however it only does so if one assumes that everything else is not varying (in other words, that there is a different value of $\alpha$ but physics is otherwise unchanged). Since here one is dealing with a chain of nuclear reactions, this is likely to be a very crude assumption, as has been amply documented in the recent literature. Indeed, the Oklo nuclear reactions are more sensitive to the analogous coupling for the strong nuclear force, $\alpha_s$. We refer the interested reader to a recent review on the subject \cite{OKLOrev} and references therein. So while it is clear that this is not as 'clean' and reliable a measurement as the atomic clock and QSO measurements, one can certainly take these constraints at face value. The current constraint, from the analysis of \cite{OKLO1}, is
\begin{equation} \label{oklo}
\frac{\Delta\alpha}{\alpha}=(0.5\pm6.1)\times10^{-8}\,,
\end{equation}
at an effective redshift $z_{\rm Oklo}=0.14$. This nominally strong bound ultimately exploits the presence of a 97.3 meV resonance in the neutron capture by the Samarium-149 isotope (whereas the typical energy scale of nuclear reactions is of order MeV). We note that even stronger bounds have been obtained in \cite{OKLO2,OKLO3}, but these rely on additional assumptions, while the bound of \cite{OKLO1} is more conservative.

Compact objects have also been the focus of significant recent studies, exploring their suitability as probes of the stability of fundamental couplings. There are many theoretical studies which quantify the effect of varying couplings on these objects and use their known properties to infer {\it a posteriori} limits on such variations. Such analyses have been carried out for Population III stars \cite{Ekstrom}, solar-type stars \cite{Vieira} and neutron stars \cite{PerezGarcia}. In all these cases current sensitivities are around the 50 ppm level, and often the limiting factor comes from nuclear physics uncertainties. 

More recently direct observational constraints have been obtained, also at about the 50 ppm level of sensitivity, for both $\alpha$ and $\mu$ using white dwarf stars \cite{WDmu,WDalpha}. These come from spectroscopic observations of highly excited metal lines (FeV and NiV) and molecular hydrogen, respectively
\begin{equation}\label{wdfe}
\left(\frac{\Delta\alpha}{\alpha}\right)_{FeV}=42\pm16\, ppm
\end{equation}
\begin{equation}\label{wdni}
\left(\frac{\Delta\alpha}{\alpha}\right)_{NiV}=-61\pm58\, ppm
\end{equation}
\begin{equation}\label{wdh1}
\left(\frac{\Delta\mu}{\mu}\right)_{GD133}=-27\pm47\, ppm
\end{equation}
\begin{equation}\label{wdh2}
\left(\frac{\Delta\mu}{\mu}\right)_{G29-38}=-58\pm38\, ppm
\end{equation}
and provide a means to constrain environmental dependencies---in this case, possible dependencies on the gravitational potential. In the case of the $\alpha$ measurements, the FeV and NiV results are inconsistent with each other at the 1.6 sigma level, the likely reason being related to uncertainties in laboratory wavelength measurements of these transitions; assuming that these can be improved, the sensitivity of these measurements should certainly reach the ppm level.

The effects of varying fundamental couplings on the white dwarf mass-radius relation were recently studied in \cite{WDMR}, both for the simple case of a polytropic stellar structure model and for a more general model. This analysis shows that independent measurements of the mass and radius, together with direct spectroscopic measurements of $\alpha$ in white dwarf atmospheres such as those discussed in the previous paragraph, lead to constraints on unification scenarios which interestingly are almost orthogonal to the ones coming from atomic clocks. Currently available measurements do not yet provide stringent constraints, but improvements in mass and radius measurements, expected for example from the Gaia satellite \cite{Gaia}, can break parameter degeneracies and lead to strong new constraints.

At higher redshifts the cosmic microwave background provides a very clean probe in principle: it is well known that varying couplings will affect the ionization history of the universe (including the energy levels, the binding energies and the Thomson cross-section), and moreover the relevant physics is to a large extent linear and well understood. Nevertheless, the sensitivity of this probe is limited by the presence of degeneracies with other cosmological parameters, so current constraints are around the 3700 ppm level for $\alpha$ \cite{Planck}, and even worse for $\mu$ and for spatial variations. Given the ppm constraints at low redshifts, CMB constraints will only be competitive for very specific classes of models that would predict strong variations in the very early universe---this would not be the case in the simplest dilaton-type (string-theory inspired) models. Conversely, the detection of $\alpha$ variations at the CMB epoch that are not detectable by more sensitive low-redshift spectroscopic methods would certainly point to new and unexpected physics. Next-generation missions such as CORE should improve these bounds by almost one order of magnitude \cite{COREcmb}.

Another recent approach is to extract constraints on $\alpha$ by comparing X-ray and Sunyaev-Zel'dovich cluster data, leading to percent-level constraints \cite{Galli}. Again the sensitivity of this method is very low compared to QSO spectroscopic measurements, though on the other hand this approach in principle has the advantage of large numbers. With moderate gains in the sensitivity of the observations in each individual cluster which should be easily achievable, if one is able to use the tens of thousands of clusters that will be observed by next generation missions such as CORE \cite{COREclu} one may be able to obtain independent competitive constraints, though at lower median redshifts (with current data, all clusters for which this technique has been used are at $z<0.5$). Another potential advantage of a large number of sources well spread on the sky is the possibility to constrain spatial variations. Indeed the recent analysis of de Martino {\it et al.} \cite{SZclusters}, which uses a larger cluster sample (an advantage that is partially offset by the fact that all clusters therein are at $z<0.3$) has improved both the constraints obtained by \cite{Galli} and the Planck constraints on spatial variations of $\alpha$ \cite{Planck}. These data sets are also useful since they can provide measurements of the CMB temperature at non-zero redshift, a topic to which we will return in Section \ref{redundancy}.

At even higher redshifts constraints can also be obtained from Big Bang Nucleosynthesis, though with the {\it caveat} that they will necessarily be model-dependent. The reason for this is that the first step in any analysis of the effects of varying fundamental couplings on BBN will be to ascertain its effect on the neutron to proton mass difference, and this can only be done through the phenomenological Gasser-Leutwyler formula \cite{Gasser}. That said, current phenomenological constraints are at around the percent level for relatively generic phenomenological models  \cite{BBN}, though much tighter constraints can be obtained for more specific choices of model, in particular by restricting oneself to the unification scenarios we mentioned in Section \ref{consts}, as was done in \cite{Coc}. Finally, it has been claimed that the Lithium problem might be removed in some GUT scenarios \cite{Stern}. This is plausible in principle, because one generically expects that varying couplings will have larger effects for heavier nuclei: in other words, they could significantly change the Lithium abundance while leaving those of lighter nuclei comparatively unaffected. A more detailed analysis of this scenario is probably warranted given recent observational progress \cite{Cyburt}.

\section{Cosmological models with varying couplings\label{models}}

Having reviewed the observational status of the tests of the stability of dimensionless fundamental couplings, we now move on to describe possible models in which these couplings do vary, as well as how they are constrained by the available data. This will be the subject of the present and the following sections, and we will again focus mostly on models for $\alpha$ variations.

From an observational point of view, scalar field based models for varying couplings can be conveniently divided into two broad classes \cite{GRG}. As we shall see later in this review, should varying couplings be detected, observational consistency tests can be done to ascertain to which of the two classes the model responsible for these variations belongs two. Attributing the variations to a specific model within the appropriate class will be a subsequent task.

The first, dubbed Class I, contains those models where the degree of freedom responsible for the varying constants also provides the dark energy. These are therefore natural and 'minimal' models, in the operational sense that there is a single new dynamical degree of freedom---in other words, a single extension of the standard model---accounting for both. In this class of models the redshift dependence of the couplings will be parametrically determined, and any available measurements of $\alpha$ (be they detections of variations or null results) can be used to set constraints on combinations of fundamental physics and cosmological parameters, such as the dark energy equation of state. We will discuss these models in more detail in the following section. 

Presently we focus instead on the opposite class, dubbed Class II models. These are the ones where the field that provides the varying couplings does not provide the dark energy (or at least does not provide all of it). In this case the aforementioned unique link with dark energy is lost, though the parameters of the underlying cosmological model will nevertheless affect the variation of the couplings, as we will see in specific examples. Moreover, note that even if the scalar field does not dominate the background cosmological dynamics, inferring its presence is still crucial since---through $\alpha$ variations themselves or through other effects---it can bias cosmological parameter estimations \cite{Tasos2,Calabrese}. We will return to this point in Section \ref{redundancy}. In this section we will discuss three representative examples of these models.

\subsection{Bekenstein models}

Arguably the simplest class of phenomenological models for varying couplings is the one first suggested by Bekenstein \cite{Bekenstein} where, by construction, the dynamical degree of freedom responsible for the varying coupling  has a negligible effect on the cosmological dynamics. This class includes the Sandvik-Barrow-Magueijo model for a varying fine-structure constant $\alpha$ \cite{SBM} and the Barrow-Magueijo model for a varying proton-to-electron mass ratio $\mu$, both of which have been studied in some detail.

These models are characterized by a single phenomenological dimensionless parameter, $\zeta$, describing the strength of the coupling of the dynamical scalar degree of freedom to the electromagnetic sector, and thus determining the amount of Weak Equivalence Principle (WEP) violation in the model. In what follows we describe the simplest version of both models. While extensions of the $\alpha$ model with additional (functional) degrees of freedom have been suggested \cite{extend1,extend2}, the quantity and quality of the available data (and the fact that no strong evidence for a nonzero coupling $\zeta$ currently exists) motivate us to focus on the simplest scenarios.

The Bekenstein-Sandvik-Barrow-Magueijo (BSBM) model for varying $\alpha$  was introduced in \cite{SBM}, drawing on Bekenstein's earlier work \cite{Bekenstein}. It is a model where the variation of $\alpha$ is due to a varying electric charge, while other parameters are assumed to remain constant. Conceptually one could say that this is a dilaton-type model (see the following sub-section), though one where the field is postulated to couple only to the electromagnetic sector of the Lagrangian. The model's dynamical equations are obtained by standard variational principles, as discussed in \cite{SBM}. Specifically, the value of the fine-structure constant is related to the scalar field $\psi$ via $\alpha/\alpha_0= e^{2(\psi-\psi_0)}$ though as usual the observational parameter of choice is the relative variation of $\alpha$, ${\Delta\alpha}/{\alpha}$, cf. Eq. \ref{defalpha}. Without loss of generality we henceforth re-define the field such that at the present day $\psi_0=0$.

Assuming a flat, homogeneous and isotropic cosmology (in agreement with the latest cosmological data \cite{Planck}), one obtains the following Friedmann equation \cite{SBM}
\be
H^2=\frac{8\pi G}{3}\left[\rho_m(1+\omega\zeta_\alpha e^{-2\psi})+\rho_r e^{-2\psi}+\rho_\Lambda+\frac{1}{2}\omega{\dot\psi}^2  \right]\,,
\ee
with the dots denoting derivatives with respect to physical time, and the $\rho_i$ respectively denoting the matter, radiation and dark energy densities. The scalar field equation is
\be
{\ddot\psi}+3H{\dot\psi}=-2\zeta_\alpha G\rho_m e^{-2\psi} \,.
\ee
Here $\omega$ is a parameter that can be defined as $\omega\sim\hbar c/\ell^2$, where $\ell$ effectively describes the scale below which one has significant deviations from standard electromagnetism. For simplicity (and consistently with the analyses in \cite{SBM,Leal}) one can take $\omega\sim1$, leaving the coupling $\zeta_\alpha$ as the only phenomenological free parameter in the model. Typical values for this parameter are discussed in some detail in \cite{SBM}, but irrespective of theoretical expectations $\zeta_\alpha$ can be taken as a free phenomenological parameter, to be constrained by observations. Note that in addition to radiation and matter the model needs a dark energy component, which for simplicity is assumed to be a cosmological constant, to match cosmological observations. It is straightforward to show that the dynamical scalar field $\psi$ is constrained to be entirely subdominant in the dynamics of the universe (one practical consequence of this being that we can assume the standard values of the cosmological parameters), and its only role is to drive a variation of the fine-structure constant.

In practice it is more convenient to write this equation as a function of redshift; recalling that
\be
\frac{dz}{dt}=-(1+z)H\,,
\ee
one finds
\be\label{dyna}
\psi''+\left(\frac{d\ln{H}}{dz}-\frac{2}{1+z}\right)\psi'=-\frac{3\zeta_\alpha \Omega_m}{4\pi}\frac{(1+z)}{E^2(z)}  e^{-2\psi}\,;
\ee
here the primes denote derivatives with respect to redshift, and for future convenience we have defined the dimensionless function
\be
E(z)=\frac{H(z)}{H_0}\,.
\ee
The above equation can be straightforwardly integrated, together with the Friedmann equation, by standard numerical methods. One finds that deep in the matter era the relative $\alpha$ variation is proportional to $\log{(1+z)}$ (just as in most dilaton-type scenarios where this behavior occurs throughout, see below), but the onset of acceleration quickly freezes the dynamics of the field and leads to comparatively smaller variations close to the present day. The current drift rate of $\alpha$, expressed in dimensionless units, is
\be
\left(\frac{1}{H}\frac{\dot\alpha}{\alpha}\right)_0=-\left(\frac{\alpha'}{\alpha}\right)_0=-2{\psi'}_0
\ee
and will be relevant for comparison to atomic clock measurements. For example the previously mentioned bound by Rosenband {\it et al.} \cite{Rosenband} expressed in these units is
\be\label{RosenH}
\left(\frac{1}{H}\frac{\dot\alpha}{\alpha}\right)_0=(-2.2\pm3.2)\times10^{-7}.
\ee
Moreover, in this type of model there are composition-dependent forces which lead to a WEP violation at a level quantified by the E\"{o}tv\"{o}s parameter, denoted $\eta$ \cite{Will}. Specifically, for this model the relation between $\eta$ and the coupling parameter is \cite{SBM}
\be
\eta_\alpha\sim3\times 10^{-9}\zeta_\alpha\,,
\ee
to be compared to current bounds coming from torsion balance experiments \cite{Torsion}
\be\label{etaTorsion}
\eta=(-0.7\pm1.3)\times 10^{-13}
\ee
and lunar laser ranging \cite{LLR}
\be\label{etaLLR}
\eta=(-0.8\pm1.2)\times 10^{-13}\,.
\ee

Constraints on this model were recently discussed in \cite{LeiteBek} and will be updated here, including the new data that became available in recent months. Specifically, the following data sets were used (differences/updates relative to the analysis of \cite{LeiteBek} are indicated in brackets)
\begin{itemize}
\item The Union2.1 data set of 580 Type Ia supernovas \cite{Union};
\item A compilation of 38 Hubble parameter measurements by Farooq {\it et al.} \cite{Farooq} (whereas \cite{LeiteBek} used the 35 measurements then available)
\item The astrophysical measurements of Webb {\it et al.} as well as the 21 dedicated measurements listed in Table \ref{tab01} (whereas \cite{LeiteBek} used the 11 dedicated measurements then available)
\item The atomic clock constraint on $\alpha$ by Rosenband {\it et al.}, cf. Eq. \ref{RosenH}, and the Oklo contraint \cite{OKLO1}, cf. Eq. \ref{oklo}. 
\end{itemize}
For future reference, note that this data set---which we henceforth refer to as the {\it Canonical Data set}---will also be used later in this review to constrain other classes of models.

For the purpose of constraining the Bekenstein-type models, the background cosmology (supernova and Hubble parameter) data will effectively provide conservative constraints on the present-day matter density, $\Omega_m$. Stronger constraints could be obtained by using for example CMB priors, but for the sake of consistency it is simpler to use only low-redshift data throughout the analysis.

\begin{figure}
\includegraphics[width=3in]{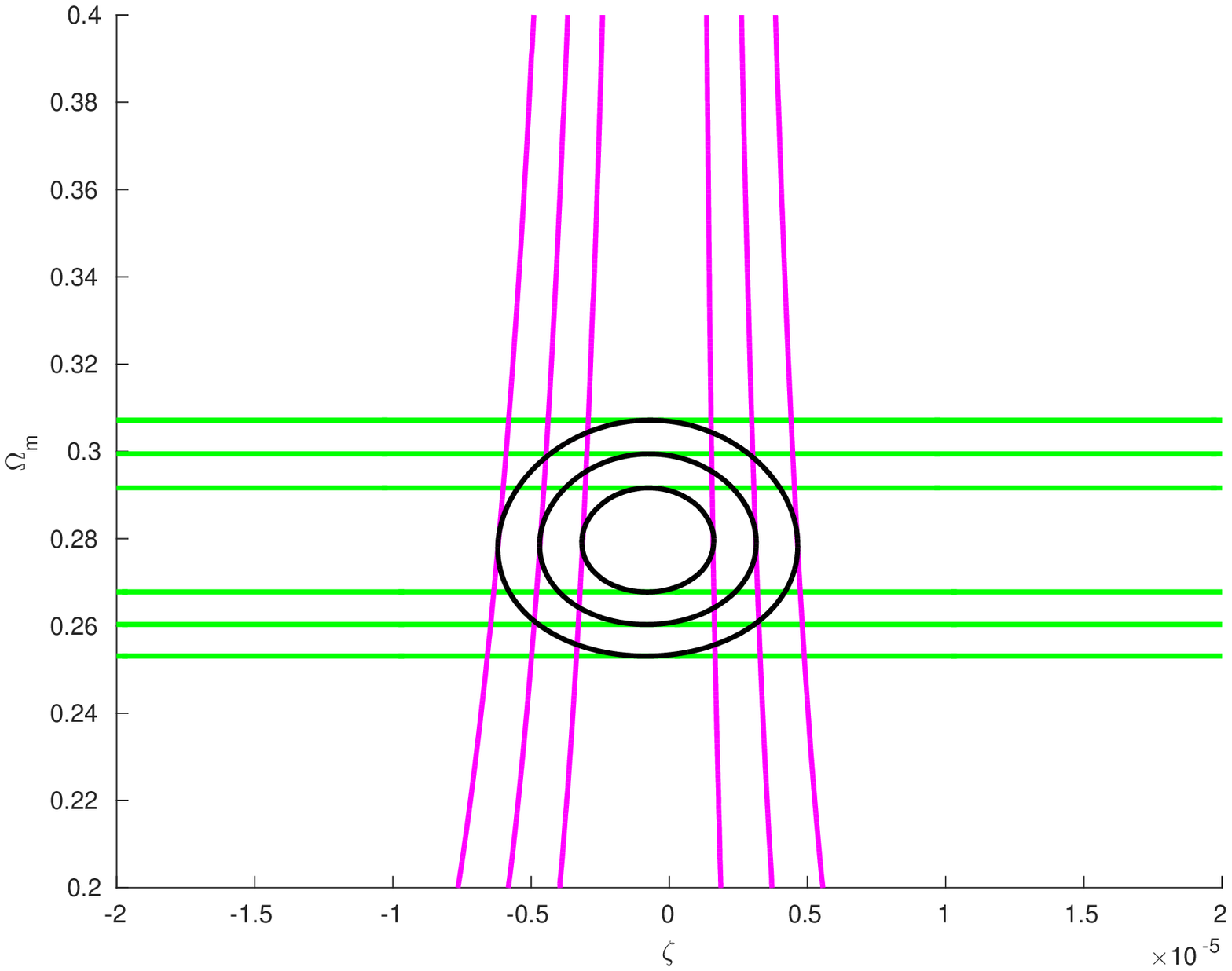}
\vskip0.2in
\includegraphics[width=3in]{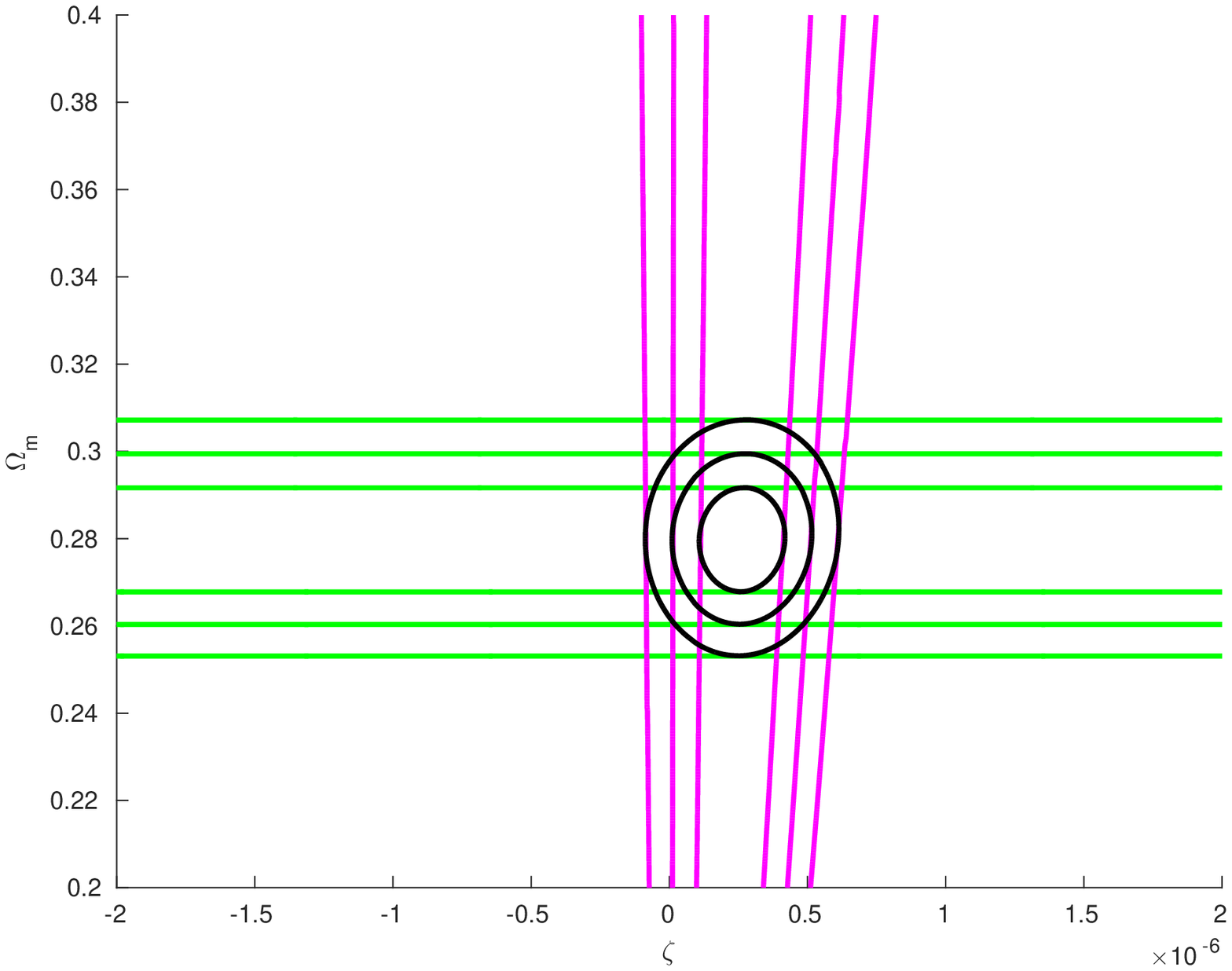}
\caption{Astrophysical and cosmological constraints on Bekenstein-type models, specifically 2D likelihood contours on the $\zeta_\alpha-\Omega_m$ plane (top panel) and the $\zeta_\mu-\Omega_m$ plane (bottom panel). Constraints from the cosmological data are shown in green (horizontal contours) and those from astrophysical data in magenta (vertical contours); the joint constraints are shown in black. One, two and three sigma contours are plotted in all cases. The reduced chi-square for the maximum likelihood value is $\chi^2_\nu=0.96$ for the $\alpha$ case and $\chi^2_\nu=0.94$ for the $\mu$ case. These plots update those presented in \protect\cite{LeiteBek}.}
\label{fig04}
\end{figure}

The resulting constraints are summarized in the top panel of Fig. \ref{fig04}. As expected the correlation between the two parameters is small, and the cosmological data sets mostly fix the matter density, while the $\alpha$ measurements constrain the coupling $\zeta_\alpha$. Specifically, marginalizing over $\zeta_\alpha$ yields
\be
\Omega_m=0.28\pm0.03\,,
\ee
at the three sigma ($99.7\%$) confidence level, which is fully compatible with other extant cosmological data sets, including the recent Planck data \cite{PlanckParams}. On the other hand, marginalizing over $\Omega_m$, we find
\be
\zeta_\alpha=(-0.8\pm1.5)\times10^{-6}\,,\quad 68.3\% C.L.
\ee
\be
\zeta_\alpha=(-0.8\pm4.5)\times10^{-6}\,,\quad 99.7\% C.L.\,.
\ee
Finally on can express this constraint on the coupling as a constraint on the E\"{o}tv\"{o}s parameter,
\be
\eta_\alpha < 1.3\times 10^{-14}\,, \quad 99.7\% C.L.\,.
\ee
These constraints slightly improve those of \cite{LeiteBek}, and Table \ref{tab07} compares the two. Note that the bound on $\eta_\alpha$ is a stronger bound than the current local experimental limits. In other words, models in this class in agreement with the $\alpha$ constraints also satisfy current WEP bounds. However, we note that the recently launched MICROSCOPE satellite is expected to improve the sensitivity of local bounds to $\eta\sim10^{-15}$ \cite{MICROSCOPE}, thus enabling additional constraints.

The Bekenstein-Barrow-Magueijo model for varying $\mu$ was introduced in \cite{BM}. It is again a dilaton-type model and to a large extent analogous to the $\alpha$ model, the main difference being that it describes a varying electron mass rather than a varying electric charge. Other parameters are again assumed to remain constant. Observationally, this leads to a varying proton-to-electron mass ratio, $\mu$. (Note that \cite{BM} uses a definition of $\mu$ which is opposite to the one we follow here, specifically it defines it as $m_e/m_p$.) As before we can assume that the field driving these variations, in this case denoted $\phi$, does not significantly contribute to the Friedmann equation. Moreover we restrict ourselves to flat, homogeneous and isotropic cosmologies, and assume that the dark energy is provided by a cosmological constant. In this case the electron mass is given by $m_e/m_{e0}=e^{\phi-\phi_0}$, while the observationally relevant parameter is ${\Delta\mu}/{\mu}(z)$. Without loss of generality we also re-define the field such that $\phi_0=0$.

In this case the dynamical equation for $\phi$, analogous to Eq. (\ref{dyna}), has the form
\be
\phi''+\left(\frac{d\ln{H}}{dz}-\frac{2}{1+z}\right)\phi'=-\zeta_\mu \Omega_m\frac{(1+z)}{E^2(z)}  e^{\phi}\,.
\ee
For simplicity (and in analogy with the $\alpha$ case discussed above), we have defined the dimensionless coupling $\zeta_\mu$. Written in terms of the parameters used in \cite{BM}, this is has the form
\be
\zeta_\mu=\frac{3\Omega_b}{8\pi\mu G \omega}\left(1-\frac{f_{He}}{2}\right)\,,
\ee
where $\Omega_b$ is the present-day baryon density and $f_{He}\sim1/12$ is the Helium-4 number fraction. Qualitatively the redshift dependence of $\mu$ in this model is quite similar to that of the $\alpha$ model, the main quantitative difference being that, as we will presently see, the allowed values of the coupling  will be smaller. In this case the current drift rate of $\mu$, expressed in dimensionless units is
\be
\left(\frac{1}{H}\frac{\dot\mu}{\mu}\right)_0=-\left(\frac{\mu'}{\mu}\right)_0={\phi'}_0\,,
\ee
which is constrained by a weaker bound from local atomic clock measurements given by Eq. \ref{atomclockmu}, which when expressed in dimensionless units becomes
\be
\left(\frac{1}{H}\frac{\dot\mu}{\mu}\right)_0=(2.0\pm1.3)\times10^{-6}\,.
\ee
Note that this bound is stronger than the one used in the analysis of \cite{LeiteBek}, but it is in any case weaker than the astrophysical measurements, so updating it will have a fairly small impact on the overall constraints. Conversely, in this case the amount of WEP violation is much stronger
\be
\eta_\mu\sim10^{-4}\zeta_\mu\,.
\ee

Naturally in this case we can use the same cosmological data sets as in the case of the $\alpha$ model, together with the $\mu$ measurements in Table \ref{tab02} and the above atomic clock bound on $\mu$. The results of the analysis are summarized in the bottom panel of Fig. \ref{fig04}. Again the correlation between the two parameters is small (though note that it is different from that in the $\alpha$ model case). Marginalizing over $\zeta_\alpha$ we unsurprisingly find the same constraint on $\Omega_m$ as in the previous case, while for the coupling $\zeta_\mu$ we find
\be
\zeta_\mu=(2.7\pm1.0)\times10^{-7}\,,\quad 68.3\% C.L.
\ee
\be
\zeta_\mu=(2.7\pm3.1)\times10^{-7}\,,\quad 99.7\% C.L.\,;
\ee
this seemingly corresponds to detection of a non-zero coupling with a statistical significance of 2.5 standard deviations. However, this value of the coupling is actually incompatible with the bound which comes from WEP violations
\be
\zeta_\mu <4\times 10^{-9}\,, \quad 99.7\% C.L.\,;
\ee
here, unlike in the case of the $\alpha$ model, this constraint is about two orders of magnitude stronger than that coming from the $\mu$ measurements.

While we will discuss future prospects for this field in Section \ref{forecasts}, here we take the opportunity to briefly discuss forthcoming improvements to these constraints, expected from the new generation of high-resolution ultra-stable optical spectrographs. The first of these, ESPRESSO \cite{ESPRESSO}, will be installed at the combined Coud\'e focus of ESO's VLT in 2017 and it will become the instrument of choice for tests of the stability of fundamental constants until the era of the Extremely Large Telescopes, and particularly its flagship spectrograph, ELT-HIRES \cite{HIRES}.

A preliminary selection of the list of $\alpha$ targets to be observed during the ESPRESSO Fundamental Physics Guaranteed Time Observations (GTO) has been recently done \cite{MSC}, identifying 14 absorption systems in the redshift range $1.35\le z\le 3.02$. For this list of ESPRESSO targets one can generate simulated measurements with the expected ESPRESSO sensitivity, assuming two different scenarios, referred to as `Baseline' and `Ideal' \cite{LeiteBek}. These are meant to represent two estimates of ESPRESSO's actual performance and sensitivity for these measurements, with the former being conservative an the latter being somewhat more optimistic (for example, it may imply additional telescope time on each target). Naturally, the actual performance of the instrument will only be known after commissioning, but one may expect it to be somewhere between the two. Further discussion of the assumptions underlying these Baseline and Ideal scenarios can be found in \cite{MSC}. The ESPRESSO target list was also used for a second forecast, in this case for ELT-HIRES, by extrapolating the gains from the increased telescope collecting area and---crucially---assuming that the wavelength coverage of ELT-HIRES is at least equal to that of ESPRESSO (this will certainly be the case at the red end, but what happens at the blue end remains to be seen at the time of writing).

These future (simulated) astrophysical data sets were used instead of the current $\alpha$ measurements to obtain constraints on the model parameters. Thus a data set of 314 current measurements was replaced by one with only 14, spanning a smaller redshift range but naturally having much better precision. The previously defined cosmological data sets as well as the atomic clocks bound of Rosenband {\it et al.} were also used. Naturally these are conservative assumptions since both of these data sets are also expected to improve, but the goal is to directly assess the impact of the improved astrophysical measurements.

\begin{table*}
\centering
\begin{tabular}{|l|c|c|}
\hline
Data set & $\sigma_\zeta$ ($68.3\%$ C.L.) & $\sigma_\zeta$ ($99.7\%$ C.L.) \\
\hline
Current (Reference \protect\cite{LeiteBek}) & $1.7\times10^{-6}$ & $4.8\times10^{-6}$ \\
Current (This review) & $1.5\times10^{-6}$ & $4.5\times10^{-6}$ \\
\hline
ESPRESSO Baseline & $6.0\times10^{-7}$ & $1.8\times10^{-6}$ \\
ESPRESSO Ideal & $2.1\times10^{-7}$ & $6.3\times10^{-7}$ \\
\hline
ELT-HIRES Baseline & $1.1\times10^{-7}$ & $3.2\times10^{-7}$ \\
ELT-HIRES Ideal & $2.3\times10^{-8}$ & $7.0\times10^{-8}$ \\
\hline
\end{tabular}
\caption{\label{tab07}Current one and three sigma uncertainties on the coupling $\zeta_\alpha$ (marginalizing over $\Omega_m$) obtained from current data, and the corresponding forecasts for the forthcoming ESPRESSO Fundamental Physics GTO target list and the next-generation ELT-HIRES (under the assumptions discussed in the text). The difference between the constraints of \protect\cite{LeiteBek} stems for the fact that new $\alpha$ and Hubble parameter measurements became available.}
\end{table*}

The forecasts for the various scenarios are compared with the current constraints in Table \ref{tab07}. These results make it clear that even a relatively small set of only 14 measurements will lead to very significant improvements. As previously mentioned the Baseline and Ideal scenarios are intended to bracket the actual performance of ESPRESSO and ELT-HIRES (with somewhat larger uncertainties on the latter, given the earlier stage of its development). We therefore expect the ESPRESSO GTO data set to improve constraints on the coupling in these models by a factor of about 5 while an analogous ELT-HIRES program can improve it by a factor of about 50, both these factors being calculated relative to the present values of the constraints.

\subsection{Runaway dilaton models}

String theory does predict the presence of a scalar partner of the spin-2 graviton, the dilaton, hereafter denoted $\phi$. A particular class of string-inspired models is the so-called runaway dilaton scenario of Damour, Piazza and Veneziano \cite{DPV1,DPV2}. In this scenario, which among other things provides a way to reconcile a massless dilaton with experimental data, the dilaton decouples while cosmologically attracted towards infinite bare coupling, and the coupling functions have a smooth finite limit
\begin{equation}\label{eq:coupfunc}
B_i(\phi)=c_i+{\cal O}(e^{-\phi})\,.
\end{equation}
As extensively discussed in \cite{DPV2}, provided there's a significant (order unity) coupling to the dark sector, the runaway of the dilaton towards strong coupling may yield violations of the Equivalence Principle and variations of the fine-structure constant $\alpha$ that are potentially measurable. We now describe this scenario, mostly following the discussion in \cite{Dilaton}.

The Einstein frame Lagrangian for this class of models is \cite{DPV1,DPV2}
\begin{equation}\label{eq:lagr}
{\cal L}=\frac{R}{16\pi G}-\frac{1}{8\pi G}\left(\nabla\phi\right)^2-\frac{1}{4}B_F(\phi)F^2+... \,.
\end{equation}
where $R$ is the Ricci scalar and $B_F$ is the gauge coupling function. From this one can show \cite{DPV2} that the corresponding Friedmann equation is as follows
\begin{equation}\label{eq:friedmann}
3H^2=8\pi G\sum_i \rho_i+H^2\phi'^2\,,
\end{equation}
where the sum is over the components of the universe, except the kinetic part of the dilaton field which is described by the last term, and the prime is the derivative with respect to the logarithm of the scale factor. The sum does include the potential part of the scalar field; the total energy density and pressure of the field are
\begin{equation}\label{phidens}
\rho_\phi=\rho_k+\rho_v=\frac{(H\phi')^2}{8\pi G}+V(\phi)
\end{equation}
\begin{equation}\label{phipres}
p_\phi=p_k+p_v=\frac{(H\phi')^2}{8\pi G}-V(\phi)\,;
\end{equation}
here $k$ and $v$ correspond to the kinetic and potential parts of the field, with the latter providing the dark energy. On the other hand, the evolution equation for the scalar field is
\begin{equation}\label{eq:field}
\frac{2}{3-\phi'^2}\phi''+\left(1-\frac{p}{\rho}\right)\phi'=-\sum_i\alpha_i(\phi)\frac{\rho_i-3p_i}{\rho}\,.
\end{equation}
Here $p=\sum_ip_i$, $\rho=\sum_i\rho_i$, and the sums are again over all components except the kinetic part of the scalar field.

The $\alpha_i(\phi)$ are the couplings of the dilaton with each component $i$, so they characterize the effect of the various components of the universe in the dynamics of the field. One may generically expect that the dilaton has different couplings to different components \cite{DPV2}, though one must bear in mind that experimental constraints impose a tiny coupling to baryonic matter, as we will discuss presently. In these models, this small coupling could naturally emerge due to a Damour-Polyakov type screening of the dilaton \cite{Polyakov}.

The relevant parameter here is the coupling of the dilaton field to hadronic matter. As discussed in \cite{Polyakov}, to a good approximation this is given by the logarithmic derivative of the QCD scale, since hadron masses are proportional to it (modulo small corrections). Assuming that all gauge fields couple, near the string cutoff, to the same $B_F(\phi)$, and in accordance with Eq. (\ref{eq:coupfunc}) which yields
\begin{equation}\label{bfhere}
B_F^{-1}(\phi)\propto (1-b_Fe^{-c\phi})\,,
\end{equation}
we can write
\begin{equation}\label{alphadefn}
\alpha_{had}(\phi)\sim 40 \frac{\partial\ln B_F^{-1}(\phi)}{\partial\phi}\,,
\end{equation}
where the numerical coefficient is further described in \cite{DPV2}, and we finally obtain
\begin{equation}\label{alphahad}
\alpha_{had}(\phi)\sim 40\, b_F c\,e^{-c\phi}\,.
\end{equation}
Note that $c$ and $b_F$ are constant free parameters: the former one is expected to be of order unity and the latter one much smaller. Moreover, if we set $c=1$ (which we will do henceforth) we can also eliminate $b_F$ by writing 
\begin{equation}\label{alphahadrel}
\frac{\alpha_{had}(\phi)}{\alpha_{had,0}}=e^{-(\phi-\phi_0)}\,,
\end{equation}
where $\phi_0$ is the value of the field today, and simultaneously writing the field equation in terms of $(\phi-\phi_0)$.

There are two local constraints. Firstly the Eddington parameter $\gamma$, which quantifies the amount of deflection of light by a gravitational source, has the value
\begin{equation}\label{eddingt}
\gamma-1=-2\alpha_{had,0}^2\,,
\end{equation}
and is constrained by the Cassini bound \cite{Cassini}
\be
\gamma-1=(2.1\pm2.3)\times10^{-5}\,.
\ee
Secondly, in this model the dimensionless E\"{o}tv\"{o}s parameter, quantifying violations to the Weak Equivalence Principle, has the value
\begin{equation}\label{eotvos}
\eta_{AB}\sim5.2\times10^{-5}\alpha_{had,0}^2\,,
\end{equation}
and is constrained by the torsion balance and Lunar Laser Ranging tests already discussed in the previous subsection, cf. Eqs. \ref{etaTorsion}--\ref{etaLLR}. From these we conservatively obtain the bound
\begin{equation}\label{boundhad}
|\alpha_{had,0}|\le 10^{-4}\,.
\end{equation}
Using Eq. (\ref{alphahad}), and still assuming that $c\sim1$, this yields a bound on the product of $b_F$ and the exponent of $\phi_0$, namely $\phi_0\ge\ln{(|b_F|/2\times10^{-6})}$. Nevertheless, this is not explicitly needed: the evolution of the system will be determined by $\alpha_{had}$ rather than by $b_F$ or $\phi_0$.

These constraints do not apply to the dark sector (in other words, to dark matter and/or dark energy) whose couplings may be stronger. There are two possible scenarios to consider. A first possibility is that the dark sector couplings (which we will denote $\alpha_m$ and $\alpha_v$ for the dark matter and dark energy respectively) are also much smaller than unity, that is $\alpha_m,\alpha_v\ll1$. In this case the small field velocity leads to violations of the Equivalence Principle and variations of the fine-structure constant that are quite small. For this case to be observationally realistic the fractions of the critical density of the universe in the kinetic and potential parts of the scalar field must be
\begin{equation}\label{lambdadil}
\Omega_k=\frac{1}{3}{\phi'}^2\ll1\,,\qquad \Omega_v\sim0.7;
\end{equation}
note that if one assumes a flat universe, then $\Omega_m+\Omega_k+\Omega_v=1$. (Do not confuse the index $k$, which refers to the kinetic part of the scalar field, with the curvature term in standard cosmology, which we are setting to zero throughout.) A more interesting possibility is that the dark couplings $\alpha_m$ and/or $\alpha_v$ are of order unity. If so, violations of the Equivalence Principle and variations of the fine-structure constant are typically larger, and may well be observable. In this case $\Omega_k$ may be more significant, and $\Omega_v$ should be correspondingly smaller \cite{quint}. Nevertheless the dark matter coupling is also constrained: during matter-domination the equation of state has the form
\begin{equation}
w_m(\phi)=\frac{1}{3} {\phi'}^2 \sim\frac{1}{3}\alpha_m^2\,,
\end{equation}
and must therefore be small.

The present value of the field derivative is also constrained if one assumes a spatially flat universe; in that case the deceleration parameter
\begin{equation}\label{deccel}
q=-\frac{a{\ddot a}}{{\dot a}^2}=-1-\frac{\dot H}{H^2}\,
\end{equation}
can be written as
\begin{equation}\label{deccelhere}
{\phi_0'}^2=(1+q_0)-\frac{3}{2}\Omega_{m0} \,
\end{equation}
and using a reasonable upper limit for the deceleration parameter, $q_0=-0.57\pm0.04$ \cite{Neben}, and a lower limit for the matter density, say from the Planck mission \cite{PlanckParams}, one conservatively obtains
\begin{equation}\label{phibound}
|\phi_0'|\le 0.3  \,,
\end{equation}
which is nevertheless almost three times tighter than the one available at the time of \cite{DPV2}. Thus in this scenario both the hadronic coupling and the field speed today are constrained.

Moreover, we can use the field equation, Eq. (\ref{eq:field}), to set a consistency condition for $\phi_0'$. For this we only need to assume that the field is moving slowly today (a good approximation given the bounds on its speed) and therefore the $\phi''$ term should be subdominant in comparison with the other two. Then we easily obtain
\begin{equation}\label{phicons}
\phi_0'=\, - \, \frac{\alpha_{had}\Omega_b+\alpha_m\Omega_c+4\alpha_v\Omega_v}{\Omega_b+\Omega_c+2\Omega_v} \,,
\end{equation}
with all quantities being evaluated at redshift $z=0$. To avoid confusion we have denoted baryonic and cold dark matter by $\Omega_b$ and $\Omega_c$ respectively; naturally $\Omega_m=\Omega_b+\Omega_c$. We choose the cosmological parameters in agreement with Planck data \cite{PlanckParams}, setting the current fractions of baryons, dark matter and dark energy to be respectively $\Omega_{b}\sim0.04$, $\Omega_{c}\sim0.27$ and $\Omega_{\phi}=\Omega_{k}+\Omega_{v}\sim0.69$. Noting that $|\alpha_{had,0}|\le 10^{-4}$, that $|\phi_0'|\le 0.3$ and that $\Omega_k={\phi_0'}^2/3$ is necessarily small, we can consider three particular cases of this relation
\begin{itemize}
\item The {\bf dark coupling} case, where $\alpha_m=\alpha_v$ (and both are assumed to be constant), leads to
\begin{equation}\label{alphadark}
|\alpha_v|< 0.3 \frac{\Omega_m+2\Omega_v}{\Omega_c+4\Omega_v}\sim0.17 \,;
\end{equation}
\item The {\bf matter coupling} case, where $\alpha_m=\alpha_{had}$ (and both are field-dependent, as in Eq. \ref{alphahadrel}), leads to
\begin{equation}\label{alphamat}
|\alpha_v|< 0.3 \frac{\Omega_m+2\Omega_v}{4\Omega_v}\sim0.18 \,;
\end{equation}
\item The {\bf field coupling} case, where $\alpha_m=-\phi'$, leads to
\begin{equation}\label{alphafield}
|\alpha_v|<0.3 \frac{\Omega_b+2\Omega_v}{4\Omega_v}\sim0.15 \,.
\end{equation}
\end{itemize}
Note that in all cases $\alpha_v$ is a constant (field-independent) parameter. Naturally these are back-of-the-envelope constraints that need to be improved by a more robust analysis, but they are enough to show that order unity couplings $\alpha_v$ will be strongly constrained. An additional constraint will come from atomic clock measurements, as we will now discuss.

Consistently with our previous assumption that all gauge fields couple to the same $B_F$, here $\alpha$ will be proportional to $B_F^{-1}(\phi)$, as given by Eq. (\ref{bfhere}). This will also imply that $\alpha$ will be related to the hadronic coupling. One can then show that the evolution of $\alpha$ is given by \cite{DPV2}
\begin{equation}\label{alphazero}
\frac{1}{H}\frac{\dot\alpha}{\alpha}=\frac{b_Fce^{-c\phi}}{1-b_Fce^{-c\phi}}\,{\phi'}\sim b_Fce^{-c\phi}{\phi'}\sim\frac{\alpha_{had}}{40}{\phi'}\,.
\end{equation}
In particular this equation applies at the present day (describing the current running of $\alpha$) and this variation is constrained by the Rosenband {\it et al.} bound \cite{Rosenband}; assuming the Planck value for the Hubble constant $H_0=(67.4\pm1.4)\, {\rm km/s/Mpc}$, we find
\begin{equation}\label{jointphibound}
|\alpha_{had,0}{\phi'}_0| \sim |b_Fce^{-c\phi_0}{\phi'}_0| \le 3\times 10^{-5}\,.
\end{equation}
Thus atomic clock experiments constrain the product of the hadronic coupling and the field speed today. It is interesting to note that this constraint---which stems from microphysics---is similar to the one obtained by multiplying the individual constraints on each of them, which are given respectively by Eq. \ref{boundhad} and Eq. \ref{phibound} and come from macrophysics (Solar System or torsion balance tests, plus a cosmology bound).

In \cite{DPV2} the authors first obtained approximate solutions for the evolution of $\alpha$ by assuming that $\phi'=const.$ in both the matter and the dark energy eras (naturally the two constants are allowed to be different). However, as pointed out in \cite{Dilaton}, by integrating Eq. (\ref{alphazero}) or by directly using the relation between $\alpha$ and $B_F(\phi)$ one can express the redshift dependence of $\alpha$ in the general form
\begin{equation}
\frac{\Delta\alpha}{\alpha}(z)=B_F^{-1}(\phi(z))-1 =b_F\left(e^{-\phi_0}-e^{-\phi(z)}\right)\,,
\end{equation}
where for simplicity we have again set $c\sim1$. This can also be recast in the more suggestive form
\begin{equation}
\frac{\Delta\alpha}{\alpha}(z)=\frac{1}{40}\alpha_{had,0}\left[1-e^{-(\phi(z)-\phi_0)}\right]\,.
\end{equation}
Thus the behavior of $\Delta\alpha/\alpha$ close to the present day depends both on $\alpha_{had,0}$ (which provides an overall normalization) and on the speed of the field, ${\phi_0'}$, which can also be related to the values of the couplings as in Eq. (\ref{phicons}). When dealing with high-resolution spectroscopic measurements one is interested in the evolution of $\alpha$ at relatively low redshifts, in which case one can linearize the field evolution
\begin{equation}\label{evollinear}
\phi\sim\phi_0+{\phi_0'}\ln{a}\,,
\end{equation}
and therefore the evolution of $\alpha$ will take the simpler form
\begin{equation}\label{evolslow}
\frac{\Delta\alpha}{\alpha}(z)\approx\, -\frac{1}{40}\alpha_{had,0} {\phi_0'}\ln{(1+z)}\,;
\end{equation}
this is indeed what is obtained with the simplifying assumptions of \cite{DPV1,DPV2}. Nevertheless, note that $\phi-\phi_0$ can still be of order unity by redshift $z=5$ for values of the coupling that saturate the current bounds, and therefore the evolution of $\alpha$ should be calculated using the full equations.

By numerically solving the previously discussed Friedmann and scalar field equations one can further study the cosmological dynamics of this model \cite{Dilaton}. Note that in this model the dark energy equation of state is
\begin{equation}
1+w_0=\frac{2\Omega_{k}}{\Omega_{k}+\Omega_{v}}=\frac{2}{3}\frac{{\phi_0'}^2}{\Omega_{k}+\Omega_{v}}\,,
\end{equation}
and the range of allowed values for ${\phi_0'}$ (specifically, $|\phi_0'|\le 0.3$) leads to $-1\le w_0<-0.91$, which is perfectly compatible with current observational bounds \cite{PlanckParams}. Using all available $\alpha$ data (both that of \cite{Webb} and the dedicated measurements of Table \ref{tab01}) one finds no significant evidence for a non-zero coupling $\alpha_{had,0}$. Note that Hubble parameter measurements do help to constrain the current speed of the field to be small.

Forecasts for future constraints on this model were discussed in \cite{Runaway}, using a combination of simulated cosmological probes and astrophysical tests of the stability of the fine-structure constant $\alpha$ expected from ELT-HIRES \cite{HIRES}. The three different scenarios for the dark sector couplings discussed above were separately considered, with the goal of identifying observational differences between
them, and the degeneracies between the parameters ruling the coupling of the dilaton field to the other components of the universe were identified and quantified. This analysis shows that if the couplings are very small (e.g., $\alpha_b=\alpha_v\sim0$) these degeneracies strongly affect the constraining power of future data, while if they are sufficiently large (say, $\alpha_b>10^{-5}$ or $\alpha_v>0.05$, both still well below current upper bounds) the degeneracies can be partially broken. The conclusion is therefore that the ELT will be able to explore some of this additional parameter space, and improve current constraints by about one order of magnitude.

\subsection{Environmental dependencies from symmetron models}

The models discussed in the two previous subsections led to redshift (in other words, time) dependencies of $\alpha$ and have spatial variations that are of second order and therefore much smaller. Given the observational indications of possible spatial variations, we now discuss a scenario where these may be larger than the time variations, in the context of symmetron models. This was first studied in \cite{Marvin}, whose discussion we now follow.

In the symmetron model \cite{symmoriginal,olivepospelov}, the vacuum expectation value (VEV) of a scalar field depends on the local mass density, becoming large in regions of low density and small in regions of high density. The coupling of the scalar to matter is proportional to the VEV, leading to a theory where the scalar can couple with gravitational strength in regions of low density, but be decoupled and screened in regions of high density. This is achieved through the interplay of a symmetry breaking potential and a universal quadratic coupling to matter. In vacuum, the scalar acquires a VEV which spontaneously breaks a $\mathcal{Z}_2$ symmetry $\phi\to-\phi$. In the regions of sufficiently high matter density, the field is confined near $\phi=0$, and the symmetry is restored. The fifth force arising from the matter coupling is proportional to $\phi$ making the effects of the scalar small in high density regions.

The symmetron model is a scalar-tensor modification of gravity described by the action
\be
S = \int dx^4\sqrt{-g}\left[\frac{R}{2}M_{\rm pl}^2 - \frac{1}{2}\left(\partial\phi\right)^2 - V(\phi)\right]
+ S_m(\Psi_m; \tilde{g}_{\mu\nu})
\ee
where $g = \det g_{\mu\nu}$, $M_{\rm pl} = 1/\sqrt{8\pi G}$, $S_m$ is the matter-action and units with $\hbar = c \equiv 1$ are being used. The matter fields $\Psi_m$ are coupled to the scalar field via a conformal coupling
\be
\tilde{g}_{\mu\nu} = g_{\mu\nu}A^2(\phi)\,.
\ee
Because of this coupling the matter fields will experience a fifth-force, which in the non-relativistic limit is given by
\be
\vec{F}_{\phi} \equiv \frac{dA(\phi)}{d\phi}\vec{\nabla}\phi = \frac{\phi\vec{\nabla}\phi}{M^2}
\ee
where the last equality only holds for the symmetron. The symmetron potential is chosen to be of the symmetry breaking form
\be
V(\phi) = - \frac{1}{2}\mu^2\phi^2 + \frac{1}{4}\lambda\phi^4
\ee
where $\mu$ (not to be confused with the proton-to-electron mass ratio discussed elsewhere in this review) is a mass-scale and the conformal coupling is chosen as the simplest coupling consistent with the potential symmetry $\phi\to-\phi$,
\be
A(\phi) =  1 + \frac{1}{2}\left(\frac{\phi}{M}\right)^2\,,
\ee
where $M$ is a mass scale and $\lambda$ a dimensionless coupling constant. A variation of the action with respect to $\phi$ leads to the following field equation
\be
\nabla^2\phi = \frac{dV_{\rm eff}}{d\phi}\,.
\ee
The dynamics of $\phi$ is determined by the effective potential
\be
V_{\rm eff} = V(\phi) + A(\phi)\rho_m = \frac{1}{2}\left[\frac{\rho_m}{\mu^2M^2} - 1\right]\mu^2\phi^2 + \frac{1}{4}\lambda\phi^4\,.
\ee
In the early Universe where the matter density is high the effective potential has a minimum at $\phi = 0$ where the field will reside. As the Universe expands the matter density dilutes until it reaches a critical density
\be
\rho_{\rm SSB} = \mu^2M^2\,,
\ee
for which the symmetry breaks and the field moves to one of the two new minima $\phi = \pm \phi_0 = \pm {\mu}/\sqrt{\lambda}$.

The fifth force between two test particles residing in a region of space where $\phi = \phi_{\rm local}$ is
\be
\frac{F_{\phi}}{F_{\rm gravity}} =  2\beta^2 \left(\frac{\phi_{\rm local}}{\phi_0}\right)^2\,,
\ee
where we have defined
\be
\beta = \frac{\phi_0 M_{\rm pl}}{M^2}\,,
\ee
for separations within the Compton wavelength $\lambda_{\rm local} = 1/\sqrt{V_{\rm eff,\phi\phi}(\phi_{\rm local})}$ of the scalar-field. For larger separations the force is suppressed by a factor $e^{-r/\lambda_{\rm local}}$. In the cosmological background before symmetry breaking $\phi_{\rm local} \approx 0$ and the force is suppressed. After symmetry breaking the field moves towards $\phi = \pm\phi_0$ and the force can be comparable with gravity for $\beta = \mathcal{O}(1)$. In high density regions, like the Sun and our Galaxy, non-linear effects in the field equation ensure that the force is effectively screened thereby evading local gravity constraints.

It is convenient to introduce the variables
\be
a_{\rm SSB} = \left(\frac{\rho_{m0}}{\rho_{\rm SSB}}\right)^{1/3}
\ee
and
\be
\lambda_{\phi 0} = \frac{1}{\sqrt{2}\mu}\,,
\ee
together with the already defined quantities $\beta$ and $ \rho_{\rm SSB}$, which are respectively the coupling strength relative to gravity and the density at which the symmetry is broken; $a_{\rm SSB}$ is the corresponding scale-factor for when this happens in the cosmological background and $\lambda_{\phi 0}$ is the range of the fifth force when the symmetry is broken. Local gravity constraints, discussed in \cite{symmoriginal,symmnbody,symmcosm1,symmcosm2} force the range of the field to satisfy
\be
\lambda_{\phi 0} < 1 {\rm Mpc}/h
\ee
for symmetry breakings close to the present day, i.e. $a_{\rm SSB} \sim 1$.

The electromagnetic field is unaffected by a conformal transformation because of the conformal invariance of the electromagnetic action,
\be
S_{\rm EM}(A_{\mu}; g_{\mu\nu}A^2(\phi)) \equiv S_{\rm EM}(A_{\mu}; g_{\mu\nu})\,.
\ee
However one can consider generalizations where the electromagnetic field is coupled to the scalar field via
\be
S_{\rm EM} =  -\int dx^4\sqrt{-g} A^{-1}_{\gamma}(\phi)\frac{1}{4}F_{\mu\nu}^2\,.
\ee
With this coupling it is still the case that perfect fluid radiation does not affect the Klein-Gordon equation for the scalar field because the stress-energy tensor of the electromagnetic field is traceless. This coupling leads to the fine-structure constant depending on $\phi$ as
\be
\alpha(\phi) = \alpha_0 A_{\gamma}(\phi)
\ee
where again $\alpha_0$ is the laboratory value.

Now, the simplest choice for $A_{\gamma}$, compatible with the $\phi\to-\phi$ symmetry of the symmetron, is
\be
A_{\gamma}(\phi) = 1 + \frac{1}{2}\left(\frac{\beta_{\gamma}\phi}{M}\right)^2
\ee
where $\beta_{\gamma}$ is the scalar-photon coupling relative to the scalar-matter coupling, i.e. a value of $\beta_\gamma = 1$ implies that the scalar-photon coupling is the same as the scalar-matter coupling. A variation of $\phi$ therefore leads to a variation of the fine-structure constant $\alpha$ with respect to the laboratory value $\alpha_0$ given by
\be
\frac{\Delta \alpha}{\alpha} = A_{\gamma}(\phi)-1 = \frac{1}{2}\left(\frac{\beta_{\gamma}\phi}{M}\right)^2\,.
\ee
For the symmetron we have \cite{Marvin}
\be
\frac{\Delta \alpha}{\alpha}\simeq \beta^2\beta_\gamma^2\left(\frac{0.5}{a_{\rm SSB}}\right)^3 \left(\frac{\phi}{\phi_0}\right)^2\left(\frac{\lambda_{\phi 0}}{{\rm Mpc}/h}\right)^2 \left(\frac{\Omega_{m0}}{0.25}\right)\, ppm\,.
\ee
For our fiducial model parameters $a_{\rm SSB} \sim 0.5$, $\beta \sim 1$, $\lambda_{\phi 0} \sim 1 {\rm Mpc}/h$ we can therefore have a maximum variation of $\alpha$, achieved in the broken phase $\phi = \phi_0$, of
\be
\left(\frac{\Delta \alpha}{\alpha}\right)_{\rm max} \simeq \beta_\gamma^2 \, ppm\,,
\ee
which for $\beta_\gamma \sim 1$ will lead to ppm-level variations.

In passing it is worth pointing out that an alternative possibility would be the well motivated exponential coupling
\be
A_{\gamma}(\phi) = e^{\frac{\beta_\gamma\phi}{M_{\rm pl}}} \simeq  1 + \frac{\beta_\gamma\phi}{M_{\rm pl}}\,,
\ee
which we have expanded as a linear function since the argument of the exponential will clearly be required by observations to be much less than unity. However, this coupling does not respect the $\phi\to-\phi$ symmetry. In this case we would find
\be
\frac{\Delta \alpha}{\alpha} = \beta_\gamma\beta\left(\frac{0.5}{a_{\rm SSB}}\right)^3\left(\frac{\phi}{\phi_0}\right)\left(\frac{\lambda_{\phi 0}}{{\rm Mpc}/h}\right)^2\left(\frac{\Omega_{m0}}{0.25}\right)\,ppm\,,
\ee
which for $\beta_\gamma \sim 1$ is again of the same order as found above for the quadratic coupling.

Note that in the last scenario the variation is proportional to $\phi$ instead of $\phi^2$. This means that the variation can have both signs if the symmetry is broken differently in different places in the Universe---in other words, if we have domain walls. At a naive, qualitative level, a domain wall based scenario capable of accounting for the claimed dipole would simultaneously require low tension walls (so they evade other cosmological constraints, in particular from the CMB \cite{Lazanu}) and presumably a number of walls per Hubble volume of order unity; those two requirements are not compatible for the simplest domain wall models \cite{Walls1,Walls2}, although they may be made so with some fine-tuning \cite{Walls3,Walls4}.

In \cite{Marvin} this scenario was further studied, using N-body simulations taken from the earlier work in \cite{symmnbody}, in which the full spatial distribution of $\alpha$ at different redshifts has been calculated. Simulated sky maps for this variation were obtained, and their power spectrum calculated. The N-body simulations confirm that in high-density regions of space (such as deep inside dark matter halos) the value of $\alpha$ approaches the value measured on Earth, while in the low-density outskirts of halos the scalar field value can approach the symmetry breaking value, leading to significantly different values of $\alpha$. Importantly, these results also show that with low-redshift symmetry breaking these models exhibit some dependence of $\alpha$ on look-back time (as opposed to a pure spatial dipole) which could in principle be detected by sufficiently accurate spectroscopic measurements.

The matter power-spectrum is a useful way to characterize the clustering scales of matter in the universe \cite{Lyth}. Likewise, a  power-spectrum of $\alpha$ will track the clustering scales of the scalar-field, since its behavior is what determines $\alpha$. As we now discuss, the $\alpha$ power-spectrum is closely related to the matter power-spectrum for the symmetron model.

At the linear level and in the quasi-static approximation, the perturbations of the scalar field in Fourier space, $\phi(k,a) = \overline{\phi}(a) + \delta\phi(k,a)$, satisfy \cite{unified}
\be
\delta\phi \simeq -\frac{\overline{\rho}_m}{M_{\rm Pl}}\frac{\beta a^2}{k^2 + a^2m_\phi^2}\left(\frac{\overline{\phi}}{\phi_0}\right)\delta_m\,,
\ee
where $m_\phi^2 = V_{\rm eff,\phi\phi}(\overline{\phi})$ is the scalar field mass in the cosmological background, $\delta_m$ is the matter density contrast and $k$ is the co-moving wavenumber. The Fourier modes of $\alpha$ at linear scales then become
\be
\frac{\alpha(k,a)}{\alpha_0} = 1 + \frac{1}{2}\left(\frac{\beta_\gamma(\overline{\phi} + \delta\phi)}{M}\right)^2 \simeq  \frac{\overline{\alpha}(a)}{\alpha_0} + \frac{\beta_\gamma^2\overline{\phi}\delta\phi}{M^2} 
\ee
which we can write
\be
\frac{\alpha(k,a)}{\alpha_0} = \frac{\overline{\alpha}(a)}{\alpha_0}  - \left(\frac{\overline{\phi}}{\phi_0}\right)^2\left(\frac{\overline{\rho}_m}{M_{\rm Pl}^2}\frac{\beta_\gamma^2\beta^2a^2}{k^2 + a^2m_\phi^2}\delta_m\right)\,,
\ee
where
\be
\overline{\alpha}(a) \equiv \alpha_0\left[1+ \frac{1}{2}\left(\frac{\beta_\gamma\overline{\phi}(a)}{M}\right)^2\right]
\ee
is the value of $\alpha$ corresponding to the scalar field value in the cosmological background. To construct a power-spectrum of $\alpha$ it is convenient to compare $\alpha(k,a)$ relative to $\overline{\alpha}(a)$ since
\be\label{aexpr}
\frac{\alpha(k,a) - \overline{\alpha}(a)}{\alpha_0} \simeq  -\beta_\gamma^2\beta^2\frac{3\Omega_m}{a}\frac{H_0^2}{k^2 + a^2m_\phi^2}\delta_m
\ee
is directly proportional to the matter perturbation $\delta_m$. We therefore define
\be
P_{\alpha}(k,a) \equiv \left|\frac{\alpha(k,a) - \overline{\alpha}}{\alpha_0}\right|^2\,.
\ee
Using Eq.~(\ref{aexpr}) we find
\be\label{apow}
P_{\alpha}(k,a) =  \left[\frac{3\Omega_mH_0^2\beta_\gamma^2\beta^2}{a(k^2 + a^2m_\phi^2)} \left(\frac{\overline{\phi}}{\phi_0}\right)^2\right]^2P_m(k,a)\,,
\ee
where $P_m(k,a) = |\delta_m(k,a)|^2$ is the matter power-spectrum. The background field value and the scalar field mass are given by \cite{symmnbody}
\be
\left(\frac{\overline{\phi(a)}}{\phi_0}\right)^2 = \left(1 - \left(\frac{a_{\rm SSB}}{a}\right)^3\right)\,,\quad a \geq a_{\rm SSB}
\ee
\be
 m_\phi^2(a) = \frac{1}{\lambda_{\phi 0}^2}\left(1 - \left(\frac{a_{\rm SSB}}{a}\right)^3\right)\,,\quad a \geq a_{\rm SSB}\,,
\ee
and by using $H_0 = \frac{h}{2.998\cdot 10^3 {\rm Mpc}}$ we get
\be
P_{\alpha}(k,a)=\left[\frac{0.33\cdot\Omega_m10^{-6}\beta_\gamma^2\beta^2}{a((k/m_\phi)^2 + a^2)}\left(\frac{\lambda_{\phi 0}}{{\rm Mpc}/h}\right)^2\right]^2P_m(k,a)\,.
\ee
The analysis in \cite{Marvin} confirms that the analytic result of Eq.~(\ref{apow}), which is based on perturbation theory, gives a remarkably good fit (modulo a constant factor) up to  $k\sim 3~h/$Mpc which coincides with the particle Nyquist frequency of the simulation and the grid used to calculate the power-spectrum (in other words we cannot trust the results for larger wavenumbers). This result implies that the perturbations in the scalar field track the matter perturbations very closely even in the non-linear regime. In modified gravity models with a screening mechanism such as the symmetron this sort of effect is expected as the scalar field will sit close to the minimum of the effective potential, which is determined by the local matter density, in most regions of space.

As we already saw in Section \ref{uveslp}, at a phenomenological level it is common to fit the astrophysical measurements with a simple dipole, with or without an additional dependence on redshift or look-back time. On the other hand, from a theoretical point of view simplistic dipole models would require significant fine-tuning to explain such a behavior, and as the symmetron example illustrates a physically motivated approach would rely on environmental dependencies \cite{olivepospelov}. This therefore calls for more robust methodologies which enable accurate comparisons between models and observations. Early work along these lines was done by Murphy {\it et al.}, who calculated the two-point correlation function of the Keck subsample of the aforementioned archival data, finding it to be consistent with zero \cite{MWF}.

A more comprehensive and robust methodology to test models with spatial variations of the fine-structure constant $\alpha$, based on the calculation of the angular power spectrum of these measurements, has been recently introduced in \cite{PowerA}. This is based on the calculation of the 2D angular power spectrum of these measurements, which can then be related to the 3D power spectrum by standard methods \cite{jeong}, including the Limber approximation  \cite{loverde}. Applying it to the case of symmetron models and using the $\alpha$ measurements already described in Section \ref{uveslp} of this review, this analysis finds no indications of deviations from the standard behavior, with current data providing an upper limit to the strength of the symmetron coupling to gravity
\be
\log{\beta^2}<-0.9
\ee
when this is the only free parameter, and not able to constrain the model when also the symmetry breaking scale factor $a_{SSB}$ is free to vary. Future more precise $\alpha$ measurements can significantly tighten this constraint.

\section{Dynamical dark energy and varying couplings\label{darkside}}

Observations suggest that the universe is dominated by an energy component whose gravitational behavior is quite similar to that of a cosmological constant. Although a cosmological constant is consistent with existing data, its value would need to be so much smaller that particle physics expectations that a dynamical scalar field is arguably a more likely explanation \cite{Peebles}. Such a field must be slow-rolling close to the present day (which is mandatory for $p<0$ and acceleration) and be dominating the dynamics, providing some $70\%$ or so of the critical density (which provides a rough normalization). It then follows that if the field couples to the rest of the model---which as previously mentioned it will naturally do, unless some new symmetry is postulated to suppress the couplings---it will lead to potentially observable long-range forces and time dependencies of the constants of nature.

In particular, a coupling to the electromagnetic sector will lead to spacetime variations of the fine-structure constant $\alpha$ \cite{Carroll,Dvali,Chiba}. Clearly in this scenario the same dynamical degree of freedom is responsible for the dark energy and the variation of $\alpha$; these are therefore Class I models, in the sense described in the previous section. Tests of the stability of fundamental couplings (whether they are detections of variations or null results) will constrain fundamental physics and cosmology. This therefore ensures a 'minimum guaranteed science': theoretical constraints will simply depend on the sensitivity of the tests.

As already emphasized in Section \ref{consts}, the importance of improved null results stems from the fact that there is no natural expectation for the scale of the putative variations, since they are controlled by an unknown parameter. But this also implies that any new, improved constraint will rule out some previously viable models. This is entirely analogous to cosmological constraints on dynamical dark energy: one is looking for deviations from the canonical behavior $w_\phi=p_\phi/\rho_\phi=-1$, without any idea of when (meaning, at what level) or if such deviations will be found.

Here we explicitly demonstrate this point, drawing on recent work and discussing specific examples, involving both canonical and non-canonical scalar fields which are constrained by a combination of background cosmology data (Type Ia supernova and Hubble parameter measurements) and astrophysical and local measurements of $\alpha$---the {\it Canonical Data set} that was already discussed in the previous section when discussing Bekenstein models. In passing we note that in all the models we will discuss the evolution of $\alpha$ is monotonic. While this is a common behavior in the simplest (and thus more natural) models, one can certainly have models where $\alpha$ displays oscillations. An example are the so-called exotic singularity models \cite{Dabrowski}, though in this case the value of the present-day drift of $\alpha$ tends to be comparatively large and therefore some fine-tuning is needed for these models to satisfy atomic clock bounds.

\subsection{Canonical scalar fields}

Dynamical scalar fields in an effective four-dimensional field theory are naturally expected to couple to the rest of the theory, unless a (still unknown) symmetry is postulated to suppress this coupling \cite{Carroll,Dvali,Chiba}. We will assume this to be the case for the dynamical degree of freedom responsible for the dark energy. Specifically we will assume a coupling between the scalar field, denoted $\phi$, and the electromagnetic sector, which stems from a gauge kinetic function $B_F(\phi)$
\begin{equation}
{\cal L}_{\phi F} = - \frac{1}{4} B_F(\phi) F_{\mu\nu}F^{\mu\nu}\,.
\end{equation}
One can assume this function to be linear,
\begin{equation}
B_F(\phi) = 1- \zeta \kappa (\phi-\phi_0)\,,
\end{equation}
(with $\kappa^2=8\pi G$) since, as was pointed out in \cite{Dvali}, the absence of such a term would require the presence a $\phi\to-\phi$ symmetry, but such a symmetry must be broken throughout most of the cosmological evolution. As is physically clear, the relevant parameter in the cosmological evolution is the field displacement relative to its present-day value (in particular $\phi_0$ could be set to zero). In these models the proton and neutron masses are also expected to vary, due to the electromagnetic corrections of their masses, and one relevant consequence of this fact is that local tests of the Equivalence Principle lead to the conservative constraint on the dimensionless coupling parameter (see \cite{Uzan} for an overview)
\begin{equation}
|\zeta_{\rm local}|<10^{-3}\,,\label{localzeta}
\end{equation}
while in \cite{Erminia1} an independent few-percent constraint on this coupling was obtained using CMB and large-scale structure data in combination with direct measurements of the present-day Hubble parameter.

We note that there is in principle an additional source term driving the evolution of the scalar field, due to a $F^2B_F'$ term. By comparison to the standard (kinetic and potential energy) terms, the contribution of this term is expected to be subdominant, both because its average is zero for a radiation fluid and because the corresponding term for the baryonic density is tightly constrained by the same reasons discussed in the previous paragraph. For these reasons, in what follows we neglect this term, which would lead to spatial/environmental dependencies. We nevertheless note that this term can play a role in scenarios where the dominant standard term is suppressed---an example is the symmetron scenario which was discussed in the previous section.

With these assumptions one can explicitly relate the evolution of $\alpha$ to that of dark energy, as in \cite{Erminia1} whose derivation we summarize here. The evolution of $\alpha$ can be written
\begin{equation}
\frac{\Delta \alpha}{\alpha} \equiv \frac{\alpha-\alpha_0}{\alpha_0} =B_F^{-1}(\phi)-1=\zeta \kappa (\phi-\phi_0) \,,
\end{equation}
and defining the fraction of the dark energy density
\begin{equation}
\Omega_\phi (z) \equiv \frac{\rho_\phi(z)}{\rho_{\rm tot}(z)} \simeq \frac{\rho_\phi(z)}{\rho_\phi(z)+\rho_m(z)} \,,
\end{equation}
where in the last step we have neglected the contribution from radiation (since we will be interested in low redshifts, $z<5$, where it is indeed negligible), the evolution of the putative scalar field can be expressed in terms of the dark energy properties $\Omega_\phi$ and $w_\phi$ as \cite{Nunes}
\begin{equation}
1+w_\phi = \frac{(\kappa\phi')^2}{3 \Omega_\phi} \,,
\end{equation}
with the prime denoting the derivative with respect to the logarithm of the scale factor. We finally obtain
\begin{equation} \label{eq:dalfa}
\frac{\Delta\alpha}{\alpha}(z) =\zeta \int_0^{z}\sqrt{3\Omega_\phi(z')\left(1+w_\phi(z')\right)}\frac{dz'}{1+z'}\,.
\end{equation}
The last relation assumes a canonical scalar field, but the argument can be repeated for phantom fields \cite{Phantom}, leading to 
\begin{equation} \label{eq:dalfa2}
\frac{\Delta\alpha}{\alpha}(z) =-\zeta \int_0^{z}\sqrt{3\Omega_\phi(z')\left|1+w_\phi(z')\right|}\frac{dz'}{1+z'}\,;
\end{equation}
the change of sign stems from the fact that one expects phantom filed to roll up the potential rather than down. As is physically clear, if one does not detect variations of $\alpha$, either the field dynamics is very slow (in other words, its equation of state is very close to $w=-1$) or the coupling is very small. Therefore astrophysical measurements mainly constrain the product of a cosmological parameter and a fundamental physics one. 

The realization that varying fundamental couplings induce violations of the universality of free fall goes back at least to the work of Dicke---we refer the reader to \cite{DamDon} for a recent thorough discussion. In our present context, the key point is that a light scalar field such as we are considering inevitably couples to nucleons due to the $\alpha$ dependence of their masses, and therefore it mediates an isotope-dependent long-range force. This can be simply quantified through the dimensionless E\"{o}tv\"{o}s parameter $\eta$, which describes the level of violation of the Weak Equivalence Principle. One can show that for the class of models we are considering $\eta$ and the dimensionless coupling $\zeta$ are simply related by \cite{Dvali,Chiba,Damour,Uzan}
\begin{equation} \label{eq:eotvos}
\eta \approx 10^{-3}\zeta^2\,;
\end{equation}
note that the relation is different from the ones obtained for Class II models in the previous section.

The first detailed analysis to explore this possibility was done in \cite{Pinho1}, constraining models with a constant equation of state $w(z)=w_0$, and using the data sets available at the time: with respect to the data sets we listed in the previous section there were only 28 Hubble parameter measurements and only 11 dedicated measurements of $\alpha$. The present-day values of the Hubble parameter and matter density were respectively fixed to $H_0=70$ km/s/Mpc and $\Omega_{m}=0.3$, and a flat universe was assumed, so $\Omega_{\phi}=0.7$. This choice of cosmological parameters is consistent with the supernova and Hubble parameter data being used, and one can verify that allowing $H_0$, $\Omega_m$ or the curvature parameter to vary (within observationally reasonable ranges) and marginalizing over these parameters does not significantly change the results. This should be intuitively clear: a ppm level variation of $\alpha$ cannot noticeably affect these cosmological parameters. It is clear that the critical cosmological parameter here is $w_0$ itself---it is the one that is correlated with $\zeta$. For the models under consideration the present-day drift rate, which is constrained by atomic clocks, is
\begin{equation} \label{clocks2}
\frac{1}{H_0}\frac{\dot\alpha}{\alpha} =\, \mp \, \zeta\sqrt{3\Omega_{\phi0}|1+w_0|}\,,
\end{equation}
with the $-$ and $+$ signs respectively corresponding to the canonical and phantom field cases.

\begin{figure}
\begin{center}
\includegraphics[width=3in]{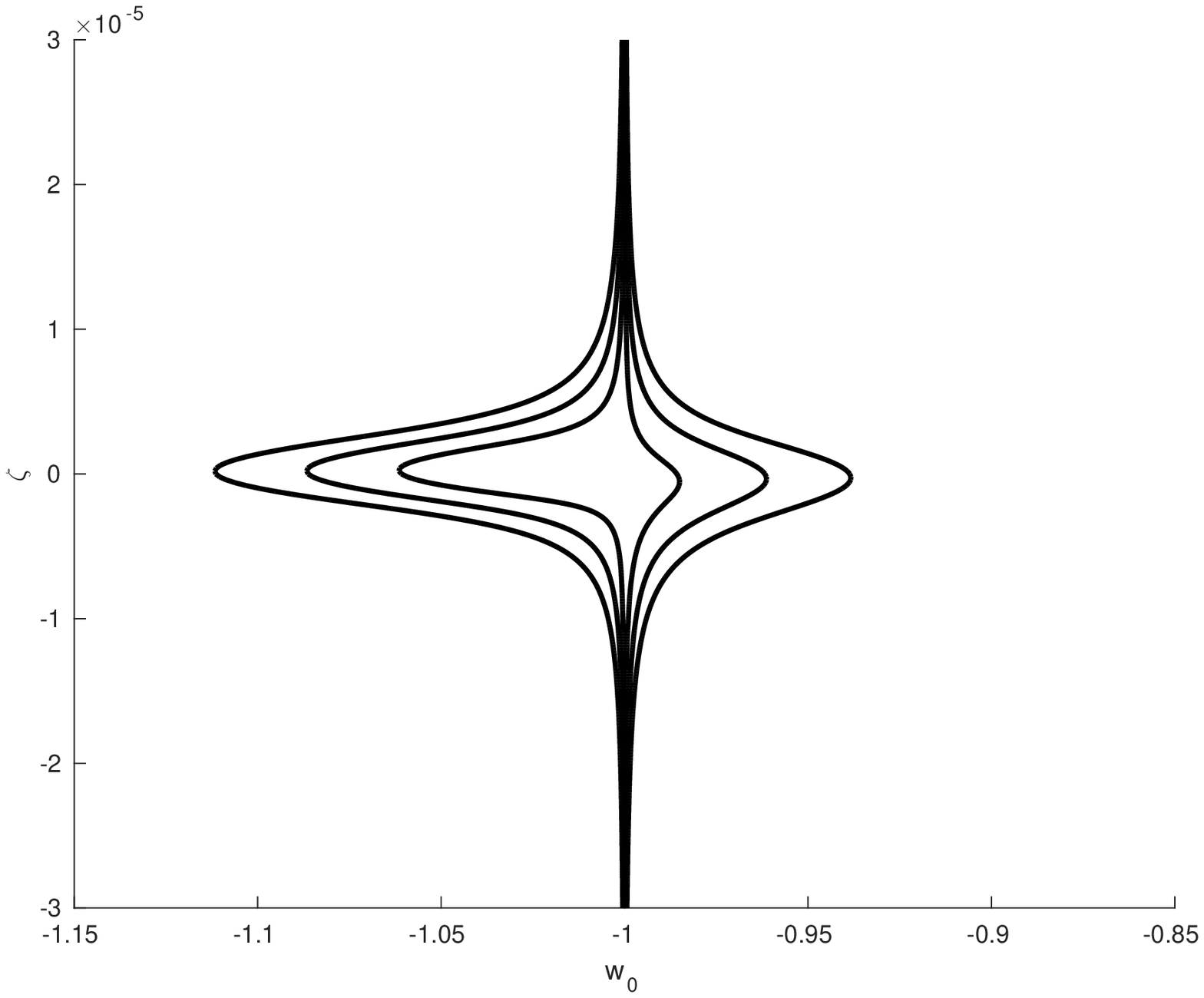}
\vskip0.2in
\includegraphics[width=3in]{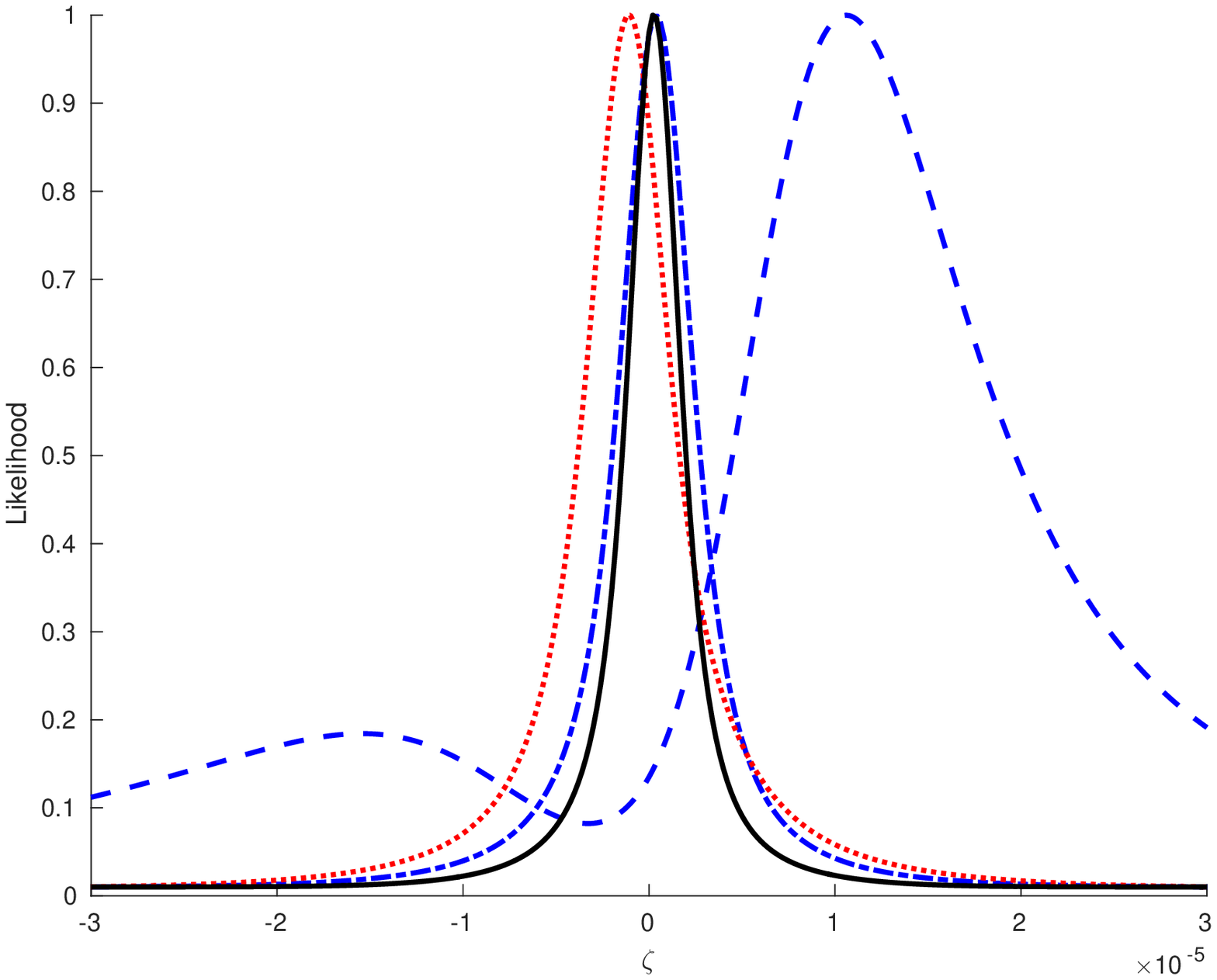}
\end{center}
\caption{\label{fig05}{\bf Top panel:} One, two and three sigma constraints on the $\zeta-w_0$ plane from the full Canonical Data set discussed in the text; the reduced chi-square at the maximum likelihood is $\chi^2_\nu=0.97$. This updates the analysis of \protect\cite{Pinho1}. {\bf Bottom panel:} 1D likelihood for $\zeta$, marginalizing over $w_0$, for cosmological $+$ Webb {\it et al.} data (blue dashed), cosmological $+$ the dedicated $\alpha$ measurements of Table \protect\ref{tab01}  and Oklo (blue dash-dotted), cosmological $+$ Rosenband {\it et al.} atomic clock bound (red dotted) and the combination of all data sets (black solid).}
\end{figure}

Figure \ref{fig05}, which updates the analysis in \cite{Pinho1}, shows the results of this analysis: in a nutshell the data is compatible with the null result, although the Webb {\it et al.} archival data does show a weak preference (at about the two sigma level) for a non-zero coupling. The cosmological data we are considering is insensitive to $\zeta$. (Strictly speaking, a varying $\alpha$ does affect the peak luminosity of Type Ia supernovas \cite{Kohri}, but as shown in \cite{Calabrese} for ppm level $\alpha$ variations the effect is too small to have an impact on current data sets, so this effect can be neglected.) Naturally, the cosmological data does constrain $w_0$, effectively providing a prior on it. The analysis of \cite{Pinho1} found the two-sigma ($95.4\%$) bound
\begin{equation} \label{newzetabound}
|\zeta|<5\times10^{-6}\,,
\end{equation}
while at three-sigma $\zeta$ is was unconstrained. With the additional measurements of the Hubble parameter and $\alpha$ we now find the two-sigma ($95.4\%$) bound
\begin{equation} \label{newzetabound2}
\zeta=(0.2\pm3.9)\times10^{-6}\,,
\end{equation}
and we can also obtain a three-sigma upper bound
\begin{equation} \label{newzetabound3}
|\zeta|<1.8\times10^{-5}\,.
\end{equation}
This leads to the two-sigma indirect bound
\begin{equation} \label{etaboundfix}
\eta<1.6\times10^{-14}\,,
\end{equation}
a $40\%$ improvement relative to the bound in \cite{Pinho1}, which as $\eta<2.5\times10^{-14}$. Again, note that this bound is much stronger than the current direct bounds that were discussed in the previous section, cf. Eqs. \ref{etaTorsion}--\ref{etaLLR}.

The above constraints were obtained by assuming a fiducial model for the dark energy with a constant equation of state, $w(z)=w_0$. This assumption was relaxed in \cite{Pinho2}, who studied more general models where the dark energy equation of state does vary with redshift, thereby assessing how said results depend on the choice of fiducial model for the dark energy. Aiming to preserve conceptual simplicity, the alternative parametrizations chosen did not increase the number of free parameters. We will briefly summarize this analysis, though note that unlike the constant equation of state case we will not update the constraints in the light of the newest data. 

The first such parametization is the one recently introduced by Slepian {\it et al.} \cite{SGZ}. The Friedmann equation has the following form
\begin{equation}\label{newsqrt}
\frac{H^2(z)}{H^2_0}=\Omega_{m}(1+z)^{3}+\Omega_{\phi}\left[\frac{(1+z)^{3}}{\Omega_{m}(1+z)^{3}+\Omega_{\phi}}\right]^{\frac{1+ w_0}{\Omega_{\phi}}} \,.
\end{equation}
One can assume flat universes, so $\Omega_m+\Omega_\phi=1$, and the model is therefore characterized by three independent parameters: $H_0$, $\Omega_m$ (both kept fixed as previously justified) and $w_0$, the latter being still the value of the dark energy equation of state today. The dark energy equation of state has the following behavior
\begin{equation}\label{darkeos}
w(z)=-1+(1+w_0)\frac{H_0^2}{H^2(z)}\,.
\end{equation}
Note that for high redshifts this always approaches -1, and it diverges from this value as the universe evolves, reaching $w_0$ today. This is therefore a parametrization for thawing models, in the classification of \cite{FrTh}. Apart from its simplicity, this choice of parametrization is also motivated by the recent result that if physical priors are used, allowed quintessence models are mostly thawing \cite{Marsh}. 

The results of this analysis, presented in \cite{Pinho2}, are qualitatively similar to those for a model with a constant equation of state. At the two-sigma ($95.4\%$) confidence level one finds
\begin{equation} \label{zetaboundSGZ}
|\zeta_{\rm SGZ}|<5.6\times10^{-6}\,,
\end{equation}
which leads to a constraint on WEP violations
\begin{equation} \label{etaboundSGZ}
\eta_{\rm SGZ}<3.1\times10^{-14}\,.
\end{equation}
Quantitatively, these constraints are slightly weaker than those obtained for the constant equation of state model, cf. Eq. \ref{newzetabound}. Physically, the reason for this is that in a thawing model with a given $w_0$ the amount of $\alpha$ variation at a given non-zero redshift will be slightly smaller than that in a constant equation of state model with the same $w_0$. In any case, the inferred indirect WEP bound is still stronger than the available direct bounds.

Having considered thawing models, we can also discuss the opposite scenario: that of freezing models where the dark energy equation of state evolves towards $-1$. A further motivation here stems from the fact that in many dilaton-type models the scalar field depends logarithmically on the scale factor
\begin{equation}
{\phi}(z)\propto \log{(1+z)}\,.
\end{equation}
The runaway dilaton scenario discussed in Section \ref{models} is an obvious example of this. (By comparison, note that in BSBM models the field departs from this behavior and freezes quite abruptly at the onset of the acceleration phase \cite{SBM}.) For a linear gauge kinetic function as we are assuming here, it follows that in that case
\be
\frac{\Delta\alpha}{\alpha}\propto\ln(1+z)\,.
\ee
It is therefore an interesting exercise to determine what condition on the dark energy equation of state for Class I models will lead to such a behavior for $\alpha(z)$, while bearing in mind that some Class II models are also known to display such a behavior. 

The answer to this question can be found with a little algebra. From Eq. \ref{eq:dalfa} we infer that the function inside the square root therein must be a constant, that is
\begin{equation}
\Omega_{\phi}(z)[1+w(z)]=const.\,;
\end{equation}
upon differentiation this can be recast into the following equation
\begin{equation}
\frac{dw}{dz}=-3(1+w_0)\frac{w}{1+z}\left[\frac{1+w}{1+w_0}-\Omega_{\phi0}\right]\,. \label{eq:diff}
\end{equation}
Note that the initial condition for the first derivative is
\begin{equation}
\left[\frac{dw}{dz}\right]_{0}=-3\Omega_m w_0(1+w_0)\,,
\end{equation}
and for the second one we could also write
\begin{equation}
\left[\frac{d^2w}{dz^2}\right]_{0}=3\Omega_m w_0(1+w_0)[1+3w_0+3\Omega_m(1+w_0)]\,,
\end{equation}
so $w'\sim3\Omega_m(1+w_0)$ and $w''\sim6\Omega_m(1+w_0)$ near the $\Lambda$CDM limit. The above equation can be easily integrated, leading to the solution \cite{Pinho2}
\begin{equation}\label{darkeos2}
w(z)=\frac{[1-\Omega_\phi(1+w_0)]w_0}{\Omega_m(1+w_0)(1+z)^{3[1-\Omega_\phi(1+w_0)]}-w_0}\,,
\end{equation}
bearing in mind that we are assuming that $\Omega_m+\Omega_\phi=1$. An analogous solution was obtained, in a different context, in \cite{Nunes}.

The Friedmann equation in this case has the explicit form
\bq\label{newsqrt2}
\frac{H^2(z)}{H^2_0}&=& \Omega_{m}(1+z)^{3}+\\
&+&\frac{\Omega_{\phi}\left[\Omega_m(1+w_0)(1+z)^3-w_0(1+z)^{3\Omega_\phi(1+w_0)}\right]}{\Omega_m(1+w_0)-w_0}\,, \nonumber
\eq
and naturally the evolution of $\alpha$ is given by
\begin{equation}
\frac{\Delta\alpha}{\alpha}(z)=\zeta\, \sqrt{3\Omega_\phi(1+w_0)}\, \ln{(1+z)}\,.
\end{equation}
One can now treat this parametrization phenomenologically, allow for values of $w_0<-1$ with a flat prior on ($1+w_0$), and fit it to the available data sets. This will lead to slightly tighter constraints, for the reason already explained: in a freezing model with a given $w_0$ the amount of $\alpha$ variation at a given non-zero redshift will be slightly larger than that in a constant equation of state model with the same $w_0$. In this case \cite{Pinho2} does find
\be
|\zeta|<4.6\times10^{-6}
\ee
at the two-sigma confidence level.

However, the above analysis may be too simplistic, since in this case one physically expects that $w_0\ge-1$. Discarding the phantom part of this parameter space one can instead use this model as a testbed for the effects of the choice of priors, and replace the flat prior on ($1+w_0$) with a logarithmic one. The analysis of \cite{Pinho2} confirms the expectation that, in principle, for a value of $w_0$ sufficiently close to $w_0=-1$ any value of the coupling would be allowed. In practice, of course, this is not so because the local WEP constraints must be satisfied. Analogously, in principle, and given the form of Eqs. \ref{eq:dalfa}-\ref{eq:dalfa2}, exactly the same could be said about the orthogonal direction (for a sufficiently small $\zeta$ any $w_0$ would be allowed), but in practice this is prevented by the strong priors on $w_0$ coming from the cosmological data sets. Nevertheless, in this case the bound on the coupling are somewhat weakened. At the one-sigma confidence level \cite{Pinho2} find for this case
\begin{equation} \label{zetaboundDIL1}
|\zeta_{\rm DIL}|<6\times10^{-6}\,,
\end{equation}
while at the two-sigma level
\begin{equation} \label{zetaboundDIL2}
|\zeta_{\rm DIL}|<2.5\times10^{-5}\,;
\end{equation}
translating these into WEP bounds, the one sigma constraint is still stronger than the direct bounds: at the one-sigma confidence level
\begin{equation} \label{etaboundDIL2}
|\eta_{\rm DIL}|<4\times10^{-14}\,.
\end{equation}
Therefore, although these constraints do exhibit some model dependence (both in terms of the class of models being assumed and in terms of the underlying priors), they are generically competitive with other existing tests of these models.

A possible caveat of the above analyses is that it is based on fiducial models where the dark energy equation of state was described by a single parameter (its present day value, $w_0$). Since there are degeneracies between the coupling $\zeta$ and $w_0$ (which are partially broken by the cosmological data sets) one may legitimately ask how robust these constraints are. A follow-up study in \cite{Pinho3} addresses this issue, extending the analysis to more general---and, arguably, more realistic---dark energy models: the well-known Chevallier-Polarski-Linder \cite{CPL1,CPL2} (hereafter CPL) and early dark energy \cite{EDE} (hereafter EDE) classes, as well as a parametrization recently discussed by Mukhanov \cite{MKH} (hereafter MKH). Compared to the models discussed above, each of these has one additional free parameter characterizing dark energy, but this extra parameter plays a different role in each of the parametrizations. Even in these extended cases it is found that the current data constrains the coupling $\zeta$ at the $10^{-6}$ level (marginalizing over other parameters), thus confirming the robustness of earlier analyses. On the other hand, the additional dark energy parameter is typically not well constrained. Again we will succinctly discuss these results. 

In the CPL parametrization the dark energy equation of state has the form
\begin{equation} \label{cpl}
w_{\rm CPL}(z)=w_0+w_a \frac{z}{1+z}\,,
\end{equation}
where $w_0$ is still its present value and $w_a$ is the coefficient of the time-dependent term. The redshift dependence of this parametrization is not intended to mimic a particular model for dark energy, but rather to enable the description of possible deviations from the $\Lambda$CDM standard paradigm without the assumption of any specific underlying theory. While it is a choice, it's one that is to some extent simple and relatively generic. Nevertheless, we can assume that also this kind of dark energy is produced by a scalar field, coupled to the electromagnetic sector. In this model the fraction of energy density provided by the scalar field is easily found to be
\begin{equation}
\Omega_{\rm CPL}(z)=\frac{1-\Omega_{\rm m}}{1-\Omega_{\rm m}+\Omega_{\rm m}(1+z)^{-3(w_0+w_a)}e^\frac{3w_az}{1+z}}\,,
\end{equation}
where $\Omega_{\rm m}$ is the present time matter density and we have also assumed a flat universe. Using the same cosmological and astrophysical data sets as before, \cite{Pinho3} find the following bounds for the coupling
\begin{equation} \label{cplzeta1}
\zeta=(1\pm3)\times10^{-6}\qquad {\rm (95.4\% C.L.)} \,
\end{equation}
\begin{equation} \label{cplzeta2}
\zeta=(1\pm8)\times10^{-6}\qquad {\rm (99.7\% C.L.)} \,,
\end{equation}
which implies a bound on the E\"{o}tv\"{o}s parameter
\begin{equation} \label{cpleta}
\eta<1.6\times10^{-14}\qquad {\rm (95.4\% C.L.)} \,.
\end{equation}
Compared to the earlier results the constraint on $\zeta$ (and consequently that on $\eta$) is now stronger. This is to be expected since $\zeta$ is correlated with the dark energy equation of state parameters: with the equation of state allowed to be further away from a cosmological constant, larger variations of $\alpha$ also become possible, and the existing $\alpha$ measurements therefore impose a tighter constraint on $\zeta$. This effect was also noticed in the case of the forecasts discussed in \cite{Calabrese}.

To assess the model-dependence of the above constraints one can again repeat the analysis for an alternative parametrization of the dark energy equation of state. The MKH parametrization provides as useful comparison point. This was introduced in an inflationary context \cite{MKH}, but it can be trivially applied for the case of the recent acceleration of the universe. Here the dark energy equation of state is
\begin{equation} \label{mkheos}
w_{\rm MKH}(z)=-1+\frac{1+w_0}{\left[1+\ln{(1+z)}\right]^\beta}\,,
\end{equation}
where $w_0$ is its present day value and the slope $\beta$ controls the overall redshift dependence. Specifically $\beta<0$ corresponds to freezing models, $\beta=0$ to a constant equation of state and $\beta>0$ to thawing models, in the classification of \cite{FrTh}. This corresponds to the following behavior of the dark energy density
\begin{eqnarray}
\frac{\rho_{MKH}(z)}{\rho_0} &=& \exp\left[3\frac{1+w_0}{1-\beta}\left([1+\ln(1+z)]^{1-\beta}-1\right) \right]\,, \beta\neq1 \\
\frac{\rho_{MKH}(z)}{\rho_0} &=& \left[1+\ln{(1+z)}\right]^{3(1+w_0)} \,, \beta=1
\label{rhomkh}
\end{eqnarray}
and it is easy to verify that this has the correct behavior in the appropriate limits. The analysis shows that freezing models (with $\beta<0$) are comparatively more constrained than thawing ones (with $\beta>0$). Physically the reason for this is again clear: for a given value of $w_0$, a freezing model leads to a larger variation of $\alpha$ than an thawing one, and is therefore more tightly constrained by current data. In this case the 1D marginalized constraint on the scalar field coupling is
\begin{equation} \label{mkhzeta}
|\zeta|<6\times10^{-6}\qquad {\rm (95.4\% C.L.)} \,
\end{equation}
leading to
\begin{equation} \label{mkheta}
\eta<3.6\times10^{-14}\qquad {\rm (95.4\% C.L.)} \,.
\end{equation}
In these case, and as compared to the CPL case, we get weaker constraints on $\zeta$.

Finally, in the EDE class of models the dark energy density fraction is
\bq
\Omega_{\rm EDE}(z) &=& \frac{1-\Omega_m - \Omega_e \left[1- (1+z)^{3 w_0}\right] }{1-\Omega_m + \Omega_m (1+z)^{-3w_0}} +\\ \nonumber
&+& \Omega_e \left[1- (1+z)^{3 w_0}\right] \label{edeomega} 
\eq
while the dark energy equation of state is
\begin{equation}
w_{\rm EDE}(z)=-\frac{1}{3[1-\Omega_{\rm EDE}]} \frac{d\ln\Omega_{\rm EDE}}{d\ln a} + \frac{a_{eq}}{3(a + a_{eq})}\,;
\label{eq:edew}
\end{equation}
here $a_{eq}$ is the scale factor at the epoch of equal matter and radiation densities. The energy density $\Omega_{\rm EDE}(z)$ has a scaling behavior evolving with time and approaching a finite constant $\Omega_e$ in the past, rather than approaching zero as was the case for the other canonical models just considered. A flat universe is also assumed. The present day value of the equation of state is $w_0$, and the equation of state follows the behavior of the dominant component at each cosmic epoch, with $w_{\rm EDE}\approx1/3$ during radiation domination and $w_{\rm EDE}\approx0$ during matter domination. Even though this is a phenomenological parametrization, we will again  assume that this kind of dark energy is the result of an underlying scalar field, which couples to the electromagnetic sector.

Using a flat prior on $w_0$ and further assuming that $w_0\ge-1$, \cite{Pinho3} obtain non-trivial constraints on the fraction of early dark energy
\begin{equation} \label{edeome}
\Omega_e<0.033\qquad {\rm (95.4\% C.L.)} \,,
\end{equation}
which is about a factor of 3 weaker than the standard one without allowing for possible $\alpha$ variations. As for the coupling one finds
\begin{equation} \label{edezeta}
\zeta=(-1\pm5)\times10^{-6}\qquad {\rm (95.4\% C.L.)} \,
\end{equation}
leading to
\begin{equation} \label{edeeta}
\eta<3.6\times10^{-14}\qquad {\rm (95.4\% C.L.)} \,.
\end{equation}
Here, by comparison to the CPL case, the slightly stronger constraints on the dark energy sector imply slightly weaker constraints on the coupling $\zeta$. If instead a logarithmic (rather than flat) prior is used for $w_0$, the 1D marginalized constraints now become
\begin{equation} \label{logome}
\Omega_e<0.030\qquad {\rm (95.4\% C.L.)} \,,
\end{equation}
which is about ten percent stronger than the flat prior case, while the constraint on the coupling becomes weaker as well as asymmetric
\begin{equation} \label{logzeta}
\zeta=(-1^{+8}_{-11})\times10^{-6}\qquad {\rm (95.4\% C.L.)} \,
\end{equation}
leading to
\begin{equation} \label{logeta}
\eta<14.4\times10^{-14}\qquad {\rm (95.4\% C.L.)} \,;
\end{equation}
note that even in this case this constraint is still marginally stronger than the current direct bounds.

\subsection{Rolling tachyons}

The previous subsection focused on canonical scalar fields. However, the aforementioned Class I is more generic, and we will illustrate this by considering one example of a different class of models. Constraints on Dirac-Born-Infeld (DBI) type dark energy models from varying $\alpha$ have first been discussed in \cite{Garousi}. This work points out that the DBI action based on string theory naturally gives rise to a coupling between gauge fields and a scalar field responsible for the universe's acceleration. In other words, the field dynamics itself leads to $\alpha$ variations. They place constraints on particular choices of potentials, finding that some fine-tuning is needed: the potentials must be quite flat. This analysis was recently extended in \cite{Moucherek} by exploiting the availability of additional data, carrying out the analysis for more generic potentials, and also providing additional insight into the physical interpretation and relevance of the resulting constraints.

A rolling tachyon is an example of a Born-Infeld scalar, and these are well motivated in string theory \cite{Sen1,Sen2}. The interaction of scalar fields with gauge fields will naturally lead to fine-structure constant variations. A further relevant difference is that whereas the coupling of a quintessence field to matter and radiation is not fixed by the standard model of particle physics, these models provide an example where the form of the couplings can be obtained more directly from a fundamental theory, specifically from an effective D-brane action \cite{Garousi}. Therefore, apart form their intrinsic interest, they are also useful as a benchmark to study the discriminating power of future facilities among different classes of models since, as we will now see, they do have some interesting distinguishing features.

The tree-level D-brane action is a Dirac-Born-Infeld type action containing both gauge fields and scalar fields such as tachyons \cite{Sen1,Sen2}, and this action naturally gives rise to the coupling of the Born-Infeld scalars to the gauge fields, which can account for a varying $\alpha$. Rolling tachyon fields have been suggested as a candidate to explain the acceleration of the universe \cite{Sen1}. The cosmology of a homogeneous tachyon scalar field as dark energy was first studied in \cite{Bagla}, and the $\alpha$ variation for a Born-Infeld scalar coupled to the gauge field has been previously discussed in \cite{Garousi}, who obtain some qualitative constraints, and further quantified by  \cite{Moucherek}, whose analysis we now summarize and update.

We start by focusing on the tachyon part of the DBI action. Generically its Lagrangian can be written
\be
{\cal L_{\rm tac}}=-V(\phi)\sqrt{1-\partial_a\phi\partial^a\phi}\,,
\ee
with the energy density and pressure being given by
\be
\rho_\phi=\frac{V(\phi)}{\sqrt{1-\partial_a\phi\partial^a\phi}}
\ee
\be
p_\phi=-V(\phi)\sqrt{1-\partial_a\phi\partial^a\phi}\,.
\ee
For a homogeneous field in a Friedmann-Lemaitre-Robertson-Walker background containing also matter, we have
\be
H^2=\frac{8\pi G}{3}(\rho_m+\rho_\phi)
\ee
and
\be
\frac{\ddot \phi}{1-{\dot\phi}^2}+3H{\dot\phi}+\frac{1}{V}\frac{dV}{d\phi}=0\,.
\ee
Note that the tachyon field equation of state and sound speed are
\be
w_\phi={\dot\phi}^2-1\ge-1\,,
\ee
\be
c^2_s=1-{\dot\phi}^2\le1\,;
\ee
it is also useful to write
\be
{\dot\rho_\phi}=-3H(1+w_\phi)\rho_\phi=-3H\rho_\phi {\dot\phi}^2\,.
\ee
As an aside, note that in the case where the tachyon is the single component (ie, neglecting matter as well as radiation) there is a well-known solution \cite{Thanu}
\be
a\propto t^n
\ee
\be
\phi=\sqrt{\frac{2}{3n}}\, t
\ee
which ensues for the potential
\be
V(\phi)=\frac{n}{4\pi G}\left(1-\frac{2}{3n}\right)^{1/2}\frac{1}{\phi^2}\,.
\ee

To address the case including matter, we start by noting that in these models the field is constrained to be slow-rolling (especially so if it induces $\alpha$ variations, as we will shortly confirm), and in that case the scalar field equation can be approximated to
\be
3H{\dot\phi}\propto -\frac{d\ln{V}}{d\phi}\,.
\ee
Moreover, the right-hand side of this equation is a function of the field $\phi$ and the field is approximately constant. We can thus Taylor-expand the field, and write the Friedmann equation as follows
\be
\frac{H^2}{H_0^2}=\Omega_m(1+z)^3+\Omega_\phi\left[1+\left(\frac{V'}{V}\right)_0(\phi-\phi_0)\right]
\ee
with, from the scalar field equation,
\be
(\phi-\phi_0)= -\frac{1}{3}\left(\frac{1}{H}\frac{V'}{V}\right)_0(t-t_0)\,.
\ee
We therefore have
\be
\frac{H^2}{H_0^2}=\Omega_m(1+z)^3+(1-\Omega_m)\left[1-\left(\frac{V'}{V}\right)^2_0\frac{(t-t_0)}{3H_0}\right]\,,
\ee
where we also used $\Omega_m+\Omega_\phi=1$. Now, given the slow-roll approximation the correction term in square brackets is expected to be small, and therefore the calculation of the $(t-t_0)$ term can be done assuming the $\Lambda$CDM limit (in other words, the differences will be of higher order), which allows an analytic calculation to be done. After some algebra we find
\be
\frac{H^2}{H_0^2}=\Omega_m(1+z)^3+(1-\Omega_m)\left[1+\frac{2}{9}\lambda^2 f(\Omega_m,z)\right]\,,
\ee
where we have defined the dynamically relevant dimensionless parameter
\be
\lambda=\frac{1}{H_0}\left(\frac{V'}{V}\right)_0\,,
\ee
and the redshift-dependent correction factor is
\be
f(\Omega_m,z)=\frac{1}{\sqrt{1-\Omega_m}}\ln{\frac{(1+\sqrt{1-\Omega_m})(1+z)^{3/2}}{\sqrt{1-\Omega_m}+\sqrt{\Omega_m(1+z)^3+1-\Omega_m}}}\,.
\ee

It is also useful to calculate the dark energy equation of state in these models. This can be straightforwardly done using the relation
\be
\frac{d\rho_\phi}{dz}=3\frac{1+w_\phi}{1+z}\rho_\phi\,,
\ee
and leads to the following result
\bq
1+w_\phi&=&{\dot\phi^2} \\ \nonumber
&=&\frac{\lambda^2}{9+2\lambda^2f(\Omega_m,z)}\frac{\sqrt{1-\Omega_m}+\sqrt{E(\Omega_m,z)}}{E(\Omega_m,z)+\sqrt{(1-\Omega_m)E(\Omega_m,z)}}\,,
\eq
where for convenience we also defined
\be
E(\Omega_m,z)=\Omega_m(1+z)^3+1-\Omega_m\,.
\ee
As expected the field speed parametrizes the deviation of the dark energy equation of state from the cosmological constant value. Note that this equation of state $(1+w_\phi)$ tends to zero at high redshifts; in other words, these are thawing dark energy models. In particular, the equation of state at the present day is
\be
1+w_0={\dot\phi^2}_0=\frac{\lambda^2}{9}\,,
\ee
providing further physical insight into the role of the parameter $\lambda$.

Now we consider the interaction part of the DBI Lagrangian which is responsible for the $\alpha$ variation. This has the form \cite{Sen1,Sen2,Garousi}
\be
{\cal L_{\rm int}}=\frac{(2\pi{\alpha_s}')^2}{4\beta^2}V(\phi)Tr(g^{-1}Fg^{-1}F)+\ldots \,,
\ee
where $g$ and $F$ are the traces of the four-dimensional metric and the Maxwell tensor respectively, ${\alpha_s}'$ (not to be confused with the fine-structure constant) is related to the string mass scale via $M_s=1/\sqrt{{\alpha_s}'}$, and $\beta$ is a warped factor. We note that the DBI Lagrangian contains further terms that are of similar order in the gauge field, but these are not relevant for our present discussion since they do not contribute to the $\alpha$ variation. (A more systematic discussion of this point can be found in \cite{Sen1,Sen2}.) This implies, by comparison to the standard Yang-Mills case, that the value of the fine-structure constant in this case is
\be
\alpha(\phi)=\frac{\beta^2M_s^4}{2\pi}\frac{1}{V(\phi)}\,,
\ee
and therefore in these models the fine-structure constant is inversely proportional to the tachyon potential. Expressing this in terms of the relative variation of $\alpha$ with respect to the present day, we finally obtain
\be
\frac{\Delta\alpha}{\alpha}(z)=\frac{V(\phi_0)}{V(\phi)}-1\,,
\ee
where as usual $\alpha_0\sim1/137$ is the present-day value. Thus a negative value of $\Delta\alpha/\alpha$ corresponds to a smaller value of $\alpha$ in the past (meaning a weaker electromagnetic interaction), which in this class of models corresponds to a larger value of the potential $V(\phi)$.

Given this explicit dependence on the scalar field potential we can now use the same Taylor expansion and re-write this as
\be
\frac{\Delta\alpha}{\alpha}\simeq -\left(\frac{V'}{V}\right)_0(\phi-\phi_0)\simeq \frac{1}{3H_0}\left(\frac{V'}{V}\right)^2_0(t-t_0)\,.
\ee
This implies that in these models the fine-structure constant is always smaller in the past (and varies approximately linearly in time). Finally we can write
\be
\frac{\Delta\alpha}{\alpha}=-\frac{2}{9}\lambda^2f(\Omega_m,z)\,,
\ee
which shows that the dimensionless parameter $\lambda$ also provides the overall normalization for this variation. We could even write the suggestive
\be
\frac{H^2}{H_0^2}=\Omega_m(1+z)^3+(1-\Omega_m)\left[1-\frac{\Delta\alpha}{\alpha}(z)\right]\,.
\ee
This makes it clear that in this class of models any deviations from the $\Lambda$CDM behavior must be small, as we now further quantify. Indeed, we can trivially write the present-day rate of change of the fine-structure constant
\be
\frac{1}{H_0}\left(\frac{\dot\alpha}{\alpha}\right)_0= \frac{1}{3H_0^2}\left(\frac{V'}{V}\right)^2_0\,,
\ee
or equivalently, in terms of the present day dark energy equation of state
\be
\frac{1}{H_0}\left(\frac{\dot\alpha}{\alpha}\right)_0=\frac{1}{3}\lambda^2=3{\dot\phi_0}^2=3(1+w_0)\,.
\ee
As usual this drift rate is constrained by laboratory measurements with atomic clocks. Taking for example Rosenband {\it et al.} \cite{Rosenband} we have 
\be
\frac{1}{H_0}\left(\frac{\dot\alpha}{\alpha}\right)_0=(-2.2\pm3.2)\times10^{-7}\,,
\ee
which immediately shows that in these models $w_0$ is effectively indistinguishable from a cosmological constant, although they can have a distinctive astrophysical variation of $\alpha$. In this strict sense these models could actually be thought of as a physical realization of the more phenomenological Bekenstein-Sandvik-Barrow-Magueijo class of models \cite{SBM}, already discussed in the previous section. This constraint also implies that the field speed today must be tiny
\be
{\dot\phi_0}\leq 10^{-3}\,,
\ee
justifying our slow-roll approximation and also motivating the choice of a logarithmic prior for $\lambda$.

The work of \cite{Bagla} does a simple comparison with early Type Ia supernova observations. The recent \cite{Moucherek} extends this, using both the more recent Union2.1 supernova data set and the set of Hubble parameter measurements and $\alpha$ measurements available at the time. The value of the Hubble parameter was fixed to be $H_0=70$ km/s/Mpc, and as previously mentioned a flat universe was assumed, so $\Omega_m+\Omega_{\phi}=1$. These choices are again consistent with the cosmological data sets being used, and also with constraints from the cosmic microwave background \cite{PlanckParams}.

\begin{figure}
\begin{center}
\includegraphics[width=3in]{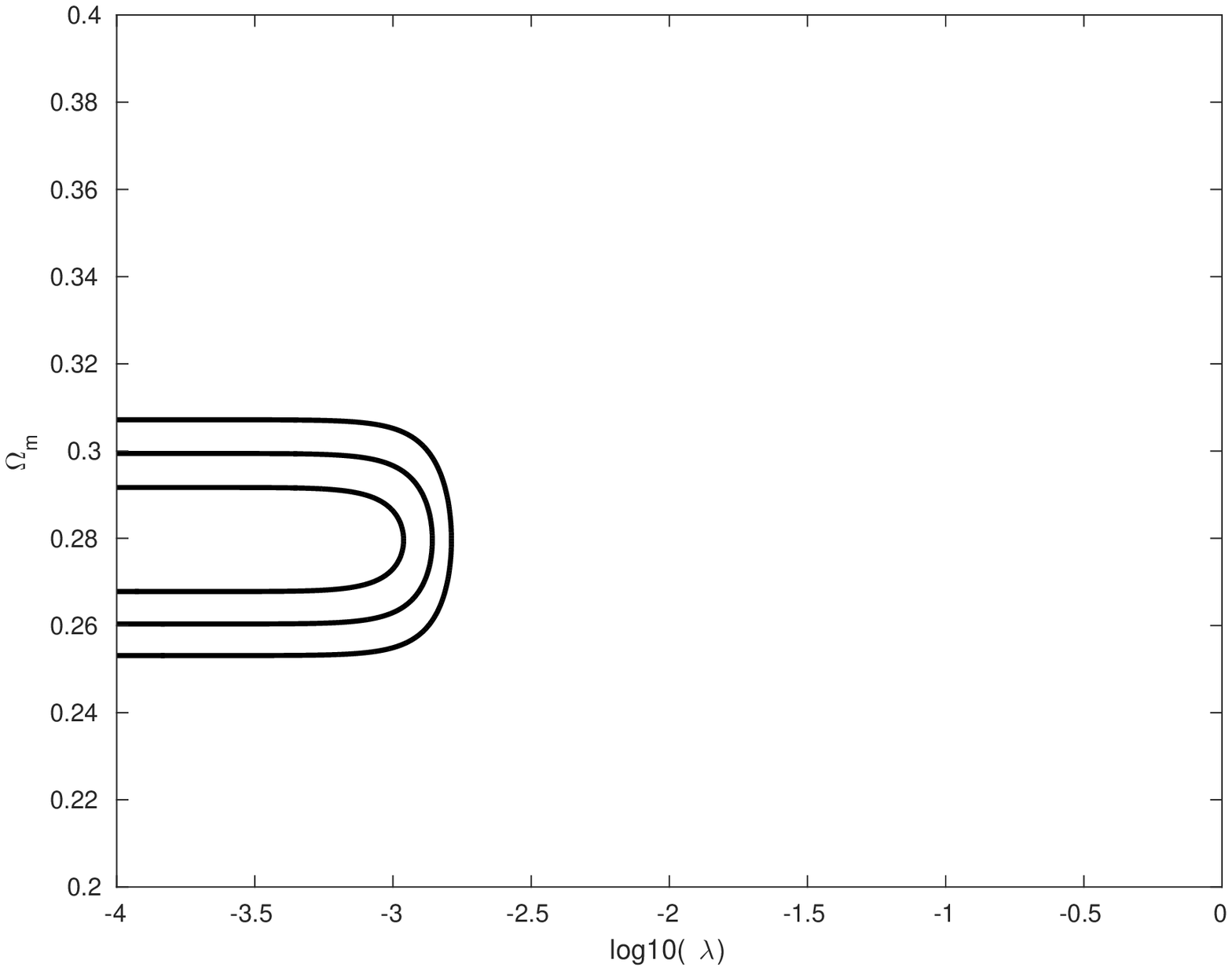}
\vskip0.2in
\includegraphics[width=3in]{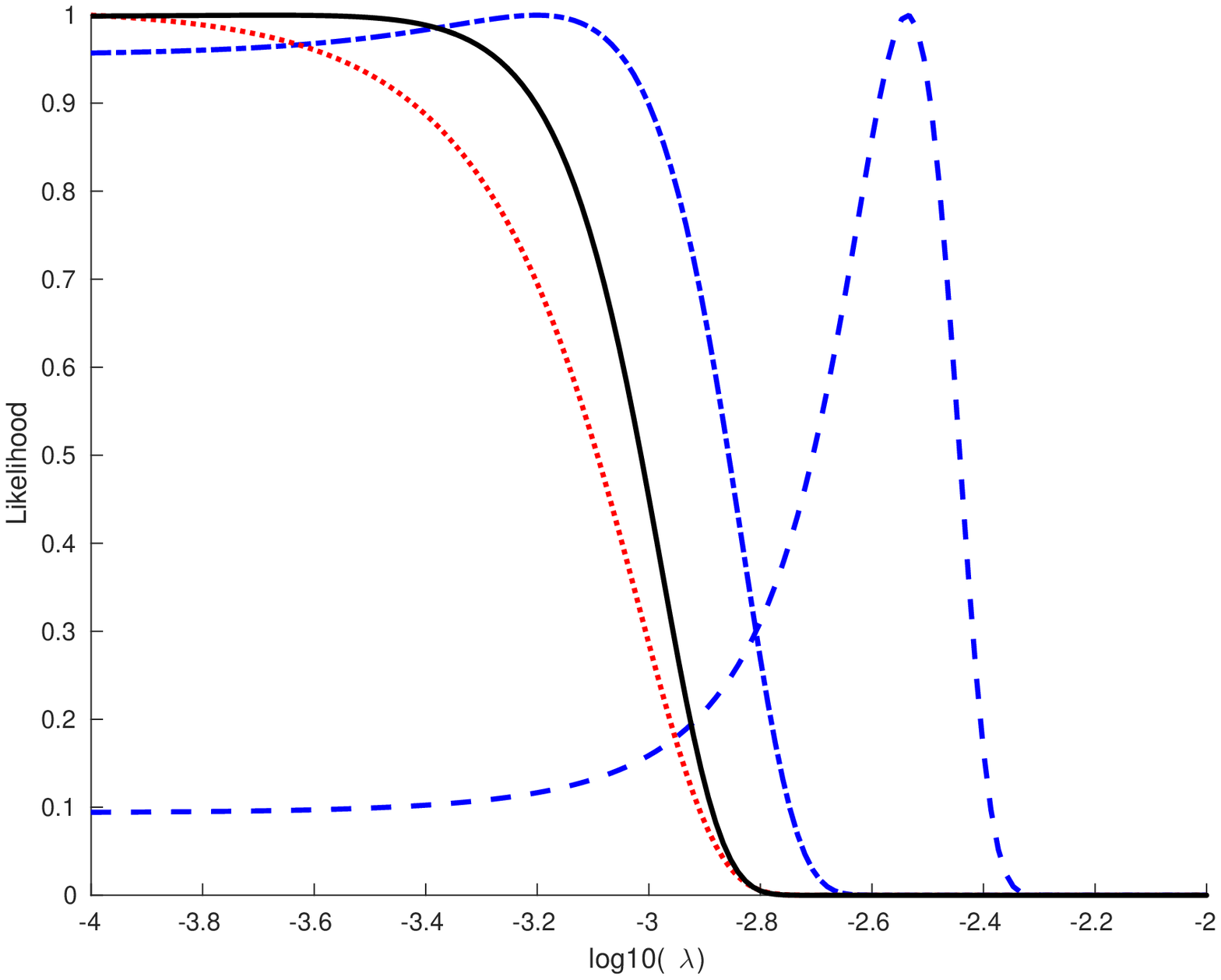}
\end{center}
\caption{\label{fig06}{\bf Top panel:} One, two and three sigma constraints on the $\lambda$-$\Omega_m$ plane from the full Canonical Data set discussed in the text; the reduced chi-square at the maximum likelihood is $\chi^2_\nu=0.96$. This updates the analysis of \protect\cite{Moucherek}. {\bf Bottom panel:} 1D likelihood for $\lambda$, marginalizing over $\Omega_m$, for cosmological $+$ Webb {\it et al.} data (blue dashed), cosmological $+$ the dedicated $\alpha$ measurements of Table \protect\ref{tab01} and Oklo (blue dash-dotted), cosmological $+$ Rosenband {\it et al.} atomic clock bound (red dotted) and the combination of all data sets (black solid).}
\end{figure}

The results of this analysis, updated for the aforementioned Canonical data set, are summarized in Fig. \ref{fig06}. As expected the cosmological data sets fix the matter density, with the $\alpha$ measurements having very little impact on it since the dependence is only logarithmic. Specifically, marginalizing over $\lambda$ one finds the following constraint
\be
\Omega_m=0.28\pm0.03\,,
\ee
at the three sigma ($99.7\%$) confidence level, which is fully compatible with other extant cosmological data sets. On the other hand, the $\alpha$ measurements strongly constrain $\lambda$, for the reasons already explained. In particular we notice that the Webb {\it et al.} data set would lead to a two-sigma detection of a non-zero $\lambda$, but the coupling is consistent with zero for other measurements of $\alpha$ and also for the combination of all the data. In this case we find, marginalizing over $\Omega_m$,
\be
\lambda<9\times10^{-4}\,,\quad 68.3\% C.L.
\ee
\be
\lambda<1.5\times10^{-3}\,,\quad 99.7\% C.L.\,.
\ee
In particular, this leads to an extremely strong constraint on the value of the present day dark energy equation of state
\be
(1+w_0)<2.6\times10^{-7}\,,\quad 99.7\% C.L.\,.
\ee
It is clear that neither current nor foreseen standard probes of background cosmology will be able to detect such a small deviation from $w_0=-1$. Thus the only possibilities to distinguish these models from the $\Lambda$CDM paradigm would be to rely on their clustering properties (a possibility that remains to be studied) or to use astrophysical measurements of the redshift dependence of $\alpha$.

At the phenomenological level the interesting feature of these models is that a single parameter---effectively the steepness of the potential, in dimensionless units---determines both the dark energy equation of state and the overall level of the $\alpha$ variations. Moreover, these are necessarily thawing models with a monotonically increasing value of $\alpha$ (in other words, they will have smaller values of $\alpha$ in the past). The current local and astrophysical tests of the stability of $\alpha$ therefore place strong constraints on the steepness of the potential, and imply that the present-day value of the dark energy equation of state, although not exactly $-1$, is effectively indistinguishable from it if one restricts oneself to standard observational probes. This highlights the importance of testing the stability of nature's fundamental couplings over a broad range of redshifts and accurately mapping their behavior. As this class of models shows, this may turn out to be the best way we have of identifying deviations from the $\Lambda$CDM paradigm, at least in the next decade. Moreover, in the event of confirmed detections of variations such a mapping is a powerful discriminator, since different classes of models lead to significantly different behaviors for the redshift dependence of $\alpha$.

\section{Complementary probes\label{redundancy}}

Whichever way one eventually finds direct evidence for new physics, it will only be trusted once it is seen through multiple independent probes. This much is clear when looking at how the events associated with the discovery of the recent acceleration of the universe unfolded: even though the supernova results were independently obtained by two different teams, they were only accepted by the wider community once they were confirmed through CMB, large-scale structure and other data. It is clear that history will repeat itself in the case of varying fundamental couplings and/or dynamical dark energy. It is therefore crucial to plan and develop consistency tests for this new physics, in other words, additional astrophysical observables whose behavior will also be non-standard (ideally in a specific and calculable way) as a consequence of either or both of the above.

An obvious example which we already discussed is that of violations of the Einstein Equivalence Principle. Varying fundamental couplings trivially violate Local Position Invariance, but it has also been shown \cite{Damour,DamDon} that variations of $\alpha$ at few ppm level naturally lead to Weak Equivalence Principle violations within one order of magnitude of current bounds on the E\"{o}tv\"{o}s parameter, cf. Eqs. \ref{etaTorsion}--\ref{etaLLR}. In that case ongoing experiments such as the MICRSOCOPE satellite, launched on 25 April 2016 and currently operating \cite{MICROSCOPE}, should find these violations. We now explore two other promising consistency tests. The first is already producing interesting (though not yet 'competitive') results, while the second is a key goal for the next generation of facilities.

\subsection{CMB temperature and distance duality}

An astrophysical consistency test is provided by the comparison of the temperature-redshift relation and the distance duality (or Etherington) relation. The temperature-redshift relation is a robust prediction of standard cosmology, based on the assumptions of adiabatic expansion and photon number conservation, but it is violated in many scenarios, including string theory inspired ones and models where $\alpha$ varies. At a phenomenological level one can parametrize deviations to this law by adding an extra parameter \cite{lima00}, say $\beta$
\begin{equation}
T_{\rm CMB}=T_0(1+z)^{1-\beta}\,, \label{tofz}
\end{equation}
with $\beta=0$ in standard cosmology. The COBE-FIRAS experiment observations provided the  most-precise blackbody spectrum ever measured, with a temperature at the present epoch, $z=0$, of \cite{fixsen09}
\be
T_0=2.7260 \pm 0.0013\,K\,. 
\ee
At higher redshifts, there are presently two main methods used to obtain direct estimates of $T_{\rm CMB}$, and from which constraints on $\beta$ can be derived. The first of these was proposed nearly 40 years ago \cite{fabbri78,rephaeli80} and is based on multi-frequency observations of the  Sunyaev-Zel'dovich (SZ) effect \cite{sunyaev70}, a distortion of the CMB spectrum produced towards galaxy clusters.

As pointed out by \cite{deMartino1}, current large galaxy cluster catalogs together with very precise CMB data should allow precisions on $\beta$ of the percent level, a notable improvement with respect to initial constrains using a few clusters \cite{battistelli02,luzzi09}. The availability of the Planck satellite data (and in particular of its cluster catalog \cite{Clusters,Cluster2}) as well as analogous catalogs from ground-based small angular scale experiments, enabled significant improvements in the precision of CMB temperature measurements from SZ clusters. Improved methodologies were suggested in \cite{deMartino1} and subsequently exploited \cite{hurier14,saro14,deMartino2,Luzzi}. Estimations of $T_{\rm CMB}(z)$ through the SZ effect are currently limited to $z<1$ due to the scarcity of galaxy clusters at high redshifts. 

Measurements of the CMB temperature at $z>1$ can be obtained through the study of quasar absorption line spectra which show energy levels that have been excited through atomic or molecular transitions after the absorption of CMB photons \cite{bahcall68}. The first constraints using this method were only obtained 17 years ago \cite{srianand00}, taking advantage of the enormous progress in high-resolution astrophysical spectroscopy; they use transitions in the UV range due to the excitation of fine-structure levels of atomic species like C{\sc i} or C{\sc ii} \cite{srianand00,ge01,molaro02,cui05}. More recently, improved constraints have been obtained from precise measurements of CO transitions and radio-mm transitions produced by the rotational excitation of molecules with a permanent dipole moment \cite{srianand08,noterdaeme10,noterdaeme11,muller13}. In passing we note that in this and the previous paragraph we are referring to direct constraints---indirect ones may also be inferred from spectral distortions \cite{Chluba}.

On the other hand the distance duality relation is an equally robust prediction of standard cosmology; it assumes a metric theory of gravity and photon number conservation, but is violated if there is photon dimming, absorption or conversion. This is also known as the Etherington relation \cite{Etherington}. At a similarly phenomenological level one can parametrize deviations to this law by adding an extra parameter, say $\epsilon$
\begin{equation}
d_L=d_A(1+z)^{2+\epsilon}\,. \label{etheringt}
\end{equation}
In fact, as shown in \cite{Tasos1} in many models where photon number is not conserved---such as those where $\alpha$ varies---the temperature-redshift relation and the distance duality relation are not independent. Assuming adiabaticity and achromaticity one can in fact show that
\be
\beta=-\frac{2}{3}\epsilon\,,
\ee
but it is easy to see that a direct relation should exist more generically. This link allows one to use distance duality measurements to improve constraints on $\beta$ \cite{Tasos2}. The combination of all currently available direct and indirect measurements of $T(z)$ constrains the phenomenological parameter $\beta$ to be, at the $68.3\%$ confidence level \cite{Subpercent}
\be
\beta=(7.6\pm8.0)\times10^{-3}\,,
\ee
which is the first sub-percent constraint on $\beta$ parameter.

In models where an evolving scalar field is coupled to the Maxwell $F^2$ term in the matter Lagrangian, photons can be converted into scalar particles violating the photon number conservation. Thus, there will be both variations of the fine-structure constant and violations of the standard $T_{CMB}(z)$ law. These can usually be written as \cite{Tasos2}
\begin{equation}\label{eq:tcmbx}
\frac{T_{CMB}(z)}{T_0}\sim(1+z)\left(1+\epsilon\frac{\Delta\alpha}{\alpha}\right)\,,
\end{equation}
or alternatively
\begin{equation}\label{eq:tcmbx2}
\frac{\Delta T_{CMB}}{T}=\frac{T_{CMB}(z)-T_{CMB, std}(z)}{T_{CMB, std}(z)}\sim\epsilon\frac{\Delta\alpha}{\alpha}\,.
\end{equation}
The coefficient $\epsilon$, which vanishes in the standard model, depends on the specific model being assumed, but it is generically expected to be of order unity. In particular, if one assumes the somewhat simplistic adiabatic limit, then one can show that $\epsilon=1/4$  \cite{Tasos2}. A subsequent analysis in \cite{Hees}, with somewhat different assumptions, confirmed these results. Therefore, if one is able to determine the CMB temperature with sufficient accuracy, this can be used as a phenomenological relation to observationally constrain time or spatial variations of $\alpha$.

This relation between $\alpha$ variations and the CMB temperature may be relevant, for example, for Planck data analysis. If the ppm $\alpha$ dipole of Webb {\it et al.} \cite{Webb} is correct, then there should be a micro-Kelvin level dipole in the CMB temperature, in addition to the standard dipole due to our motion relative to the CMB frame, which if unaccounted for could bias the analysis and in particular the ensuing cosmological parameter estimations. As briefly mentioned in Section \ref{uveslp}, SZ cluster measurements have recently been used to constrain possible spatial variations of $\alpha$ and $T(z)$ \cite{SZclusters}. Naturally, forthcoming spectroscopic measurements from ESPRESSO and ELT-HIRES and SZ cluster measurements from CORE and ground-based CMB experiments \cite{COREclu} can significantly improve current constraints \cite{Tasos1}.

Photon number non-conservation not only changes observables such as $T(z)$ and the distance duality relation, but may also lead to additional biases, for example for cosmology and fundamental physics constraints from Euclid \cite{Euclid}. A study of how these models weaken cosmological parameter constraints from Euclid (specifically those characterizing the dark energy equation of state) was done in \cite{Tasos2,Calabrese}. The results show that Euclid can, even on its own, constrain dark energy while allowing for photon number non-conservation, but stronger constraints can be obtained in combination with other probes. Interestingly, the ideal way to break a degeneracy involving the scalar-photon coupling is to use $T(z)$ measurements to be obtained with ALMA, ESPRESSO and ELT-HIRES \cite{Tasos2}. These three facilities may nicely complement each other in terms of the redshift coverage for these measurements with ALMA probing the low-redshift acceleration phase while ESPRESSO and ELT-HIRES will probe the deep matter era.

\subsection{Redshift drift}

When doing cosmological parameter estimation or model selection one almost always makes a certain number of explicit or implicit assumptions. For example one may be assuming a Friedmann-Lemaitre-Robertson-Walker (homogeneous and isotropic) background and the validity of General Relativity, in addition to more specific assumptions such as a flat universe or particular classes of models for dark energy. We also saw in the previous sections that there is also a broad range of models for varying couplings, relying on different underlying fundamental physics and cosmological mechanisms. Thus it is crucial for this analysis that in-built consistency checks exist, so that inconsistent assumptions can be identified and corrected. Explicit examples of incorrect assumptions in this context that can lead to observational inconsistencies, for example assuming that $\alpha$ variations are due to a Class I model when they are in fact due to a Class II model, have been discussed in \cite{Pauline1}.

It is precisely in closing the loop of consistency tests that the detection of the redshift drift signal, also known as the Sandage test \cite{Sandage,Loeb}, plays a key role. The expected signal is
\begin{equation}
\frac{\Delta z}{\Delta t}=H_0(1+z)-H(z)\,, \label{zdrift}
\end{equation}
and this is a direct probe of the dynamics of the universe, without assumptions on gravity, geometry or clustering. Unlike all the other observations of the universe we have done so far, it does not map out our (present-day) past light-cone. Instead, it directly measures evolution by comparing past light cones at different times. Therefore it provides an ideal probe of the evolution of the universe, and in particular of its dark sector. In practice the observable will be a spectroscopic velocity
\begin{equation}
\frac{\Delta v}{v}=\frac{\Delta z}{1+z}\,. \label{specvel}
\end{equation}
The redshift drift is a key driver for ELT-HIRES, and indeed---at a fundamental level---a key ELT deliverable. ELT-HIRES can measure the redshift drift signal deep in the matter era, using the Ly-$\alpha$ forest and various additional metal absorption lines \cite{Liske}. 

Apart from the fundamental aspect of being able to watch the expansion of the universe in real time, one should note that when using these observations to constrain specific models their importance is not so much that they are more constraining than other observational probes but that they tend to probe orthogonal directions in the relevant model parameter spaces, thereby breaking limiting degeneracies. A study of synergies between redshift drift and CMB measurements \cite{Pauline2}, assuming Planck-like CMB measurements and a dark energy model parametrized by a constant equation of state $w_0$, shows that the forecasted redshift drift measurements of Liske {\it et al.} \cite{Liske} for the ELT-HIRES can improve CMB results on $w_0$, $H_0$ and $\Omega_m$ by factors of  $3.3$, $3$ and $2.2$ respectively. If one further enlarges the parameter space, by assuming the CPL dark energy parametrization (with the additional parameter $w_a$), ELT-HIRES redshift drift measurements should not able to remove the degeneracy between $w_0$ and $w_a$ and therefore there is no significant improvement on the CMB constraints on these parameters; nevertheless, significantly tighter constraints on $H_0$ and $\Omega_m$ are again achieved.

More recently it has been realized that other facilities such as the SKA \cite{Kloeckner}, ALMA and intensity mapping experiments \cite{Yu} may also be able do measure the redshift drift, though this remains to be fully demonstrated. These will typically do it at lower redshifts. In the case of the SKA, suggestions have been put forth to do it using neutral Hydrogen both at $z<1$ in emission and at $z>8$ in absorption; while the former should be easily within the reach of SKA-Phase 2, the latter will certainly be much harder. Such low-redshift measurements can directly probe the accelerating phase of the universe (at redshifts that overlap with Euclid, for example), as well as provide much-needed clarification on the issue of the present-day value of the Hubble parameter, $H_0$. On the other hand, these measurements will have a smaller lever arm---only ELT-HIRES can really probe the deep matter era, roughly $2<z<5$. 

Naturally the combination of low and high redshift measurements will lead to optimal constraints and will enable the discrimination between models that would otherwise be indistinguishable. Indeed, the prospect of a model-independent mapping of the expansion of the universe from $z\sim0$ to $z\sim5$, by a combination of SKA, intensity mapping and ELT-HIRES data is a particularly exciting prospect. Last bust not least, the cosmological relevance of measurements of time or redshift derivatives of this drift has recently been highlighted \cite{Ramos}. The combination of first and second redshift derivatives is a powerful test of the $\Lambda$CDM cosmological model---and therefore of any deviations from it. In particular, the second derivative can be obtained numerically from a set of measurements of the drift at different redshifts. Such a measurement is well within the reach of the ELT-HIRES and SKA Phase 2 array surveys.

\section{The road ahead\label{forecasts}}

In previous sections we discussed how current data already provides useful constraints on fundamental physics and cosmology. Still, the imminent availability of more precise measurements will have a dramatic impact in the field. The ESPRESSO spectrograph will be commissioned at the VLT in 2017, and since it will be located at the combined Coud\'e focus it will be able to incoherently combine light from the four VLT Unit Telescopes \cite{ESPRESSO}. Looking further ahead the European Extremely Large Telescope, with first light expected in 2024, will have a 39.3m primary mirror. The larger telescope collecting areas are one of the reasons behind the expected improvements in the sensitivity of these measurements (which are photon-starved), the other such reason pertaining to technological improvements in its high-resolution spectrograph ELT-HIRES enabling, among others, higher resolution and stability \cite{HIRES}. We will now describe how current constraints are expected to improve, and some of the potential impact of these improvements on fundamental cosmology. 

We start by looking at $\alpha$ measurements by themselves and considering three Class I fiducial dynamical dark energy models where the scalar field also leads to $\alpha$ variations according to Eq. \ref{eq:dalfa}, all of which were already introduced in previous sections:
\begin{itemize}
\item A constant equation of state, $w_0=const.$
\item A dilaton-type model where the scalar field behaves as $\phi(z)\propto(1+z)$, leading to a relatively complicated dark energy equation of state
\begin{equation}
w(z)=\frac{[1-\Omega_\phi(1+w_0)]w_0}{\Omega_m(1+w_0)(1+z)^{3[1-\Omega_\phi(1+w_0)]}-w_0}\,,
\end{equation}
(where as usual we are assuming flat universes, so $\Omega_m+\Omega_\phi=1$) but simpler behavior for $\alpha$
\begin{equation}\label{dilalpha}
\frac{\Delta\alpha}{\alpha}(z)=\zeta\, \sqrt{3\Omega_\phi(1+w_0)}\, \ln{(1+z)}\,.
\end{equation}
This case is also useful since it allows analytic calculations which can be used to validate numerical codes.
\item The well-known Chevallier-Polarski-Linder (CPL) parametrization \cite{CPL1,CPL2},
\be
w(z)=w_0+w_a\frac{z}{1+z}\,.
\ee
\end{itemize}

Forecasts can be done with a Fisher Matrix analysis \cite{FMA1,FMA2}. If we have a set of M model parameters $(p_1, p_2, ..., p_M)$ and N observables---that is, measured quantities---$(f_1, f_2, ..., f_N)$, then the Fisher matrix is
\be
F_{ij}=\sum_{a=1}^N\frac{\partial f_a}{\partial p_i}\frac{1}{\sigma^2_a}\frac{\partial f_a}{\partial p_j}\,.
\ee
For an unbiased estimator, if we don't marginalize over any other parameters (meaning that all are assumed to be perfectly known) then the minimal expected error is $\theta=1/\sqrt{F_{ii}}$. The inverse of the Fisher matrix provides an estimate of the parameter covariance matrix: its diagonal elements are the squares of the uncertainties in each parameter marginalizing over the others, while the off-diagonal terms yield the correlation coefficients between parameters. The marginalized uncertainty is always greater than (or at most equal to) the non-marginalized one: marginalization can't decrease the error, and only has no effect if all other parameters are uncorrelated with it.

Previously known uncertainties on the parameters, known as priors, can be trivially added to the calculated Fisher matrix. This is manifestly the case here: a plethora of standard cosmological data sets provide priors on our cosmological parameters $(\Omega_m,w_0,w_a)$, while local constraints on the  E\"{o}tv\"{o}s parameter from torsion balance and lunar laser ranging experiments \cite{Torsion,LLR}, see Eqs. \ref{etaTorsion}--\ref{etaLLR}, provide priors on the dimensionless coupling $\zeta$. Here, following \cite{AlphaFish}, we assume the following fiducial values and prior uncertainties for our cosmological parameters
\be
\Omega_{m,fid}=0.3\,, \quad \sigma_{\Omega_m}=0.03
\ee
\be
w_{0,fid}=-0.9\,, \quad \sigma_{w_0}=0.1
\ee
\be
w_{a,fid}=0.3\,, \quad \sigma_{w_a}=0.3\,,
\ee
while for the coupling $\zeta$ we will consider three different scenarios
\be
\zeta_{fid}=0\,,\quad \zeta_{fid}=5\times10^{-7}\,,\quad \zeta_{fid}=5\times10^{-6}\,,
\ee
always with the same prior uncertainty
\be
\sigma_\zeta=10^{-4}\,.
\ee
Thus we will consider both the case where there are no $\alpha$ variations ($\zeta=0$), and the case where they exist: the case $\zeta=5\times10^{-6}$ corresponds to a coupling which saturates constraints from current data (as discussed in Sections \ref{models} and \ref{darkside}), while $\zeta=5\times10^{-7}$ illustrates an intermediate scenario.

The first ESPRESSO measurements of $\alpha$ should be obtained in the context of the consortium's Guaranteed Time Observations (GTO). The target list for these measurements has recently been selected \cite{MSC}. Note that a key limitation of ESPRESSO which must be taken into account in the target selection is its wavelength coverage range, which is narrower than the ones of its predecessors (HARPS, UVES and Keck-HIRES). Other than that, the basic selection criteria are for targets that
\begin{itemize}
\item can be observed from the VLT site (Cerro Paranal in Chile, implying declination $\delta < 30$ degrees);
\item contain transitions that allow a high sensitivity (specifically, with $\Delta q>2000$);
\item have a reported uncertainty of $\sigma_{\Delta\alpha/\alpha}<5ppm$.
\end{itemize} 
The last of these comes from the fact that simple spectra should have already produced measurements with statistically lower uncertainties. Strictly speaking there is also the possibility that new bright quasars are discovered, but since the GTO targets should be fixed soon the probability of such an occurrence is low. Additional criteria that are relevant for prioritizing the targets are:
\begin{itemize}
\item QSO brightness;
\item high number of transitions available in the system, which leads to smaller overall uncertainties and also allows for several independent measurements using different sets of transitions (an important test of possible systematics);
\item presence of at least one red shifter, one blue shifter, and one anchor; this is partially ensured by the requirement of a large $\Delta q$;
\item simpler velocity structure systems (strong but not saturated absorption features; narrow lines and large number of components, provided these are resolved or at least partially resolved);
\item systems for which the dipole model of Webb {\it et al.} \cite{Webb} predicts a higher variation of $\alpha$, that is, the ones closer to the poles of the putative dipole;
\item possibility to perform in the same system additional measurements, such as $\mu$ or $T_{CMB}$, enabling key tests of many theoretical paradigms (as explained in previous sections).
\end{itemize}
Full details of this process, which leads to the target list presented in Table \ref{tab08} can be found in \cite{MSC,GTO}. Note that the order in which they are presented should not be seen as any ranking among them: they are simply ordered according to their Right Ascension. A more detailed prioritization will require the generation of simulated ESPRESSO-like spectra of these targets, and is currently ongoing. Indeed, the first listed target does not fulfill all the criteria, but it is the only system accessible to ESPRESSO where the proton-to-electron mass ratio and the temperature-redshift relation can also be measured. This fact makes it a theoretically interesting target for testing theories where a relation between these three parameters is predicted.

\begin{table*}
\begin{center}
\begin{tabular}{|l|c c c|c c|c|}
\hline 	
 	Name & $z_{abs}$ & $\frac{\Delta \alpha}{\alpha}$ (ppm) & Max($\Delta q$) & \# trans. & Transitions & Reference \\
 	\hline 		 
J0350-3811 & 3.02 & $-27.9\pm 34.2$ & 1350& 2& SiII,\textcolor{blue}{FeII} & \cite{PHDmurphy} \\ 
J0407-4410 &  2.59  & $5.7\pm 3.4$* & 2984& 13 & AlII,AlIII,SiII,\textcolor{blue}{CrII},\textcolor{blue}{FeII},\textcolor{red}{FeII},\textcolor{blue}{NiII},\textcolor{red}{ZnII}  & \cite{PHDking} \\
J0431-4855 &  1.35 &  $-4.0\pm 2.3$* & 2990& 17 & MgI,AlII,SiII,\textcolor{blue}{CrII},\textcolor{red}{MnII},\textcolor{red}{FeII},\textcolor{blue}{NiII} & \cite{PHDking}\\
 J0530-2503 & 2.14   & $6.7\pm3.5$* & 2990 & 7 & AlII,\textcolor{blue}{CrII},\textcolor{blue}{FeII},\textcolor{red}{FeII},\textcolor{blue}{NiII}  & \cite{PHDking}\\
 J1103-2645 & 1.84 & $3.5\pm2.5$  & 2890 & 4 & SiII,\textcolor{blue}{FeII},\textcolor{red}{FeII}  & \cite{Bainbridge,lev2} \\
 J1159+0112 & 1.94   & $5.1\pm4.4$*  & 2990 & 12 &  SiII,\textcolor{blue}{CrII},\textcolor{red}{MnII},\textcolor{blue}{FeII},\textcolor{red}{FeII},\textcolor{blue}{NiII} & \cite{PHDking} \\
J1334+1649 & 1.77   & $8.4\pm4.4$ &2990 & 15 &  MgII,AlII,SiII,\textcolor{blue}{CrII},\textcolor{red}{MnII},\textcolor{blue}{FeII},\textcolor{red}{FeII},\textcolor{blue}{NiII},\textcolor{red}{ZnII}& \cite{PHDking} \\
HE1347-2457 	& 1.43  & $-21.3\pm 3.6$   & 2790 & 3 & \textcolor{blue}{FeII},\textcolor{red}{FeII}  & \cite{lev2}\\
J2209-1944 & 1.92 & $8.5\pm 3.8$ &3879& 16 &AlII,SiII,\textcolor{blue}{CrII},\textcolor{red}{MnII},\textcolor{blue}{FeII},\textcolor{red}{FeII},\textcolor{blue}{NiII},\textcolor{red}{ZnII}  & \cite{MalecNew,PHDking} \\
HE2217-2818 & 	1.69     & $1.3\pm 2.4$  &2890 & 6 & AlIII,\textcolor{blue}{FeII},\textcolor{red}{FeII} & \cite{LP1}\\
Q2230+0232 & 	1.86    & $-9.9\pm 4.9$ &3879 & 14 & SiII,\textcolor{blue}{CrII},\textcolor{blue}{FeII},\textcolor{red}{FeII},\textcolor{blue}{NiII},\textcolor{red}{ZnII} & \cite{PHDmurphy} \\
J2335-0908 & 2.15  & $5.2\pm 4.3$* &3879& 16 &  AlIII,\textcolor{blue}{CrII},\textcolor{blue}{FeII},\textcolor{red}{FeII},\textcolor{blue}{NiII},\textcolor{red}{ZnII}  & \cite{PHDking}\\
J2335-0908 & 2.28  & $7.5\pm 3.7$*   &2610& 7 &  SiIV,\textcolor{blue}{CrII},\textcolor{blue}{FeII},\textcolor{red}{FeII},\textcolor{blue}{NiII}  & \cite{PHDking}\\
Q2343+1232 & 	2.43     & $-12.2\pm 3.8$* & 3879 & 11 & AlII,SiII,\textcolor{blue}{CrII},\textcolor{blue}{FeII}   \textcolor{blue}{NiII},\textcolor{red}{ZnII}& \cite{PHDmurphy}\\
\hline 
\end{tabular}
\caption{\label{tab08}The best currently available measurements of $\alpha$, among the targets accessible to ESPRESSO. Column 1 gives the quasar name; the redshifts of the absorption system are given in Column 2; Column 3 gives the current measurement. Column 4 gives the ranges of sensitivity coefficients associated with the transitions of the absorption systems. Column 5 gives the number of transitions in each absorption system and column 6 the elements that can be detected, colored differently according to whether they are an anchor (black), a blue shifter (blue) or a red shifter (red). The last column gives the references for each measurement. Measurements flagged with a * identify targets for which some of the transitions used in the current measurement are outside the wavelength range of ESPRESSO.}
\end{center}
\end{table*}

Bearing this target list in mind one can consider the following three scenarios:
\begin{itemize}
\item {\bf ESPRESSO Baseline}: this assumes that each of the targets on the list can be measured by ESPRESSO with an uncertainty of $\sigma_{\Delta\alpha / \alpha}=0.6\,ppm$; this represents what one can currently expect to achieve on a time scale of 3-5 years (though this expectation needs to be confirmed at the time of commissioning of the instrument);
\item {\bf ESPRESSO Ideal}: this case assumes a factor of three improvement in the uncertainty, $\sigma_{\Delta\alpha / \alpha}=0.2\,ppm$; this represents somewhat optimistic uncertainties but provides a useful comparison point. Nevertheless, such an improved uncertainty should be achievable with additional integration time;
\item {\bf ELT-HIRES}: this is representative of a longer-term data set, on the assumption that the same targets are observed with the ELT-HIRES spectrograph \cite{HIRES}. An improvement in sensitivity by a factor of six relative to the ESPRESSO baseline scenario is assumed, coming from the larger collecting area of the telescope and additional improvements at the level of the spectrograph. Although at present not all details of the instrument and the telescope have been fixed, this scenario is meant to be representative of the expected sensitivity of measurements on a 10-15 year time scale.
\end{itemize}

These choices of possible theoretical and observational parameters span a broad range of possible scenarios. As a simple illustration of this point, let us consider a single measurement of $\alpha$ at redshift $z=2$. In the case of the dilaton model we have the simple relation $\Delta\alpha/\alpha(z=2)\sim0.5\zeta$. Thus if $\zeta=5\times10^{-7}$ a single precise and accurate measurement of $\alpha$ with ESPRESSO baseline sensitivity would not detect its variation, while ELT-HIRES would detect it at 2.5 standard deviations. On the other hand, for  $\zeta=5\times10^{-6}$ (which as previously mentioned saturates current bounds) a single $z=2$ ESPRESSO baseline measurement would detect a variation at $4\sigma$ and ELT-HIRES would detect it at $25\sigma$.

It is instructive to provide a discussion of the analytic result of the Fisher Matrix analysis for the dilaton model. For simplicity let's further assume that $\Omega_\phi$ (or equivalently, assuming a flat universe, $\Omega_m$) is perfectly known, so we are left with a two-dimensional the parameter space $(\zeta,w_0)$. Including priors on both, the Fisher matrix is
\[ [F(\zeta,w_0)] = \left[ \begin{array}{cc}
Q^2(1+w_0)+\frac{1}{\sigma^2_\zeta} & \frac{1}{2}Q^2\zeta \\
\frac{1}{2}Q^2\zeta & \frac{Q^2\zeta^2}{4(1+w_0)}+\frac{1}{\sigma^2_w} \end{array} \right]\,,\] 
where for convenience we have defined
\be
Q^2=3\Omega_\phi\sum_i\left[\frac{\log(1+z_i)}{\sigma_{\alpha i}}\right]^2\,.
\ee
The un-marginalized uncertainties are
\be
\theta_\zeta=\frac{\sigma_\zeta}{\sqrt{1+(1+w_0)Q^2\sigma^2_\zeta}}
\ee
\be
\theta_w=\frac{\sigma_w}{\sqrt{1+\frac{\zeta^2}{4(1+w_0)}Q^2\sigma^2_w}}\,,
\ee
while the determinant of F is
\be
det F=Q^2\left[\frac{1+w_0}{\sigma^2_w}+\frac{\zeta^2}{4(1+w_0)\sigma^2_\zeta} \right]+\frac{1}{\sigma^2_w\sigma^2_\zeta}\,;
\ee
this would be zero in the absence of priors---a point already discussed in \cite{Calabrese}---but as mentioned above cosmological data and local tests of the WEP do provide us with these priors. As expected, if $\zeta=0$ the two parameters decorrelate, and there is no new information on the equation of state ($\theta_w=\sigma_w$): if $\zeta=0$ we will always measure $\Delta\alpha/\alpha=0$ regardless of the experimental sensitivity.

Now we can calculate the covariance matrix
\[ [C(\zeta,w_0)]=\frac{1}{det F} \left[ \begin{array}{cc}
\frac{Q^2\zeta^2}{4(1+w_0)}+\frac{1}{\sigma^2_w} & -\frac{1}{2}Q^2\zeta \\
-\frac{1}{2}Q^2\zeta & Q^2(1+w_0)+\frac{1}{\sigma^2_\zeta} \end{array} \right]\,,\] 
and the correlation coefficient is
\be
\rho=\left[1+\frac{4(1+w_0)}{Q^2\zeta^2\sigma^2_w} +\frac{1}{(1+w_0)Q^2\sigma^2_\zeta}+\frac{4}{\zeta^2Q^4\sigma^2_\zeta\sigma^2_w} \right]^{-1/2}\,.
\ee
We thus confirm the physical intuition that in the limit $\zeta\to 0$, the two parameters become independent ($\rho\to0$). The general marginalized uncertainties are
\be
\frac{1}{\sigma^2_{\zeta,new}}=\frac{1}{\sigma^2_\zeta}+\frac{1}{\sigma^2_w}\frac{(1+w_0)Q^2}{\frac{\zeta^2Q^2}{4(1+w_0)}+\frac{1}{\sigma^2_w}}
\ee
\be
\frac{1}{\sigma^2_{w,new}}=\frac{1}{\sigma^2_w}+\frac{1}{\sigma^2_\zeta}\frac{\zeta^2Q^2}{4(1+w_0)}\frac{1}{(1+w_0)Q^2+\frac{1}{\sigma^2_\zeta}}\,;
\ee
In the particular case where the fiducial model is $\zeta=0$ the former becomes
\be
\frac{1}{\sigma^2_{\zeta,new}}=\frac{1}{\sigma^2_\zeta}+(1+w_0)Q^2
\ee
while the latter trivially gives $\sigma_{w,new}=\sigma_w$. These analytic results have been used to validate a more generic numerical code (where furthermore $\Omega_m$ will also be allowed to vary), which in turn was used for a more detailed discussion of the forecasts for the various cases which can be found in \cite{AlphaFish}. In what follows we briefly summarize these results.

In the case where there is no coupling between the scalar field and the electromagnetic sector of the theory (such that  $\zeta_{fid}=0$) precise $\alpha$ measurements will find null results which can be translated into bounds on $\zeta$, whose one-sigma uncertainties, marginalized over $\Omega_m$, $w_0$ and (for the case of the CPL model) $w_a$, are displayed in Table \ref{tab09}. Comparison with current bounds on $\zeta$ \cite{Pinho1,Pinho2,Pinho3}, cf. Section \ref{darkside}, shows that in this case we expect ESPRESSO to improve current bounds by about one order of magnitude. Naturally these improvements also lead to stronger bounds on the  E\"{o}tv\"{o}s parameter: indeed constraints from ESPRESSO should be stronger than those expected from the ongoing tests with the MICROSCOPE satellite \cite{MICROSCOPE}, whose sensitivity is expected to be
\be
\sigma_\eta\sim10^{-15}\,.
\ee
Looking further ahead, those from ELT-HIRES should be competitive with those of the proposed STEP satellite \cite{STEP} (though at present the sensitivity of the latter is relatively uncertain).

Table \ref{tab09} also shows that there is a mild dependence on the choice of underlying dark energy model. This has been previously studied, and is well understood---refer to Section \ref{darkside}, or to \cite{Pinho2,Pinho3,Calabrese} for further discussion of this point. The dilaton model is a 'freezing' dark energy model. Thus, according to Eq. \ref{eq:dalfa}, a dilaton model with a given value of $w_0$ will have a value of $\Delta\alpha/\alpha(z)$ that is larger than the corresponding value for a model with a constant equation of state with the same value of $w_0$. Thus, for similar cosmological priors, null measurements of $\alpha$ will provide slightly stronger constraints for the dilaton case. The same argument applies for the CPL case, where the additional free parameter $w_a$ further enlarges the range of possible values of $\alpha$.

\begin{table*}
\begin{center}
\begin{tabular}{| c | c | c | c |}
\hline
Model & ESPRESSO baseline & ESPRESSO ideal & ELT-HIRES \\
\hline
$w_0=const.$ & $4.6\times10^{-7}$ & $1.5\times10^{-7}$ & $7.6\times10^{-8}$ \\
Dilaton & $3.2\times10^{-7}$ & $1.1\times10^{-7}$ & $5.3\times10^{-8}$  \\
CPL & $3.1\times10^{-7}$ & $1.0\times10^{-7}$ & $5.1\times10^{-8}$  \\
\hline
$\eta$ & $2.1\times10^{-16}$ & $2.3\times10^{-17}$ & $5.8\times10^{-18}$ \\
\hline
\end{tabular}
\caption{\label{tab09}The first three lines show the one sigma forecasted uncertainties on the dimensionless coupling parameter $\zeta$, marginalizing over the remaining model parameters, for the various choices of fiducial cosmological model and data set of $\alpha$ measurements. The fiducial value of the coupling is $\zeta_{fid}=0$ in all cases. The last line shows the corresponding one-sigma uncertainty on the E\"{o}tv\"{o}s parameter $\eta$, in the least constraining case of the $w_0=const.$ model. See \protect\cite{AlphaFish} for further details on these results.}
\end{center}
\end{table*}

\begin{table*}
\begin{center}
\begin{tabular}{| c | c c | c c | c c |}
\hline
{ } & ESPRESSO & baseline &  ESPRESSO & ideal & ELT-HIRES  & { } \\
Parameter & $\zeta=5\times10^{-7}$ & $\zeta=5\times10^{-6}$ & $\zeta=5\times10^{-7}$ & $\zeta=5\times10^{-6}$ & $\zeta=5\times10^{-7}$ & $\zeta=5\times10^{-6}$   \\
\hline
$\rho(\zeta,w_0)$      & -0.412 & -0.728 & -0.650 & -0.822 & -0.705 & -0.914 \\
$\rho(\Omega_m,w_0)$   & $1.6\times10^{-7}$ & $1.6\times10^{-5}$ & $4.0\times10^{-6}$ & $3.3\times10^{-4}$ & $1.7\times10^{-5}$ & $1.2\times10^{-3}$ \\
$\rho(w_0,w_a)$        & $6.2\times10^{-9}$ & $4.6\times10^{-7}$ & $1.8\times10^{-5}$ &$ 1.3\times10^{-3}$ & $7.9\times10^{-5}$ & $3.4\times10^{-3}$ \\
$\rho(\zeta,\Omega_m)$ & -0.057 & -0.095 & -0.089 & -0.080 & -0.095 & -0.067 \\
$\rho(\zeta,w_a)$      & -0.395 & -0.663 & -0.620 & -0.557 & -0.663 & -0.387 \\
$\rho(\Omega_m,w_a)$   & $-9.3\times10^{-5}$ & $-8.9\times10^{-3}$ & $-8.4\times10^{-4}$ & $-6.0\times10^{-2}$ & $-3.3\times10^{-3}$ & $-1.5\times10^{-1}$ \\
\hline
$\sigma(\zeta)$ & $3.8\times10^{-7}$ & $2.1\times10^{-6}$ & $2.4\times10^{-7}$ & $1.9\times10^{-6}$ & $2.2\times10^{-7}$ & $1.7\times10^{-6}$ \\
$\sigma(w_0)$   & 0.100 & 0.100 & 0.100 & 0.100 & 0.100 & 0.100 \\
$\sigma(w_a)$   & 0.300 & 0.285 & 0.299 & 0.214 & 0.294 & 0.137 \\
\hline
\end{tabular}
\caption{\label{tab10}Results of the Fisher Matrix analysis for the case of the CPL parametrization---see \protect\cite{AlphaFish} for further details. The first six lines show the correlation coefficients $\rho$ for each pair of parameters and the last three the one-sigma marginalized uncertainties for the coupling $\zeta$ and the dark energy equation of state parameters $w_0$ and $w_a$.}
\end{center}
\end{table*}

Now consider the case where an $\alpha$ variation does exist, corresponding to a non-zero fiducial value of the dimensionless coupling $\zeta$. In this case the marginalized sensitivity on the parameter $\zeta$ will be weakened due to its correlations with other parameters. On the other hand, the $\alpha$ measurements can themselves help in constraining the cosmological parameters. Here we only consider the CPL case, whose results are summarized in Table \ref{tab10}, referring the reader to \cite{AlphaFish} for a more detailed discussion as well as a comparison of the results obtained for the various fiducial models.

The strong anticorrelation between $\zeta$ and $w_0$ (which naturally is weaker for smaller values of the coupling) is confirmed by the analysis, as is a similar anticorrelation between $\zeta$ and $w_a$. On the other hand, the present-day value of the matter density is not significantly correlated with the other parameters. Overall, with the range of assumed couplings the ESPRESSO GTO measurements would detect a non-zero $\zeta$ at between one and two standard deviations, while the same observations with the foreseen ELT-HIRES would ensure a two-sigma detection. We also note that for the largest permissible values of the coupling, ELT-HIRES measurements can improve constraints on the dark energy equation of state $w_0$ by up to ten percent. In the case of the largest currently allowed value $\zeta=5\times10^{-6}$ ELT-HIRES observations of the ESPRESSO GTO sample would detect a non-zero $\zeta$ at the $99.7\%$ (3$\sigma$) confidence level.

It is particularly worthy of note that the two dark energy equation of state parameters, $w_0$ and $w_a$, are not significantly correlated. This occurs because measurements of $\alpha$ typically span a sufficiently large redshift range (in the case of the simulated data set under consideration, roughly $1<z<3$) to make the roles of both in the redshift dependence of $\alpha$ sufficiently distinct. The practical result of this is that in the case of large values of $\zeta$ these measurements can significantly improve constraints on $w_a$---by more than a factor of two for the case of ELT-HIRES, and by about $30\%$ for the ESPRESSO ideal scenario, in the case of a large coupling---see the last line in Table \ref{tab10}. Thus $\alpha$ measurements can ideally complement cosmological probes in mapping the behavior of dynamical dark energy.

Note that the above analysis is conservative in at least one sense: the sample of $\alpha$ measurements consisted only of the 14 measurements in the range $1<z<3$ foreseen for the fundamental physics part of the ESPRESSO GTO \cite{MSC}. This is to be compared to the 293 archival measurements of Webb {\it et al.}, in the approximate redshift range $0.5<z<4.2$; while the latter contains data gathered over a period of about ten years from two of the world's largest telescopes, and the 14 GTO targets were chosen on the grounds that they are the best currently known targets for these measurements (and are visible from the location of the VLT, at Cerro Paranal in Chile) it is clear that in a time scale of 5-10 years a significantly larger data set could be obtained, also including contributions from the other extremely large telescopes, TMT \cite{TMT} and GMT \cite{GMT}.

However, this is not all. Standard observables such as supernovae are of limited use as dark energy probes \cite{Maor,Upadhye}, both because they probe relatively low redshifts (at least at the present time---future facilities may be able to detect and characterize them at higher redshifts \cite{Hook}) and because to ultimately obtain the required cosmological parameters one effectively needs to take second derivatives of noisy data. A clear detection of varying $w(z)$ is crucial, given that we know that $w\sim -1$ today. Since the field is slow-rolling when dynamically important (once the acceleration epoch has started, close to the present day), a convincing detection of a varying w(z) will be tough at low redshift, and we must probe the deep matter era regime, where the dynamics of the hypothetical scalar field is fastest.

Varying fundamental couplings are ideal for probing scalar field dynamics beyond the domination regime \cite{Nunes}: as we saw, in the case of $\alpha$ such measurements can presently be made up to redshift $z=4.2$, and future facilities such as the ELT should be able to further extend this redshift range. Thus ALMA, ESPRESSO and ELT-HIRES can realize the prospect of a detailed characterization of dark energy properties all the way until $z=4$, and possibly beyond. This possibility was first discussed in \cite{PCA,Leite1}, and further explored in \cite{Leite2,GTO}. These works also studied synergies between these measurements and those of Type Ia supernovas. In the case of ELT-HIRES a reconstruction of the dark energy equation of state using quasar absorption lines can be more accurate than using supernova data, its key advantage being huge redshift lever arm. Importantly, these measurements have the additional key role of breaking degeneracies, when combined with more 'classical' probes, for constraining dynamical dark energy models. A case in point is that of ESA's Euclid mission \cite{Euclid}, as was studied in \cite{Calabrese}. These degeneracies are broken not necessarily because measurements of varying couplings are intrinsically more constraining (that regime will only ensue for sufficiently large samples, corresponding to large amounts of telescope time) but because the extended redshift lever arm effectively makes these data sets sensitive to different directions in the relevant parameter space.

In the interest of brevity we will only provide a simple 'straw man' illustration of how a reconstruction of the dark energy equation of state using measurements of the fine-structure constant $\alpha$ compares with a reconstruction using type Ia supernovas. This mostly follows \cite{Leite1}, whose analysis is based on PCA techniques, the formalism having first been described in \cite{PCA}. One should bear in mind that PCA is a non-parametric method for constraining the dark energy equation of state. In assessing its performance, one should not compare it to parametric methods. Indeed, no such comparison is possible (even in principle), since the two methods are addressing different questions. Instead one should compare it with another non-parametric reconstruction, and for our purposes with varying couplings the type Ia supernovae provide a relevant comparison.

One considers Class I quintessence type models, as discussed in Section \ref{darkside}, with the simplest (linear) coupling to the electromagnetic sector. Then the evolution of $\alpha$ is as given above, and in this case three fiducial forms for the equation of state parameter were considered
\begin{itemize}
\item a {\bf constant} one, which remains close to a cosmological constant throughout the probed redshift range
\begin{equation}
w_{c}(z)=-0.9,
\end{equation}
\item a {\bf step} one, which evolves towards a matter-like behavior by the highest redshifts probed
\begin{equation}
w_{s}(z)=-0.5+0.5 \tanh \left(z-1.5 \right),
\end{equation}
\item a {\bf bump} one, which  has non-trivial features over a limited redshift range, perhaps due to a low-redshift phase transition associated with the onset of acceleration \cite{Mortonson}
\begin{equation}
w_{b}(z)=-0.9+1.3 \exp{ \left[-\frac{(z-1.5)^2}{0.1} \right]}\,.
\end{equation}
\end{itemize}
At a phenomenological level, these therefore aim to describe the three qualitatively different interesting scenarios. A flat universe is assumed, and the matter density is fixed at $\Omega_m=0.3$. (This is a standard procedure in dark energy PCA studies, and this specific choice of $\Omega_m$ has a negligible effect on the main result of the analysis.) For simplicity it is also assumed that the $\alpha$ measurements are uniformly distributed in the redshift range under consideration.

\begin{table}
\begin{center}
\begin{tabular}{|c|c|c|}
\hline
Model & ESPRESSO & ELT-HIRES \\
\hline
Constant & 649.8 & 19.5 \\
Step & 2231.6 & 66.9 \\
Bump & 1420.1 & 42.6 \\
\hline
\end{tabular}
\caption{\label{tab11}Number of nights of telescope time needed to achieve, with $\alpha$ measurements uniformly spaced in redshift, an uncertainty in the best-determined PCA mode equal to that expected from a SNAP-like data set of 3000 Type Ia supernovas, for the ESPRESSO and ELT-HIRES spectrograph and the various fiducial models discussed in the text. Further details can be found in \protect\cite{Leite1}.}
\end{center}
\end{table}

Finally one assumes 20 PCA bins and $\alpha$ measurements uniformly distributed in the redshift range $0.5<z<4.0$, and estimates the number of observation nights needed to obtain the same sensitivity on the first PCA mode as `classical' data set of 3000 supernovas (in this case assumed to be uniformly distributed up to $z\sim1.7$). Unsurprisingly we find that this is not possible at all with current UVES data (and the same should apply to current spectrographs at Keck or Subaru), while our estimates for ESPRESSO and ELT-HIRES are listed in Table \ref{tab11}. We thus see that a few tens of nights are sufficient for ELT-HIRES: this would fit comfortably within a GTO program, further highlighting the key role that the ELT will be able to play on fundamental cosmology.

The table also shows a strong dependence of the required number of nights on the underlying fiducial model. This is mostly due to our assumption of a uniform redshift distribution of the $\alpha$ measurements (which was chosen partially out of simplicity but partially also precisely to flesh out this dependence). To a first approximation (ie, ignoring astrophysical factors such as the fact that not all QSOs have the same brightness) uniformly sampling the redshift range where measurements can be made turns out to be the ideal observational strategy for the case of a constant equation of state. On the other hand this is far from ideal in the case of the 'step' model where the equation of state of dark energy has larger deviations from $w=-1$ at higher redshifts (in that case a better strategy would be to spend more time at higher redshifts, where for a given instrumental sensitivity the larger variations can be detected at higher statistical significance). As expected the 'bump' case leads to intermediate values between the previous two fiducials. Further discussion of these points can be found in \cite{Leite1,Leite2}.

For ESPRESSO, something of the order of a thousand nights would be needed---a large but not unrealistic number as VLT time will become progressively 'cheaper' (and more focused on cutting-edge surveys) in the ELT era. In terms of cost, a back-of-the-envelope estimate would indicate comparable numbers in the two cases---something of order 60 MEuro, even including the cost of building a specific instrument. This is incomparably cheaper than any space-based facility.

The range of redshifts considered for the measurements also plays a role: since one will effectively be calculating derivatives of the $\alpha$ data (note that this is only first derivatives, to be contrasted with the case of type Ia supernovas where second derivatives are needed) one needs the range of redshifts probed to be as wide as possible. Since different transitions sensitive to $\alpha$ will fall within the range of the spectrograph at different redshifts, one also needs a spectrograph with a relatively broad wavelength coverage: specifically for UV/optical measurements, one ideally wants to start at the atmospheric cutoff. (This is an even more pressing problem for optical/UV measurements of $\mu$ or the CMB temperature since in that case the number of known targets is much smaller and the critical absorption lines fall on the blue part of the spectrum.) Indeed, should ESPRESSO confirm variations of $\alpha$, the construction of a UV/blue-optimized new generation high-resolution spectrograph for the VLT to map out the redshift dependence of $\alpha$ in combination with ELT-HIRES would be a compelling project.

Obviously, in addition to the reconstruction of the dark energy equation of state using fundamental couplings, supernovas and other cosmological observables will still provide reconstructions at lower redshifts on their own, so one can combine the two reconstructions, as discussed in \cite{PCA,GTO}. Alternatively, one can simply compare the two types of reconstructions, which will be a test of whether or not they are consistent with one another (at least in the intermediate redshift range where the two will overlap), and assuming that they are consistent one can also infer a posterior likelihood for the coupling $\zeta$, since the fundamental couplings reconstructions depends on it but the one based on supernovas doesn't. Finally, this reconstruction will also provide a prediction for the value of the redshift drift signal at various redshifts, enabling a model-independent consistency test with the ELT or the SKA (depending on the redshift in question), as discussed in the previous section.

\section{Conclusions\label{concl}}

Tests of the stability of fundamental couplings are crossing a threshold. The first Large Program dedicated to them is drawing to a close, and while its results were limited by the spectral distortions inherent to current spectrographs it provided key information on what constitutes a good target for these measurements and how to improve analysis pipelines. The lessons thus learned will be valuable as a new generation of high-resolution ultra-stable spectrographs becomes available. So far everyone agrees that nothing is varying at the few ppm level out to redshifts $z\sim4$, with weaker constraints at higher redshifts and somewhat stronger ones within the Galaxy ($z\sim0$). Local tests with atomic clocks also provide very tight constraints. Note that a few ppm constraint is already a very tight one: to give just two examples, it is stronger than the Cassini bound on the Eddington PPN parameter \cite{Cassini}, and as we discussed in Sections \ref{models} and \ref{darkside} it leads to indirect constraints on WEP violations that are about one order of magnitude more stringent that the local direct ones. Improvements in sensitivity of more than one order of magnitude may be foreseen in the coming years.

Whether these forthcoming measurements will lead to detections of variations or to improved null results, they will have important implications for cosmology as well as for fundamental physics. In the scenario where there are no $\alpha$ variations, ESPRESSO can improve current bounds on Weak Equivalence Principle violations by up to two orders of magnitude: such bounds would be stronger than those expected from the MICROSCOPE satellite. Similarly, constraints from the high-resolution spectrograph at the ELT should be competitive with those of the proposed STEP satellite (although in this case one should be mindful of the caveat that both facilities are currently still in early stages of development). Thus astrophysical and local tests will nicely complement each other: should one detect violations in astrophysical tests while local ones give null results at the same level of sensitivity, this could be an indication for scenarios with environmental dependencies, or screening mechanisms \cite{Screening}.

In the opposite case where $\alpha$ variation be detected, and quite apart from the direct implications (direct evidence of Einstein Equivalence Principle violation, falsifying the notion of gravity as a purely geometric phenomenon, and exhibiting the presence of a fifth interaction in nature \cite{Uzan}), one will be able to map and constrain additional dynamical degrees of freedom not only through the acceleration phase of the universe (using supernovae and other standard probes) but also deep in the matter era---out to redshift $z\sim4$, and possibly beyond. Indeed in the context of the ELT there is at least in principle no reason why similarly tight measurements can't be made well beyond $z=4$. This can be achieved either by doing the measurement in the infrared (though at present it is not clear what sensitivity can be achieved, due for example to contamination from telluric lines) or by using transitions whose lab wavelengths are shorter than 1600 A (the bottleneck here being that currently the wavelengths of these transitions are not well know in the laboratory).

Finally, let us again stress the crucial role of consistency tests when one is searching for new physics. Taken together, tests of the stability of fundamental couplings, the redshift drift and constraints on the temperature-redshift relation and the distance duality relation will provide a unique opportunity for a precision mapping of the dark side of the universe. The realization of the deep connections between these various seemingly unrelated observational probes is one of the major developments in the field in recent years. The ELT will enable further relevant tests, including tests of strong gravity around the galactic black hole \cite{Hair,Trippe}, which were not discussed in this review. Last but not least, this new probe of the dark universe has many interesting synergies with other facilities, particularly ALMA, Euclid and the SKA, some of which remain to be fully explored.

Let us conclude by stressing again that the observational evidence for the recent acceleration of the universe demonstrates that our canonical theories of cosmology and particle physics are incomplete---if not incorrect---and that new physics is out there, waiting to be discovered. This review has highlighted the key role of astrophysical and local tests of the stability of fundamental couplings in this quest for new physics, discussing both the main developments in the past few years and also the ones that can be foreseen in the coming years, enabled by forthcoming high-resolution ultra-stable spectrographs. As is often the case in science, the most exciting developments may well be the ones we can not foresee.

\ack

The idea behind such a review probably emerged after my plenary talk at COSMO-15 in Warsaw. I thank Leszek Roszkowski for his invitation to write it (and his patience with the time it took me to do it), as well as John Beacom, Joe Liske, Alessandro Marconi, Michael Murphy and Wim Ubachs who at various points expressed interest---and encouragement---in such a review.

Many interesting discussions with current and past members of the CAUP Dark Side team (Ana Catarina Leite, Ana Mafalda Monteiro, Ana Marta Pinho, Catarina Alves, David Corre, Duarte Magano, Fernando Moucherek, In\^es Mota, Jo\~ao Vilas Boas, Jos\'e Pedro Vieira, Lu\'{\i}s Ventura, Maria Ramos, Mariana Juli\~ao, Marvin Silva, Miguel Ferreira, Pauline Vielzeuf, Pedro Leal, Pedro Pedrosa, Pedro Vianez, Rui Alves, Sim\~ao Jo\~ao and Tom\'as Silva) and our Dark Side summer interns (Arnau Gusart, Catarina Rocha, Joan Sol\`a, J\'ulia L\'opez, Mar Pino, Maximilian von Wietersheim, Oriol Frigola and Patr\'{\i}cia Carreira) as well as with many other colleagues and collaborators in the work discussed herein (Erminia Calabrese, Gemma Luzzi, Hugo Messias, Joe Liske, John Webb, Luca Amendola, Matteo Martinelli, Michael Murphy, Nelson Nunes, Paolo Molaro, Ricardo G\'enova-Santos, Stefania Pandolfi, Stefano Cristiani, Tasos Avgoustidis and Wim Ubachs), have shaped my views on this subject and are gratefully acknowledged.

This work was done in the context of project PTDC/FIS/111725/2009, The Dark Side of the Universe, from FCT (Portugal). The author is also supported by an FCT Research Professorship, contract reference IF/00064/2012, funded by FCT/MCTES (Portugal) and POPH/FSE (EC).

\section*{References}
\bibliography{martins}

\end{document}